%% file: main.tex
\documentclass{vldb}
\usepackage{graphicx}
\usepackage{balance}  

\usepackage{times}

\usepackage{bbding}
\usepackage{latexsym,amssymb,amsmath}
\usepackage{xspace}
\usepackage{comment}
\usepackage{url}
\usepackage{enumerate}
\usepackage[defblank]{paralist}
\usepackage{graphicx}
\usepackage{subfigure}
\graphicspath{{images/}}
\usepackage{calc}
\usepackage{mathtools}
\usepackage{microtype}
\usepackage{tabularx} 
\usepackage{booktabs} 
\usepackage{empheq}
\usepackage{tikz}
\usepackage{hyperref}
\usepackage{threeparttable}
\usepackage{booktabs}
\usepackage{tabularx}
\usepackage{rotating}
\newtheorem{theorem}{Theorem}[section]
\newtheorem{exampleAux}{Example}[section]
\newenvironment{example}{\begin{exampleAux}\upshape}{\qedfull\end{exampleAux}}
\newtheorem{lemma}{Lemma}[section]

\usepackage{tikz}
\usetikzlibrary{shapes}
\usetikzlibrary{shapes.multipart}
\usetikzlibrary{positioning}
\usetikzlibrary{decorations}
\usetikzlibrary{decorations.pathmorphing} 
\usetikzlibrary{fit}					
\usetikzlibrary{backgrounds}	
\usetikzlibrary{matrix,arrows}
\usetikzlibrary{calc}
\usetikzlibrary{through}

\allowdisplaybreaks[1]

\input{macros}

\input{macros-specific}

\newcommand{\statew}{1.2 cm}
\newcommand{\stateh}{1.2 cm}
\newcommand{\displx}{1.7 cm}
\newcommand{\disply}{1.5 cm}

\begin{document}

\title{Verification of Relational Data-Centric Dynamic Systems with
  External Services
}

\numberofauthors{3} 
\author{
\alignauthor
Babak Bagheri Hariri\\
Diego Calvanese\\
Marco Montali\\
       \affaddr{Free Univ. of Bozen/Bolzano}\\
       \email{{\small lastname@inf.unibz.it}}
\alignauthor
Giuseppe De Giacomo\\
       \affaddr{Sapienza Universit\`a di Roma}\\
       \email{{\small degiacomo@dis.uniroma1.it}}
\alignauthor 
Alin Deutsch\\
       \affaddr{UC San Diego}\\
      \email{{\small deutsch@cs.ucsd.edu}}
}
\additionalauthors{Additional authors: John Smith (The Th{\o}rv{\"a}ld Group,
email: {\texttt{jsmith@affiliation.org}}) and Julius P.~Kumquat
(The Kumquat Consortium, email: {\texttt{jpkumquat@consortium.net}}).}
\date{30 July 1999}

\maketitle
\begin{abstract}
\input{abstract}

\end{abstract}




\input{introduction}
\input{dcds}

\input{verification}

\input{mula}

\input{mulp}

\input{deterministic}

\input{det-bounded}
\input{det-wa} 
\input{nondeterministic}

\input{discussion}

\input{relatedwork}

\input{conclusions}

\bibliographystyle{abbrv}
\bibliography{dcds-bib}

\newpage
\appendix

\input{appendix-figures}
\input{appendix-mula}

\input{appendix-mulp}

\input{appendix-det-verification-results}
\input{appendix-nondet-verification-results}

\input{appendix-discussion}

\input{appendix-example}

\end{document}

%% file: macros.tex
\def\qedfull{\hfill{\qedboxfull}}
\def\qedboxfull{\vrule height 5pt width 5pt depth 0pt}

\newcommand{\A}{\mathcal{A}} \newcommand{\B}{\mathcal{B}}
\newcommand{\C}{\mathcal{C}} \newcommand{\D}{\mathcal{D}}
\newcommand{\E}{\mathcal{E}} \newcommand{\F}{\mathcal{F}}
 \renewcommand{\H}{\mathcal{H}}
\newcommand{\I}{\mathcal{I}} 
 \renewcommand{\L}{\mathcal{L}}
\newcommand{\M}{\mathcal{M}} 
 \renewcommand{\P}{\mathcal{P}}
 \newcommand{\R}{\mathcal{R}}
\renewcommand{\S}{\mathcal{S}} \newcommand{\T}{\mathcal{T}}
 \newcommand{\V}{\mathcal{V}}

\newcommand{\mf}{\mathfrak}

\newcommand{\ra}{\rightarrow}

\newcommand{\set}[1]{\{#1\}}                      

\newcommand{\tup}[1]{\langle#1\rangle}            

\newcommand{\myi}{\emph{(i)}\xspace}
\newcommand{\myii}{\emph{(ii)}\xspace}

\newcommand{\vfo}{\ensuremath{v}}
\newcommand{\vso}{\ensuremath{V}}

\newcommand{\BOX}[1]{ [-] #1}
\newcommand{\DIAM}[1]{\langle - \rangle #1}

\newcommand{\limp}{\rightarrow}

\newcommand{\MOD}[1]{(#1)^\mf{A}}

\newcommand{\MODA}[1]{(#1)_{\vfo,\vso}^\mf{A}}   

\newcommand{\MODAfirst}[1]{(#1)_{\vfo}^\mf{A}}   

\newcommand{\MODAX}[2]{(#1)_{\vfo #2,\vso}^\mf{A}}   

\newcommand{\MODAZ}[2]{(#1)_{\vfo,\vso #2}^\mf{A}}   

\newcommand{\true}{\mathsf{true}}
\newcommand{\false}{\mathsf{false}}

\newcommand{\ans}[2][]{\mathit{ans}_{#1}(#2)}

\newcommand{\map}[2]{#1 \rightsquigarrow #2}
\newcommand{\carule}[2]{#1 \mapsto #2}

\newcommand{\actex}[1]{\mathsf{#1}}


%% file: macros-specific.tex
\newcommand{\CONST}{\C}
\newcommand{\adom}[1]{\textsc{adom}(#1)}

\newcommand{\FUNC}{\ensuremath{\F}\xspace}   

\newcommand{\HERBRAND}{\ensuremath{\mathbb{U}}\xspace}

\newcommand{\dl}{\ensuremath{\D}\xspace}
\newcommand{\schema}{\ensuremath{\R}\xspace}
\newcommand{\idb}{\ensuremath{\I_0}\xspace}
\newcommand{\pl}{\ensuremath{\P}\xspace}
\newcommand{\aset}{\ensuremath{\A}\xspace}
\newcommand{\rset}{\ensuremath{\varrho}\xspace}
\newcommand{\pos}[1]{\ensuremath{#1^+}\xspace}

\newcommand{\sys}{\ensuremath{\S}\xspace}
\newcommand{\dcds}{\mbox{DCDS}\xspace}
\newcommand{\rexec}[1]{\textsc{exec}_{#1}\xspace}

\newcommand{\muladom}{\ensuremath{\muL_{A}}\xspace}
\newcommand{\mulpers}{\ensuremath{\muL_{P}}\xspace}
\newcommand{\hmadom}{\ensuremath{\L_{A}}\xspace}
\newcommand{\hmpers}{\ensuremath{\L_{P}}\xspace}

\newcommand{\hbsim}{\approx}
\newcommand{\pbsim}{\sim}

\newcommand{\inadom}{\ensuremath{\textsc{live}}}

\newcommand{\restrict}[2]{\ensuremath{#1 \vert_{#2}}}

\newcommand{\prop}[1]{\ensuremath{\textsc{prop}(#1)}}

\newcommand{\muL}{\mu\L} 

\newcommand{\rmap}{\ensuremath{\M}\xspace}

\newcommand{\EC}{\ensuremath{\E}}       

\newcommand{\doo}[3]{\textsc{do}^{#1}_{#2}(#3)}
\newcommand{\effect}[1]{\textsc{effect}(#1)}

\newcommand{\RCALLC}{\textsc{serviceCalls}}

\newcommand{\pickVal}{\textsc{pickValue}}

\newcommand{\eSysStatus}{\ensuremath{\mathsf{Status}}\xspace}

\newcommand{\eTravel}{\ensuremath{\mathsf{Travel}}\xspace}
\newcommand{\eHotel}{\ensuremath{\mathsf{Hotel}}\xspace}
\newcommand{\eFlight}{\ensuremath{\mathsf{Flight}}\xspace}

\newcommand{\eId}{\ensuremath{\mathsf{id}}\xspace}
\newcommand{\eClientName}{\ensuremath{\mathsf{eName}}\xspace}
\newcommand{\eHotelName}{\ensuremath{\mathsf{hName}}\xspace}

\newcommand{\eTrId}{\ensuremath{\mathsf{trId}}\xspace}
\newcommand{\eDate}{\ensuremath{\mathsf{date}}\xspace}
\newcommand{\eStatus}{\ensuremath{\mathsf{status}}\xspace}

\newcommand{\ePrice}{\ensuremath{\mathsf{price}}\xspace}
\newcommand{\eCurrency}{\ensuremath{\mathsf{currency}}\xspace}
\newcommand{\eUSDPrice}{\ensuremath{\mathsf{priceInUSD}}\xspace}

\newcommand{\eFNumber}{\ensuremath{\mathsf{fNum}}\xspace}

\newcommand{\ePassed}{\ensuremath{\mathsf{passed}}\xspace}
\newcommand{\eRFR}{\textit{`readyForRequest'}}
\newcommand{\eRTV}{\textit{`readyToVerify'}}
\newcommand{\eRTU}{\textit{`readyToUpdate'}}
\newcommand{\eRC}{\textit{`requestConfirmed'}}

\newcommand{\ePhaseC}{\textit{`checkPrice'}\xspace}
\newcommand{\ePhaseD}{\textit{`checkTravel'}\xspace}

\newcommand{\eVarPrice}{\mathit{price}\xspace}
\newcommand{\eVarCurrency}{\mathit{currency}\xspace}

\newcommand{\eVarDate}{\mathit{date}\xspace}
\newcommand{\eVarPriceInUSD}{\mathit{priceInUSD}\xspace}

\newcommand{\escClientName}{\textsc{inEName}() \xspace}

\newcommand{\escHName}{\textsc{inHName}() \xspace}
\newcommand{\escHDate}{\textsc{inHDate}() \xspace}
\newcommand{\escHPricec}{\textsc{inHPrice}() \xspace}
\newcommand{\escHCurrency}{\textsc{inHCurrency}() \xspace}
\newcommand{\escHPUSD}{\textsc{inHPInUSD}() \xspace}
\newcommand{\escFDate}{\textsc{inFDate}() \xspace}
\newcommand{\escFNumber}{\textsc{inFNum}() \xspace}
\newcommand{\escFprice}{\textsc{inFPrice}() \xspace}
\newcommand{\escFlightCurrency}{\textsc{inFCurrency}() \xspace}
\newcommand{\escFPUSD}{\textsc{inFPUSD}() \xspace}

\newcommand{\escMakeDecision}[1]{\textsc{makeDecision}(#1) \xspace}

 \newcommand{\rsa}{\rightsquigarrow}

\newcommand{\escConvertAndCheck}{\textsc{convertAndCheck}}

\newcommand{\eAReimburseRequest}{\textit{InitiateRequest}\xspace}
\newcommand{\ePriliminaryCheck}{\textit{VerifyRequest}\xspace}
\newcommand{\eUpdateRequest}{\textit{UpdateRequest}\xspace}
\newcommand{\eAcceptRequest}{\textit{AcceptRequest}\xspace}

\newcommand{\eCheckPrice}{\textit{CheckPrice}\xspace}
\newcommand{\eCheckTravel}{\textit{CheckTravel}\xspace}

\newcommand{\cts}[1]{\ensuremath{\Upsilon_#1}\xspace}
\newcommand{\sts}[1]{\ensuremath{\Theta_#1}\xspace}
\newcommand{\ncts}[1]{\cts{#1}}
\newcommand{\nsts}[1]{\sts{#1}}

\renewcommand{\mf}[1]{\Upsilon}
\newcommand{\db}{\mathit{db}}

\newcolumntype{R}{>{\raggedleft\arraybackslash}X} 
\newcolumntype{L}{>{\raggedright\arraybackslash}X} 
\newcolumntype{C}{>{\centering\arraybackslash}X}

\newcommand{\scset}{\ensuremath{\mathbb{SC}}\xspace}

\newcommand{\domain}[1]{\ensuremath{\textsc{dom}(#1)}\xspace}
\newcommand{\image}[1]{\ensuremath{\textsc{im}(#1)}\xspace}

\usepackage{color}

\newcommand{\groundexec}[2]{{\textsc{evals}_{#1}(#2)}}
\newcommand{\nrexec}[1]{{\textsc{n-exec}_{#1}\xspace}}

\newcommand{\skolems}[1]{{\textsc{calls}({#1})}}
\newcommand{\returns}{{\textsc{evals}}}
\newcommand{\prunings}{{\textsc{prunings}}}
\newcommand{\arem}[1]{{\textcolor{blue}{\small\it [[{#1} --alin]]}}}

\newcommand{\mrem}[1]{{\textcolor{darkgreen}{\small\it [[{#1}
      --marco]]}}}
\newcommand{\eat}[1]{}
\newcommand{\ERP}{{\textsc{Rcycl}}}
\newcommand{\recyclable}{{\ensuremath \mathit{RecyclableValues}}}
\newcommand{\used}{{\ensuremath \mathit{UsedValues}}}
\newcommand{\visited}{{\ensuremath \mathit{Visited}}}
\newcommand{\fresh}{{\V}}
\newcommand{\trans}{\Longrightarrow}

\newcommand{\cstate}{\mbox{state}\xspace}

\definecolor{acsiorange}{rgb}{.957,.494,.102}
\definecolor{acsired}{rgb}{.996,0,0}
\definecolor{acsigreen}{rgb}{.502,.773,.212}
\definecolor{acsiblue}{rgb}{.169,.510,.788}

\definecolor{darkblue}{rgb}{0.2,0.2,0.6}
\definecolor{darkred}{rgb}{0.6,0.1,0.1}
\definecolor{darkgreen}{rgb}{0.2,0.6,0.2}

\tikzstyle{background}=[rectangle,
                                                fill=gray!10,
                                                inner sep=0.2cm,
                                                rounded corners=5mm]
\tikzstyle{backgroundR}=[rectangle,
                                                fill=gray!10,
                                                inner sep=0.2cm]

\tikzstyle{backgroundtight}=[rectangle,
                                                fill=gray!10,
                                                inner sep=0cm,
                                                rounded corners=5mm]

\def\tikzSimpleState[#1,#2,#3,#4];{ \node at (#3,#4) [ rectangle,
  rounded corners=3mm, thick, draw, minimum
  height=1cm, minimum width=1cm] (#1) { #2}; }

\def\tikzState[#1,#2,#3,#4,#5];{ \node at (#4,#5) [ rectangle split,
  rounded corners=3mm, rectangle split parts=2, thick, draw, minimum
  height=4cm] (#1) { #2
   \nodepart{second} #3}; }

\def\state[#1,#2,#3,#4,#5,#6];{\node at (#4,#5) [draw,rounded
  corners=3mm, thick] (#1) {#2}
}

%% file: abstract.tex
Data-centric dynamic systems are systems where both the process controlling the
dynamics and the manipulation of data are equally central. Recently such kinds
of systems are increasingly attracting the interest of the scientific
community, especially in their variant called artifact-centric business
processes.
In this paper we study verification of (first-order) $\mu$-calculus variants over
\emph{relational data-centric dynamic systems}, where data are represented by a
full-fledged relational database, and the process is described in terms of
atomic actions that evolve the database.  The execution of such actions may
involve calls to external services, providing fresh data inserted into the
system.  As a result such systems are typically infinite-state.
We show that verification is undecidable in general, and we isolate
notable cases, where decidability is achieved.
Specifically we start by considering service calls that return values
deterministically (depending only on passed parameters).  We show that
in a $\mu$-calculus variant that preserves knowledge of objects
appeared along a run we get decidability under the assumption that the
fresh data introduced along a run are bounded, though they might not
be bounded in the overall system.  In fact we tie such a result to a
notion related to weak acyclicity studied in data exchange.
Then, we move to nondeterministic services where the assumption of data bounded
run would result in a bound on the service calls that can be invoked during the
execution and hence would be too restrictive. So we investigate decidability
under the assumption that knowledge of objects is preserved only if they are
continuously present. 
We show that if infinitely many values occur in a run but do not accumulate in the same state,
then we get again decidability.  We give syntactic conditions to avoid
this accumulation through the novel notion of ``generate-recall acyclicity'',
which takes into consideration
that every service call activation generates new values that cannot be
accumulated indefinitely.

%% file: introduction.tex
\renewcommand{\arem}[1]{}
\renewcommand{\mrem}[1]{}

\arem{Without all my comments, we are within space budget (I even
  created a little bit of spare room for one paragraph, just in case
  :) ). To remove my comments, no need to go manually through each
  file. Just look at the first commented line of file
  ``introduction.tex'', and uncomment it.}

\arem{In case you need to create even more space tomorrow: cutting the
  abstract in half is a good candidate.  It is long for an abstract,
  redundant wrt introduction, and even contains cut-and-pasted text
  from intro, which is always best avoided. We need not list our
  results in the abstract, we can just keep the first half ending in
  ``... such systems are typically infinite state.''  Then we say that
  in this challenging setting, we study verification under two
  semantics for services, and for two expressive logics.}

\arem{I have cut the terms and categories. They are required for
  camera-ready, not for submission. This gains us significant space,
  as the Introduction section can then start in the first column.}

\section{Introduction}
\label{sec:introduction}
\arem{Guys, I have performed surgery on the intro in a few places.  I
  have tried to achieve two things: 1) most important, be more
  explicit about our not being finite-state and about our decidability
  results being non-obious; and 2) claim our results more
  comprehensively (I mention that the semantic conditions are
  undecidable and that we give sufficient syntactic conditions). I did
  so in a ``space-bounded'' way: if you disable comments, I am using
  the same space as before (even one line less).  To make it easy for
  you to undo my edits, I am placing in blue comments the original
  text I cut, and I am marking what I added.}

Data-centric dynamic systems ($\dcds$s) are systems where both the
process controlling the dynamics and the manipulated data are equally
central. Recently such kinds of systems are increasingly attracting
the interest of the scientific community.  In particular, the so
called artifact-centric approach to modeling business processes has
emerged, with the fundamental characteristic of considering both data
and processes as first-class citizens in service design and
analysis~\cite{Nigam03:artifacts,Hull2008:Artifact,DBLP:journals/debu/CohnH09,CGHS:ICSOC:09,DBLP:journals/ijcis/AalstBEW01,DBLP:conf/time/AbiteboulBGM09}.
This holistic view of data and processes together promises to avoid
the notorious discrepancy between data modeling and process modeling
of more traditional approaches that consider these two aspects
separately~\cite{BGHLS-2007:artifacts-analysis,Bhatt-2005:Artifacts-pharm}.

$\dcds$s are constituted by 
\myi  a \emph{data layer}, which is used to hold  the relevant
information to be manipulated by the system, and 
\myii a \emph{process layer} formed by the invokable \emph{(atomic)
  actions} and a process based on them. Such a process characterizes
the dynamic behavior of the system. Executing an action has effects on
the data manipulated by the system, on the process state, and on the
information exchanged with the external world.

$\dcds$s deeply challenge formal verification by requiring
simultaneous attention to both data and processes: indeed, on the one
hand they deal with full-fledged processes and require analysis in
terms of sophisticated temporal
properties~\cite{Clarke1999:ModelChecking}; on the other hand, the
presence of possibly unbounded data \arem{we don't need citation here:
  \cite{AbHV95}} makes the usual analysis based on model checking of
finite-state systems impossible in general, since, when data evolution
is taken into account, the whole system becomes infinite-state.

In this paper we study \emph{relational} $\dcds$s, where data are
represented by a full-fledged relational database, and the process is
described in terms of atomic actions that evolve the database.  The
execution of such actions may involve calls to external services,
providing fresh data inserted into the system.  As a result such
systems are infinite-state in general.
In particular, actions are characterized using conditional effects.
\arem{we used to call conditional effects post-conditions, but I
  dropped this, it isn't quite true, as we spend ample space in the
  discussion section to explain.}  Effects are specified using
first-order (FO) queries to extract from the current database the
objects we want to persist in the next state, and using conjunctive
queries on these objects to generate the facts that are true in the
next state. In addition, to finalize the next state we call external
services (function calls) that provide new information and objects
coming from the external world.
\arem{ As written, the statement below may be confusing to the
  uninitiated reader.  First, because the discussion on disjunction
  seems to come out of the blue, and without having the intuition of
  artifact post-conditions, I personally couldn't have followed it,
  having no clue what ``disjunctive information'' stands for.  Second,
  because for instance in artifact systems with post-conditions that
  are disjunctive, the next state after executing an action is ALSO a
  single relational database. Disjunction only says that there is
  another source of non-determinism, besides the non-determinism of
  choosing the fresh input values. Notice though that in \dcds, we can
  use the non-deterministic choice of action parameters --which is
  disjunctive-- to simulate the kind of disjunction we get in
  disjunctive artifact post-conditions.
  \\BEGIN\\
  ``This implies that we do not get disjunctive information as the
  effect of executing an action, though we get fresh values as the
  result of introducing new input from outside.
This form of effects assures that the resulting state after the
execution of an action is still a relational DB. This is an assumption
that can be made often, and in particular, in all those applications
in which we have complete information, possibly after calls to the
external world, about the result of executing an action.''\\
END\\
We also have complete information if we use under-specified
post-conditions a la artifacts.  We have complete information at *run*
time, when we get just a regular db.  We do not have complete
information at *specification* time.  Either we dedicate the
additional space to discuss this and avoid confusions, or we just
cut. I am for the latter.  I think that the above text, extended with
the points I am making above, is a digression that comes too early
here, breaks the flow of the story, and is anyway too subtle for an
introduction even if we had the space.  I suggest we drop it to
streamline the story, space considerations aside. If we really want to
keep it, then it belongs in the DCDS section, as a motivation of why
we pick this model.  Or maybe the Discussion section, because we keep
implicitly referring to post-conditions, so we are implicitly
comparing to the artifact model.}

On top of such a framework, we introduce powerful verification logics,
which are FO variants of $\mu$-calculus
\cite{DBLP:journals/jcss/LuckhamPP70,Park76:Mu,Emerson96,BrSt07}. $\mu$-calculus
is well known to be more expressive than virtually all temporal logics
used in verification, including CTL, LTL, CTL*, PDL, and many others.
\arem{ ``For this reason $\mu$-calculus is often used in theoretical
  investigation, though in applications simpler logics like CTL or LTL
  are preferred.''\\
This reads as if we are doing ivory-tower research on logics nobody
uses in practice. I propose to replace with below, which makes us look
better:} Our approach is remarkably robust: while it is common to use
simpler logics like CTL and LTL towards verification decidability, our
decidability results hold for significantly more expressive
$\mu$-calculus variants, and thus carry over to all these other
logics.
Our variants of $\mu$-calculus are based on first-order queries over
data in the states of the \dcds, and allow for first-order
quantification across states (within and across runs), though in a
controlled way.  No limitations whatsoever are instead put on the
fixpoint formulae, which are the key element of the $\mu$-calculus.

In particular we consider two variants of $\mu$-calculus. The first
variant is called $\muladom$, and requires that first-order
quantification across states be always bounded to the active domain of
the state where the quantification is evaluated. This quantification
mechanism indirectly preserves, at any point, knowledge of objects
that appeared in the history so far, even if they disappeared in the
meantime.
The second variant, called $\mulpers$, restricts the first-order
quantification in \muladom by requiring that only quantified object
that are still present in the current domain are of interest as we
move from one state to the next . That is, knowledge of objects is
preserved only if they are continuously present.
For these two logics we define novel notions of bisimulation, which we
exploit to prove our results.

We show that verification of both $\muladom$ and $\mulpers$ is
undecidable in general. In fact we get undecidability even ruling out
first-order quantification and branching time.  However we isolate two
notable decidable cases.
Specifically we start by considering service calls that return values
deterministically (depending only on passed parameters).  We show that
verification of $\muladom$ properties is decidable under the
assumption that the cardinality of fresh data introduced along each
run is bounded (\emph{run-bounded} $\dcds$s), though it need not be
bounded across runs.  \arem{added this. BEGIN:} Decidability is
therefore not obvious, given that the logic permits quantification
over values occurring across (potentially infinitely many) branching
run continuations.  Run-boundedness is a semantic property which we
show undecidable to check, but for which we propose a sufficient
syntactic condition related to the notion of weak acyclicity studied
in data exchange \cite{Kolaitis05:Exchange}. \arem{END.}
Then, we move to nondeterministic services where same-argument service
calls possibly return different values at different moments in time.
To exploit the results on run-bounded $\dcds$s in this case we would
have to limit the number of service calls that can be invoked during
the execution, which would be a too restrictive condition on the form
of $\dcds$s. So we focus on the above $\mulpers$ fragment of
$\muladom$.  \arem{Cut this because it is already said above: ``where
  first-order quantification ranges across states only over objects
  that remain continuously present.''}  We show that if infinitely
many values occur in a run but do not accumulate in the same state
(our system is then called \emph{state-bounded}) then $\mulpers$
verification is decidable.  \arem{added this. BEGIN:} This comes as a
pleasant surprise, given that when compared to run-boundedness,
state-boundedness permits an additional kind of data unboundedness
({\em within} the run, as opposed to only {\em across} runs).
State-boundedness is a semantic property as well, and we show that
checking it is undecidable.  \arem{END.}  We then give a novel
syntactic condition, ``generate-recall acyclicity'', which suffices to
enforce that if a service generates new values by being called an
unbounded number of times, then these values cannot be accumulated
(``recalled'') indefinitely.

 The rest of the paper is organized as follows.
 Sec.~\ref{sec:dcds} introduces (relational) \dcds's.
 Sec.~\ref{sec:verif} introduces verification of \dcds's and the two
 variants of $\mu$-calculus that we consider.
 Sec.~\ref{sec:determ} focus the analysis of \dcds's under the
 assumption that external service calls behave deterministically.
 Sec.~\ref{sec:nondet} consider the case in which external service
 calls behave nondeterministically.
 Sec.~\ref{sec:discussion} discusses the various notions introduced
 %
 Sec.~\ref{sec:related} reports on related work.
 Finally, Sec.~\ref{sec:conclusions} concludes the paper.
All proofs are given in the appendix, which also includes a full-fledged example of a \dcds.


%% file: dcds.tex
\section{Data-Centric Dynamic Systems}\label{sec:dcds}

In this section, we introduce the notion of \emph{(relational) data-centric dynamic
 system}, or simply \dcds. A \dcds is a pair $\sys=\tup{\dl,\pl}$
formed by two interacting layers: a \emph{data layer} $\dl$ and a \emph{process
layer} $\pl$  over it.
Intuitively, the data layer keeps all the data of
 interest, while the process layer modifies and evolves such data.  We
 keep the structure of both layers to the minimum,
in particular we do not
 distinguish between various possible components providing the data,
 nor those providing the subprocesses running concurrently.  Indeed
 the framework can be further detailed in several directions, while
 keeping the results obtained here (cf.\ Section~\ref{sec:discussion}).

\subsection{Data Layer}
\label{sec:datalayer}

The data layer represents the information of interest in our application.  It
is constituted by a relational schema $\schema$ equipped with equality
constraints%
\footnote{Other kinds of constraints    can also be included without affecting the
 results reported here (cf.\ Section~\ref{sec:discussion}).}
\mrem{In the footnote we mention denials, but now Alin has
broaden the discussion}
$\EC$, e.g., to state keys of relations, and an initial database
instance $\idb$, which conforms to the relational schema and the
equality constraints. The values stored in this database belong to a
predefined, possibly infinite, set $\CONST$ of constants. These
constants are interpreted as themselves, blurring the distinction
between constants and values. We will use the two terms
interchangeably.

Given a database instance $\I$, its active domain $\adom{\I}$ is the
subset of $\CONST$ such that $c\in\adom{\I}$ if and only if $c$ occurs in
$\I$. 

A \emph{data layer} is a tuple $\dl = \tup{\CONST,\schema,\EC,\idb}$ where:
\begin{compactitem}
\item $\CONST$ is a countably infinite set of constants/values.
\item $\schema =\set{R_1, \ldots, R_n}$ is a database schema, constituted by a
  finite set of relation schemas.
\item $\EC$ is a finite set  $\set{\EC_1, \dots, \EC_m}$ of
  equality constraints. Each $\EC_i$ has the form
  \[
    Q_i \ra \textstyle \bigwedge_{j=1,\dots, k} z_{ij}=y_{ij},
  \]
  where $Q_i$ is a domain independent FO query over $\schema$ using constants
  from the active domain 
 $\adom{\idb}$ of $\idb$ and whose free variables are
  $\vec{x}$, and $z_{ij}$ and $y_{ij}$ are either variables in $\vec{x}$ or
  constants in $\adom{\idb}$.\footnote{For convenience, and without loss of
   generality, we assume that all constants used inside formulae appear in
   $\idb$.}
\item \idb is a database instance that represents the initial state of the data
  layer, which conforms to the schema $\schema$ and \emph{satisfies} the
  constraints $\EC$: namely, for each constraint $Q_i \ra \bigwedge_{j=1,\dots,
   k} z_{ij}=y_{ij}$ and for each tuple (i.e., substitution for the free
  variables) $\theta \in \ans[]{Q_i,\I}$, it holds that
  $z_{ij}\theta=y_{ij}\theta$.%
  \footnote{We use the notation $t\theta$ (resp., $\varphi\theta$) to denote
   the term (resp., the formula) obtained by applying the substitution $\theta$
   to $t$ (resp., $\varphi$). Furthermore, given a FO query $Q$ and a database
   instance $\I$, the \emph{answer} $\ans{Q,\I}$ to $Q$ over $\I$
 is the set of assignments $\theta$ from the free variables of $Q$ to
the domain of $\I$, such that $\I\models Q\theta$. We treat $Q\theta$ as
a boolean query, and with some abuse of notation, we say
$\ans{Q\theta,\I} \equiv \true$ if and only if $\I\models Q\theta$.
}\mrem{I have added the notion of ans(Q,I)}
\end{compactitem}



\subsection{Process Layer}
\label{sec:processlayer}
The process layer constitutes the progression mechanism for the \dcds.
We assume that at every time the current instance of the data layer
can be arbitrarily queried, and can be updated through action
executions, possibly involving external service calls to get new
values form the environment.
Hence the process layer is composed of three main notions: actions,
which are the atomic progression steps for the data layer; external
services, which can be called during the execution of actions; and
processes, which are essentially nondeterministic programs that use
actions as atomic instructions.
While we require the execution of actions to be
sequential, we do not impose any such constraints on processes, which
in principle can be formed by several concurrent branches, including
fork, join, and so on. Concurrency is to be interpreted by
interleaving and hence reduced to nondeterminism, as often done in
formal verification \cite{Baier2008:ModelChecking,Emerson96}. There can be many ways to provide
the control flow specification for processes. Here we adopt a simple
rule-based mechanism, but our results can be
immediately generalized to any process formalism whose processes
control flow is finite-state. Notice that this does not imply that the
transition system associated to a process over the data layer is
finite-state as well, since the data manipulated in the data layer
may grow over time in an unbounded way.

Formally, a process layer $\pl$ over a data layer $\dl =
\tup{\CONST,\schema,\EC,\idb}$,  is a tuple
$\pl=\tup{\FUNC,\aset,\rset}$ where:
\begin{compactitem}
\item $\FUNC$ is a finite set of \emph{functions}, each representing the
  interface to an \emph{external service}. Such services can be called, and as
  a result the function is activated and the answer is produced. How the result
  is actually computed is \emph{unknown} to the \dcds since the services are
  indeed external.

\item $\aset$ is a finite set of \emph{actions}, whose execution progresses the
  data layer, and may involve external service calls.

\item $\rset$ is a finite set of \emph{condition-action rules} that form the
  specification of the overall \emph{process}, which tells at any moment which
  actions can be executed.
\end{compactitem}

An \emph{action} $\alpha\in\aset$ has the form
\[
  \actex{\alpha}(p_1,\ldots,p_n): \set{e_1,\ldots,e_m},
\]
where:
\begin{inparaenum}
\item[\myi] $\actex{\alpha}(p_1, \ldots, p_n)$ is the \emph{signature} of the action,
  constituted by a name $\actex{\alpha}$ and a sequence $p_1,\ldots,p_n$ of
  \emph{input parameters} that need to be substituted with values for the
  execution of the action, and
\item[\myii] $\set{e_1,\ldots,e_m}$, also denoted as $\effect{\alpha}$, is a set of
  \emph{effect specifications}, whose specified effects are assumed to take
  place simultaneously.
\end{inparaenum}
Each $e_i$ has the form
$\map{q_i^+\land   Q_i^-}{E_i}$,
where:
\begin{compactitem}
\item $q_i^+\land Q_i^-$ is a query over $\schema$ whose terms are variables
  $\vec{x}$, action parameters, and constants from $\adom{\idb}$.
  The query $q_i^+$ is a UCQ, and the query $Q_i^-$
  is an arbitrary FO formula whose free variables are included in
  those of $q_i^+$.  Intuitively, $q_i^+$ selects the tuples to
  instantiate the effect, and $Q_i^-$ filters away some of them.
\item $E_i$ is the effect, i.e., a set of facts for $\schema$, which
  includes as terms: terms in $\adom{\idb}$, input parameters, free
  variables of $q_i^+$, and in addition Skolem terms formed by
  applying a function $f\in\FUNC$ to one of the previous kinds of
  terms.  Such Skolem terms involving functions represent external service
  calls and are interpreted so as to return a value chosen by an external
  user/environment when executing the action.
\end{compactitem}
The \emph{process} $\rset$ is a finite set
of \emph{condition-action rules}, of the form $\carule{Q}{\alpha}$, where
$\alpha$ is an action in $\aset$ and $Q$ is a FO query over $\schema$ whose
free variables are exactly the parameters of $\alpha$, and whose other terms
can be either quantified variables or constants in $\adom{\idb}$.

For a detailed example of a \dcds we refer to
Appendix~\ref{ex:travelReimburseSystem}.

\subsection{Semantics via Transition System}

The semantics of a \dcds is defined in terms of a possibly infinite transition
system whose states are labeled by databases. Such a transition system
represents all possible computations that the process layer can do on the data
layer.  A transition system $\mf{}$ is a tuple of the form
$\tup{\Delta,\R,\Sigma,s_0,\db,{\Rightarrow}}$, where:
\begin{compactitem}
\item $\Delta$ is a countably infinite set of values;
\item $\R$ is a database schema;
\item $\Sigma$ is a set of states;
\item $s_0 \in \Sigma$ is the initial state;
\item $\db$ is a function that, given a state $s \in \Sigma$, returns the
  database associated to $s$, which is made up of values in $\Delta$ and
  conforms to $\R$;
\item ${\Rightarrow} \subseteq \Sigma\times\Sigma$ is a transition relation
  between pairs of states.
\end{compactitem}
In order to precisely build the transition system associated to a \dcds, we
need to better characterize the behavior of the external services, which are
called in the effects of actions. This is done in Sections~\ref{sec:determ}
and~\ref{sec:nondet}.

%% file: verification.tex
\section{Verification}\label{sec:verif}

To specify dynamic properties over
a \dcds, we use $\mu$-calculus \cite{Emerson96,Stirling2001:Modal,BrSt07},
one of the most powerful temporal logics for which model checking has
been investigated in the finite-state setting. Indeed, such a logic is
able to express both linear time logics such as LTL and PSL, and
branching time logics such as CTL and CTL* \cite{Clarke1999:ModelChecking}.
The main characteristic of $\mu$-calculus is the ability of expressing
directly least and greatest fixpoints of (predicate-transformer)
operators formed using formulae relating the current state to the next
one. By using such fixpoint constructs one can easily express
sophisticated properties defined by induction or co-induction. This is
the reason why virtually all logics used in verification can be
considered as fragments of $\mu$-calculus.
From a technical viewpoint, $\mu$-calculus separates local properties,
i.e., properties asserted on the current state or on states that are
immediate successors of the current one, and properties that talk about
states that are arbitrarily far away from the current one~\cite{BrSt07}. The
latter are expressed through the use of fixpoints.

In this work, we use a first-order variant of the $\mu$-calculus
\cite{Park76:Mu},  called $\mu\L$ and defined as follows:
\[ \Phi ::= Q \mid \lnot \Phi \mid \Phi_1 \land \Phi_2 \mid \exists
x.\Phi \mid \DIAM{\Phi} \mid Z \mid \mu Z.\Phi
\]
where $Q$ is a possibly open FO query, and $Z$ is a
second order predicate variable (of arity 0).
We make use of the following abbreviations: $\forall x.
\Phi = \neg (\exists x.\neg \Phi)$, $\Phi_1 \lor \Phi_2 =
\neg (\neg\Phi_1 \land \neg \Phi_2)$, $\BOX{\Phi} = \neg \DIAM{\neg
\Phi}$, and $\nu Z. \Phi = \neg \mu Z. \neg \Phi[Z/\neg Z]$.

As usual in $\mu$-calculus, formulae of the form $\mu Z.\Phi$ (and $\nu
Z.\Phi$) must obey to the \emph{syntactic monotonicity} of $\Phi$ wrt
$Z$, which states that every occurrence of the variable $Z$ in $\Phi$
must be within the scope of an even number of negation symbols.  This
ensures that the least fixpoint $\mu Z.\Phi$ (as well as the greatest fixpoint
$\nu Z.\Phi$) always exists.

Since $\mu\L$ also contains formulae with both individual and
predicate free variables, given a transition system $\mf{}$, we introduce an individual
variable valuation $\vfo$, i.e., a
mapping from individual variables $x$ to $\Delta$, and a
predicate variable valuation $\vso$, i.e., a mapping from predicate
variables $Z$ to subsets of $\Sigma$.
With these three notions in place, we assign meaning to formulae by associating
to $\mf{}$, $\vfo$,
and $\vso$ an \emph{extension function} $\MODA{\cdot}$, which maps
formulae to subsets of $\Sigma$.
Formally, the extension function $\MODA{\cdot}$ is defined inductively
as shown in Figure~\ref{fig:mula-semantics}.

\begin{figure}[t]
\begin{small}
\begin{align*}
\MODA{Q} = & \{s \in \Sigma\mid \ans{Q\vfo,\db(s)} \}\\
\MODA{\lnot \Phi} = & \Sigma - \MODA{\Phi} \\
\MODA{\Phi_1 \land \Phi_2} = &\MODA{\Phi_1}\cap\MODA{\Phi_2}\\
 \MODA{\exists x. \Phi} =& \{ s \in \Sigma \mid\exists t. t \in
\Delta\mbox{ and } s \in\MODAX{\Phi}{[x/t]}\}\\
 \MODA{\DIAM{\Phi}} = & \{s \in \Sigma\mid\exists s'.s \Rightarrow s' \text{ and } s' \in
\MODA{\Phi}\} \\
\MODA{Z} = & \vso(Z)\\
\MODA{\mu
Z.\Phi} = & \bigcap\{\S\subseteq\Sigma \mid \MODAZ{\Phi}{[Z/\S]}
\subseteq \S \}
\end{align*}\caption{\label{fig:mula-semantics} Semantics of $\mu\L$.}
\end{small}
\end{figure}

\begin{example}
An example of $\mu\L$ formula is:
%
\begin{equation}
  \label{formula:exists-n}
  \exists x_1,\ldots,x_n. \bigwedge_{i\neq j} x_i \neq x_j \land
  \bigwedge_{i\in\{1,\ldots,n\}}\mu Z. [Stud(x_i) \lor \DIAM{Z}]
\end{equation}
The formula asserts that there are at least $n$ distinct objects/values, each
of which eventually denotes a student along some execution path. Notice  that
the formula does not imply that all of these students will be in the same
state, nor that they will all occur in a single run. It only says that in the
entire transition systems there are (at least) $n$ distinct students.
\end{example}

When $\Phi$ is a closed formula, $\MODA{\Phi}$ depends neither on
$\vfo$ nor on $\vso$, and we denote the extension of $\Phi$ simply by
$\MOD{\Phi}$.  We say that a
closed formula $\Phi$ holds in a state $s \in \Sigma$ if $s \in
\MOD{\Phi}$. In this case, we write $\mf{},s \models \Phi$.
We say that a closed formula $\Phi$ holds in $\mf{}$, denoted by
$\mf{} \models \Phi$, if $\mf{},s_0\models \Phi$, where $s_0$ is the initial
state of $\mf{}$.
We call \emph{model checking} verifying whether $\mf{}\models \Phi$ holds.

In particular we are interested in formally verifying properties of a
\dcds.  Given the transition system $\cts{\sys}$ of a \dcds \sys and a
dynamic property $\Phi$ expressed in $\mu\L$,\footnote{We remind
  the reader that, without loss of generality, we assume that all
  constants used inside formulae $\Phi$ appear in the initial database
  instance of the \dcds.} we say that \sys
\emph{verifies} $\Phi$ if
\[
  \cts{\sys} \models \Phi.
\]
The challenging point is that $\cts{\sys}$ is in general-infinite
state, so we would like to devise a finite-state transition system which
is a faithful abstraction of $\cts{\sys}$, in the sense that it
preserves the truth value of all $\mu\L$ formulae.  Unfortunately,
this program is doomed right from the start if we insist on using
full $\mu\L$ as the verification formalism. Indeed formulae of the
form~\eqref{formula:exists-n} defeat any kind of finite-state
transition system.  So next we introduce two interesting sublogics
of $\mu\L$ that serve better our objective.

%% file: mula.tex
\subsection{History Preserving Mu-Calculus}

The first fragment of $\mu\L$ that we consider is $\muladom$, which is
characterized by the assumption that quantification over individuals
is restricted to individuals that are present in the current
database. To enforce such a restriction, we introduce a special
predicate $\inadom(x)$, which states that $x$ belongs to the current
active domain.
The logic \muladom is defined as follows:
\[ \Phi ::= Q \mid \lnot \Phi \mid \Phi_1 \land \Phi_2 \mid \exists
x.\inadom(x) \land \Phi \mid \DIAM{\Phi} \mid Z \mid \mu Z.\Phi
\]
We make use of the usual abbreviation, including $\forall x.\inadom(x) \limp
\Phi = \neg (\exists x.\inadom(x) \land \neg \Phi)$.
Formally, the extension function $\MODA{\cdot}$ is defined inductively
as in Figure~\ref{fig:mula-semantics}, with the new special predicate
$\inadom(x)$ interpreted as follows:
\[
\MODA{\inadom(x)} =  \{s \in \Sigma\mid x/d \in \vfo \text{ implies }
d \in \adom{\db(s)}\}
\]

\begin{example}
As an example, consider the following \muladom formula:
\begin{align*}
  \nu X.(\forall x. &\inadom(x) \land Stud(x)\limp\\
  &\mu Y. (\exists
  y. \inadom(y) \land Grad(x,y)
  \lor \DIAM{Y}) \land \BOX{X}),
\end{align*}
which states that, along every path, it is always true, for each
student $x$, that there exists an evolution that eventually leads to a
graduation of the student (with some final mark $y$).
\end{example}

We are going to show that under suitable conditions we can get a
faithful finite abstraction for a DCDS that preserves all formulae of
\muladom, and hence enables us in principle to use standard model
checking techniques.
Towards this goal, we introduce a notion of bisimulation that is
suitable for the kind of transition systems we consider here.  In
particular, we have to take into account that the two transition
systems are over different data domains, and hence we have to consider
the correspondence between the data in the two transition systems and
how such data evolve over time. To do so, we introduce the following
notions.

Given two domains $\Delta_1$ and $\Delta_2$, a \emph{partial
  bijection} $h$ between $\Delta_1$ and $\Delta_2$ is a bijection
between a subset of $\Delta_1$ and $\Delta_2$.  
Given a partial function $f:
  S\rightarrow S'$, we denote with $\domain{f}$ the domain of $f$,
  i.e., the set of elements in $S$ on which $f$ is defined, and with
  $\image{f}$ the image of $f$, i.e., the set of elements $s'$ in $S'$
  such that $s'=f(s)$ for some $s\in S$.
A partial bijection
$h'$ \emph{extends} $h$ if $\domain{h} \subseteq \domain{h'}$ (or
equivalently $\image{h} \subseteq \image{h'}$) and $h'(x) = h(x)$ for
all $x \in \domain{h}$ (or equivalently $h'^{-1}(y) = h^{-1}(y)$ for
all $y \in \image{h}$).
Let $db_1$
and $db_2$ be two databases over two domains $\Delta_1$ and $\Delta_2$
respectively, both conforming to the
same schema $\R$. We say that a partial bijection $h$ \emph{induces an
isomorphism} between $db_1$ and $db_2$ if $\adom{db_1}\subseteq
\domain{h} $, $\adom{db_2} \subseteq \image{h}$, and $h$ projected on
$\adom{db_1}$ is an isomorphism between $db_1$ and $db_2$.

Given two transition systems $\mf{}_1 =
\tup{\Delta_1,\R,\Sigma_1,s_{01},\db_1,\Rightarrow_1}$ and $\mf{}_2 =
\tup{\Delta_2,\R,\Sigma_2,s_{02},\db_2,\Rightarrow_2}$, and  the set   $H$  of partial bijections between
$\Delta_1$ and $\Delta_2$,  a \emph{history preserving bisimulation} between
$\mf{}_1$ and $\mf{}_2$ is a relation $\B \subseteq
  \Sigma_1 \times H \times\Sigma_2$ such that $\tup{s_1,h,s_2} \in \B$ implies that:
  \begin{enumerate}
   \item $h$ is a partial bijection between $\Delta_1$ and $\Delta_2$ that induces an isomorphism between $db_1(s_1)$ and $db_2(s_2)$;
 \item for each $s_1'$, if $s_1 \Rightarrow_1 s_1'$ then there is
    an $s_2'$ with $s_2 \Rightarrow_2 s_2'$ and a bijection
    $h'$ that extends $h$, such that $\tup{s_1',h',s_2'}\in\B$.
  \item for each $s_2'$, if $s_2 \Rightarrow_2 s_2'$ then there is
    an $s_1'$ with $s_1 \Rightarrow_1 s_1'$ and a bijection
    $h'$ that extends $h$, such that $\tup{s_1',h',s_2'}\in\B$.
 \end{enumerate}
A state $s_1 \in \Sigma_1$ is \emph{history preserving bisimilar} to $s_2 \in
  \Sigma_2$ \emph{wrt a partial bijection} $h$, written $s_1 \hbsim_h s_2$, if
  there exists a history preserving bisimulation $\B$ between $\mf{}_1$ and
  $\mf{}_2$ such that
  $\tup{s_1,h,s_2}\in\B$.
A state $s_1 \in \Sigma_1$ is \emph{history preserving bisimilar} to $s_2 \in
  \Sigma_2$, written $s_1 \hbsim s_2$, if there exists a partial bijection $h$
  and a history preserving bisimulation $\B$ between $\mf{}_1$ and
  $\mf{}_2$ such that
  $\tup{s_1,h,s_2}\in\B$.
A transition system $\mf{}_1$ is \emph{history preserving bisimilar} to
  $\mf{}_2$, written $\mf{}_1 \hbsim \mf{}_2$, if there exists a partial
  bijection $h_0$ and a history preserving bisimulation $\B$ between $\mf{}_1$
  and $\mf{}_2$ such that $\tup{s_{01},h_0,s_{02}}\in\B$.
The next theorem gives us the classical invariance result of $\mu$-calculus wrt
bisimulation, in our setting.
\begin{theorem}
\label{theorem:muladom-bisimulation}
Consider two transition systems $\mf{}_1$ and $\mf{}_2$ such that
$\mf{}_1 \hbsim \mf{}_2$.
Then for every $\muladom$ closed formula $\Phi$ , we have:
\[\mf{}_1 \models \Phi \textrm{ if and only if } \mf{}_2 \models \Phi. \]
\end{theorem}

%% file: mulp.tex
\subsection{Persistence Preserving Mu-Calculus}\label{sec:mulp}

The second fragment of $\mu\L$ that we consider is $\mulpers$, which further
restricts $\muladom$ by requiring that individuals over which we quantify must
continuously persist along the system evolution for the quantification to take
effect.

With a slight abuse of notation, in the following we write
$\inadom(x_1,\ldots,x_n) = \bigwedge_{i \in \{1,\ldots,n\}}
\inadom(x_i)$.

The logic $\mulpers$ is defined as follows:
\begin{align*}
  \Phi ::=\ & Q \mid \lnot \Phi \mid \Phi_1 \land \Phi_2 \mid \exists
  x. \inadom(x) \land \Phi \mid \DIAM{( \inadom(\vec{x}) \land
    \Phi)}  \mid \\ & \BOX{( \inadom(\vec{x}) \land \Phi)}\mid  Z \mid \mu
  Z.\Phi
\end{align*}
where $Q$ is a possibly open FO query, $Z$ is a
second order predicate variable, and the following assumption holds:
in 
$\DIAM{(\inadom(\vec{x}) \land \Phi)}$ and $\BOX{(\inadom(\vec{x}) \land \Phi)}$,
the variables $\vec{x}$ are exactly
the free variables of $\Phi$, with the proviso that we substitute to each
bounded predicate variable $Z$ in $\Phi$ its bounding formula $\mu Z.\Phi'$.
We  use the usual abbreviations, including:
$\DIAM{(\inadom(\vec{x}) \limp \Phi)} = \neg
\BOX{(\inadom(\vec{x}) \land \neg \Phi)}$ and $\BOX{(\inadom(\vec{x}) \limp \Phi)} = \neg
\DIAM{(\inadom(\vec{x}) \land \neg \Phi)}$.
Intuitively, the use of $\inadom(\cdot)$ in
\mulpers ensures that individuals are only
considered if they persist along the system evolution, while the
evaluation of a formula with individuals that are not present in the
current database trivially leads to false or true (depending on the
use of negation).

\begin{example}
Getting back to the example above, its variant in \mulpers is
\begin{align*}
\nu & X. (\forall x.\inadom(x) \land Stud(x)\limp {}\\
& \mu Y.(\exists y. \inadom(y) \land Grad(x,y)
\lor \DIAM{(\inadom(x) \land  Y)}) \land \BOX{X})
\end{align*}
which states that, along every path, it is always true, for each
student $x$, that there exists an evolution in which $x$ persists in
the database until she eventually graduates (with some final mark $y$).
Formula
\begin{align*}
\nu & X. (\forall x.\inadom(x) \land Stud(x)\limp {}\\
&\mu Y.(\exists
y. \inadom(y) \land Grad(x,y)
\lor \DIAM{(\inadom(x) \limp  Y)}) \land \BOX{X})
\end{align*}
instead states that, along every path, it is always true, for each
student $x$, that there exists an evolution in which either $x$ is not
persisted, or becomes eventually graduated (with final mark $y$).
\end{example}



The bisimulation relation that captures \mulpers is as follows.
Given two transition systems $\mf{}_1 =
\tup{\Delta_1,\R,\Sigma_1,s_{01},\db_1,\Rightarrow_1}$ and $\mf{}_2 =
\tup{\Delta_2,\R,\Sigma_2,s_{02},\db_2,\Rightarrow_2}$, and
the set $H$ of partial bijections between
$\Delta_1$ and $\Delta_2$, a \emph{persistence preserving bisimulation} between
$\mf{}_1$ and $\mf{}_2$ is a relation $\B \subseteq
\Sigma_1 \times H \times\Sigma_2$ such that $\tup{s_1,h,s_2} \in \B$ implies
that:
\begin{enumerate}
\item $h$ is an isomorphism between $db_1(s_1)$ and
  $db_2(s_2)$;\footnote{Notice that this implies $\domain{h} =
   \adom{\db_1(s_1)}$ and $\image{h} = \adom{\db_2(s_2)}$.}
\item for each $s_1'$, if $s_1 \Rightarrow_1 s_1'$ then there exists an $s_2'$
  with $s_2 \Rightarrow_2 s_2'$ and a bijection $h'$ that extends
  $\restrict{h}{\adom{db_1(s_1)} \cap \adom{db_1(s'_1)}}$, such that
  $\tup{s_1',h',s_2'}\in\B$;\footnote{ Given a set
   $D$, we denote by $\restrict{f}{D}$ the \emph{restriction} of $f$ to
   $D$, i.e., $\domain{\restrict{f}{D}} = \domain{f} \cap D$, and
   $\restrict{f}{D}(x) = f(x)$ for every $x \in \domain{f} \cap D$.}
\item for each $s_2'$, if $s_2 \Rightarrow_2 s_2'$ then there exists an $s_1'$
  with $s_1 \Rightarrow_1 s_1'$ and a bijection $h'$ that extends
  $\restrict{h}{\adom{db_1(s_1)} \cap \adom{db_1(s'_1)}}$, such that
  $\tup{s_1',h',s_2'}\in\B$.
\end{enumerate}
  We say that a state $s_1 \in \Sigma_1$ is \emph{persistence preserving bisimilar} to $s_2 \in
  \Sigma_2$ \emph{wrt a partial bijection} $h$, written $s_1 \pbsim_h s_2$, if
  there exists a persistence preserving bisimulation $\B$ between $\mf{}_1$ and
  $\mf{}_2$ such that
  $\tup{s_1,h,s_2}\in\B$.
A state $s_1 \in \Sigma_1$ is \emph{persistence preserving bisimilar} to $s_2 \in
  \Sigma_2$, written $s_1 \pbsim s_2$, if there exists a partial bijection $h$
  and a persistence preserving bisimulation $\B$ between $\mf{}_1$ and
  $\mf{}_2$ such that
  $\tup{s_1,h,s_2}\in\B$.
A transition system $\mf{}_1$ is \emph{persistence preserving bisimilar} to
$\mf{}_2$, written $\mf{}_1 \pbsim \mf{}_2$, if there exists a partial
bijection $h_0$ and a persistence preserving bisimulation $\B$ between
$\mf{}_1$ and $\mf{}_2$ such that $\tup{s_{01},h_0,s_{02}}\in\B$.
The next theorem shows the invariance of \mulpers under this notion of
bisimulation.
\begin{theorem}
\label{theorem:mulpers-bisimulation}
Consider two transition systems $\mf{}_1$ and $\mf{}_2$ such that
$\mf{}_1 \pbsim \mf{}_2$.
Then for every $\mulpers$ closed formula $\Phi$, we have:
\[\mf{}_1 \models \Phi \textrm{ if and only if } \mf{}_2 \models \Phi. \]
\end{theorem}

%% file: deterministic.tex
\section{Deterministic Services}\label{sec:determ}

Now we turn back to the semantics of $\dcds$s, and analyze them under
the assumption that external services behave
deterministically.  This means that the evaluation of
functions $f\in\FUNC$, representing the service interfaces in the
process layer, is independent from the moment in which the function is
called: whenever an external service is called twice with the same
parameters, it must return the same value. So, for example, if the
function invocation $f(a)$ returned $b$ at a certain time, then in all
successive moments the call $f(a)$ will return $b$ again. In
particular, \emph{stateless} services can be modeled with
deterministic service calls.

Under this characterization of the services we can now define the
transition system of a \dcds. We call such a transition system
``concrete'' transition system to avoid confusion with an ``abstract''
transition system that we are going to introduce for our verification
technique.

\subsection{Semantics}\label{sec:determ-cts}

Let $\S = \tup{\dl,\pl}$ be a \dcds with data layer $\dl =
\tup{\CONST,\schema,\EC,\idb}$ and process layer
$\pl=\tup{\FUNC,\aset,\rset}$.

First we focus on what is needed to characterize the states of the
concrete transition system.
One such state obviously needs to maintain the current instance of the data
layer. This instance is a database made up of constants in $\CONST$,
which conforms to the schema $\schema$ and satisfies the equality
constraints in $\EC$.
Together with the current instance, however, we also need to  remember
all answers we had so far when calling the external services.

To meet the requirement that service calls behave deterministically,
the states of the transition system keep track of all results of the
service calls made so far, in the form of equalities between Skolem
terms involving functions in $\FUNC$ and having as arguments constants
and returned values in $\CONST$.\footnote{Notice that, we have no
  knowledge of the specific functions adopted by the external
  services, and we simply assume that such functions return some value
  from $\CONST$.  We are going to have different executions of the
  system corresponding to each way to assign values to the Skolem
  terms representing the service calls.}
More precisely, we define the set of (Skolem terms representing)
service calls as $\scset = \{f(v_1,\ldots,v_n) \mid f/n \in \FUNC
\textrm{ and } \{v_1,\ldots,v_n\} \subseteq \CONST \}$, where
  $f/n$ stands for a function $f$ arity $n$.
Then we introduce a \emph{service call map}, which is a partial function
$\rmap:\scset\ra\CONST$.

Now we are ready to formally define states of the concrete transition system.
A  \emph{concrete state}, or simply \cstate, is a pair $\tup{\I, \rmap}$,
where $\I$  is a relational instance of $\schema$ over $\CONST$
satisfying each equality constraint in $\EC$, and \rmap is a  service call map.
The \emph{initial concrete state}  is $\tup{\idb, \emptyset}$.

Next we look at the result of executing an action in a \cstate.
For this it is convenient to denote the database instance $\rmap(E)$ obtained
by applying  a service call map $\rmap$ to a set  $E$ of facts including only
constants in $\CONST$ or  terms in $\domain{\rmap}$. Namely,  we define
$\rmap(E)$ as the application of $\rmap$ to all the terms appearing in $E$ where
constants are preserved. Formally,  
$\rmap(E) = \{R(c_1\ldots, c_n)  \mid R(t_1,\ldots,t_n)\in E  \mbox{
  and } c_i = t_i \mbox { if } t_i\in \CONST \mbox{ and }
\rmap(t_i)=c_i \mbox{ if } t_i\in \domain{\rmap} \mbox{ for } i\in\{1,\ldots,n\}\}$.


Let $\alpha$ be an action in $\aset$ of the form
$\actex{\alpha}(p_1,\ldots,p_m): \set{e_1,\ldots,e_m}$ with
$e_i=\map{q_i^+\land Q_i^-}{E_i}$. The parameters for $\alpha$ are
guarded by the condition-action rule $Q \mapsto \alpha$ in $\rset$. Let
$\sigma$ be a substitution for the input parameters $p_1,\ldots,p_m$
with values taken from $\CONST$.  We say that $\sigma$ is \emph{legal}
for $\alpha$ in state $\tup{\I,\rmap}$ if $\tup{p_1,\ldots,p_m}\sigma
\in \ans{Q,\I}$.
\\

\noindent
{\bf Concrete action execution.}
To capture what happens when $\alpha$ is executed in a \cstate using a
substitution $\sigma$ for its parameters, we
introduce a transition relation $\rexec{\sys}$ between
$\cstate$s, called \emph{concrete execution} of $\alpha\sigma$, such
that $\tup{\tup{\I,\rmap},\alpha\sigma,\tup{\I',\rmap'}}
\in \rexec{\sys}$ if the following holds:
\begin{compactenum} 
\item $\sigma$ is a legal parameter assignment for $\alpha$ in state
  $\tup{\I,\rmap}$,
\item $\rmap' = \RCALLC(\I, \alpha\sigma,\rmap)$,
\item $\I' = \rmap'(\doo{}{}{\I, \alpha\sigma})$, and
\item $\I'$ satisfies $\EC$,
\end{compactenum}
where $\doo{}{}{}$ and $\RCALLC()$ are defined as follows.
\[
  \doo{}{}{\I, \alpha\sigma} = \bigcup_{\map{q_i^+\land
    Q_i^-}{E_i}\in\effect{\alpha}\;}
  \bigcup_{\theta\in\ans[]{(q_i^+\land Q_i^-)\sigma,\I}} E_i\sigma\theta
\]
  applies the action
    $\alpha$ to $\I$, using $\sigma$ as the assignment for its
    parameters. The returned instance is the union of the results of
    applying the effects specifications $\effect{\alpha}$, where the
    result of each effect specification $\map{q_i^+\land Q_i^-}{E_i}$ is,
    in turn, the set of facts $E_i\sigma\theta$ obtained from
    $E_i\sigma$ grounded on all the assignments $\theta$ that satisfy
    the query $q_i^+\land Q_i^-$ over $\I$.
\[\begin{array}{l@{}l}
\RCALLC (\I, \alpha\sigma, \rmap) = {}\\
\qquad\rmap \cup \{t \mapsto \pickVal(\C)
\mid ~& t \textrm{ occurring in } \doo{}{}{\I, \alpha\sigma}\\
& \textrm{and not in } \domain{\rmap}\}
\end{array}
\]
nondeterministically
generates all possible values that can be returned by the service
calls, guaranteeing that external services behave in a deterministic
manner. More specifically, all the service calls already contained in
$\rmap$ are maintained, while new service calls are
nondeterministically bound to an arbitrary value $\pickVal(\CONST)$
taken from $\CONST$ (which will be the values assumed by such service
calls in $\rmap$ from now on in the execution).
\\

\noindent {\bf Concrete transition system.}
  The \emph{concrete transition system} $\cts{\sys}$ for $\sys$  is a
  possibly infinite-state transition system $\tup{\CONST, \R, \Sigma,
    s_0, \db,
    \Longrightarrow}$ where $s_0 = \tup{\idb,\emptyset}$ and $\db$ is
  such that $\db(\tup{\I,\rmap}) =\I$.
Specifically, we define by simultaneous induction $\Sigma$ and
$\Longrightarrow$ as the smallest sets satisfying the following properties:
\myi $s_0 \in \Sigma$; \myii if $\tup{\I,\rmap} \in \Sigma$ , then for
all
substitutions $\sigma$ for the input parameters
  of $\alpha$
and for every
  $\tup{\I',\rmap'}$ such that
  $\tup{\tup{\I,\rmap},\alpha\sigma,\tup{\I',\rmap'}}\in\rexec{\sys}$,
  we have $\tup{\I',\rmap'}\in\Sigma$ and $\tup{\I,\rmap} \Longrightarrow \tup{\I',\rmap'}$.



Intuitively, to define the concrete transition system of the \dcds
\sys we start from the initial \cstate $s_0=\tup{\idb,\emptyset}$, and for each rule
$\carule{Q}{\alpha}$ in $\pl$, we evaluate $Q$ over $\idb$, and calculate all $\cstate$s
$s$ such that $\tup{s_0,\alpha\sigma,s} \in
\rexec{\sys}$. 
Then we repeat the same steps considering each $s$,
and so on.
The computation of successor states can be done by
picking all the possible combinations of resulting values for the
newly introduced service calls, then checking if the successor
obtained for a combination satisfies the equality constraints,
filtering it away if this is not the case. It is worth noting that
when new service calls are considered, the successors can be
countably infinite.

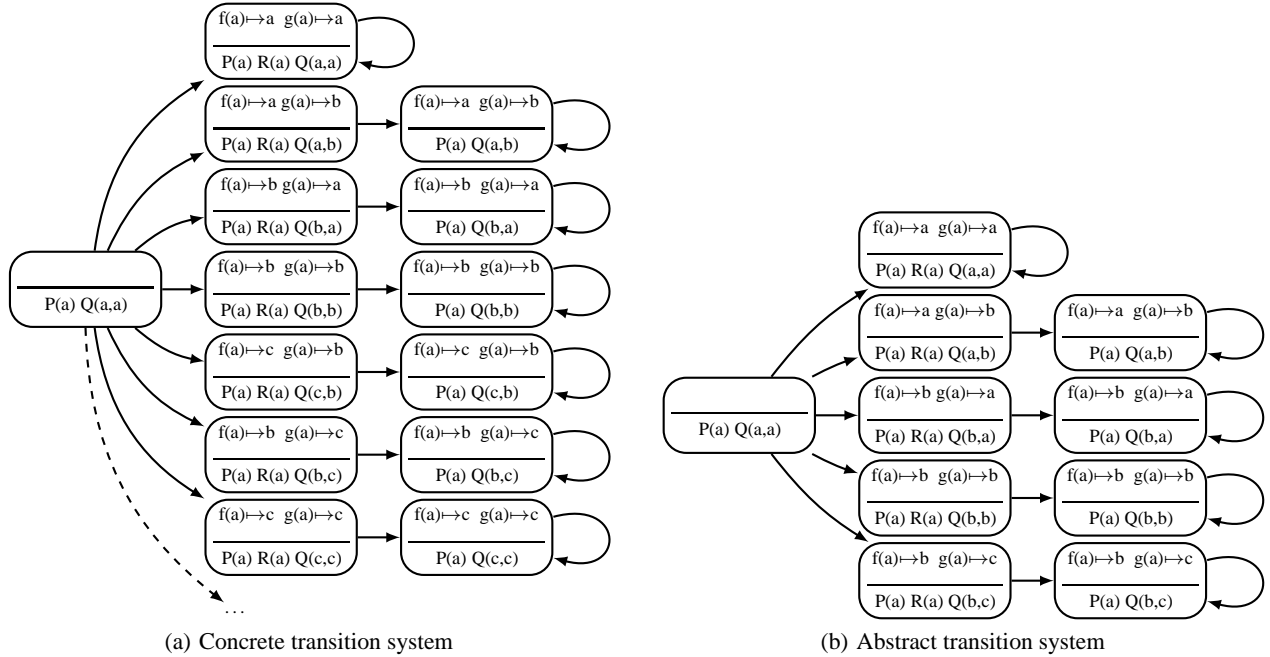
\begin{figure*}[!thb]
\centering
\subfigure[Concrete transition system\label{fig:openRS}]{\input{detCS}}
\subfigure[Abstract transition system\label{fig:openRS-abstract}]{\input{detAS}}
\caption{Concrete and abstract transition systems of the \dcds
  with deterministic services described in Example~\ref{ex:openruntime};
  special relations that store the service calls results are
  represented using a $call\mapsto value$ notation}
\end{figure*}

\begin{example}
\label{ex:openruntime}
Let $\sys = \tup{\dl,\pl}$ be a \dcds with data layer
$\dl=\tup{\CONST, \schema, \EC, \idb}$ and process layer $\pl=\tup{\FUNC,\aset,\rset}$,
where $\FUNC = \{f/1,g/1\}$, $\schema = \{Q/2, P/1,R/1\}$, $\EC = \emptyset$, $\idb =
\{P(a),Q(a,a)\}$, $ \rset = \{true
\mapsto \alpha\}$, $\aset = \{\alpha\}$, and
\begin{align*}
  \alpha: \{{Q(a,a) \land P(x)} \rightsquigarrow \{R(x)\},
  {P(x)} \rightsquigarrow{\{P(x),Q(f(x),g(x))\}} \}
\end{align*}
The concrete
transition system $\cts{\sys}$ contains infinitely many successors
connected to the initial \cstate.  These successors result from the
assignment of each possible pair of values to $f(a)$ and $g(a)$ (see
also Figure~\ref{fig:openRS}.
\end{example}

\begin{figure*}[!thb]
  \centering \subfigure[Concrete transition
  system\label{fig:openRS-ec}]{\input{detCS-ec}} \subfigure[Abstract
  transition system\label{fig:openRS-abstract-ec}]{\input{detAS-ec}}
\caption{Concrete and abstract transition systems of the \dcds with
  deterministic services described in Example~\ref{ex:openruntime-ec};
  special relations that store the service calls results are
  represented using a $call\mapsto value$ notation}
\end{figure*}

\begin{example}
\label{ex:openruntime-ec}
Consider a variation of the \dcds described in Example~\ref{ex:openruntime}, where the
data layer is equipped with an equality constraint, i.e., $\EC =
\{P(x) \land Q(y,z) \rightarrow x = y \}$.
The resulting concrete transition system has still infinitely many
successors of the initial \cstate, but the presence of the equality
constraint requires to keep only those successors in which $f(a)$ returns
$a$
(see also Figure~\ref{fig:openRS-ec}.
\end{example}

%% file: detCS-ec.tex
\renewcommand{\statew}{1.8 cm}
\renewcommand{\stateh}{1 cm}
\renewcommand{\displx}{2.5 cm}
\renewcommand{\disply}{1 cm}
\begin{tikzpicture}[node distance=.1 cm, >=latex, minimum height=\stateh,font=\scriptsize, rounded corners=3mm,thick]

\node at (0,0) [draw] (s0) {
\begin{minipage}{\statew}
\centering
~\\
~\\
\hrule
~\\
P(a)~Q(a,a)
\end{minipage}
};

\node [draw,right of=s0,xshift=\displx] (s2) {
\begin{minipage}{\statew}
\centering
f(a)$\mapsto$a~
g(a)$\mapsto$b\\
~\\
\hrule
~\\
P(a)~R(a)~Q(a,b)
\end{minipage}
};

\node [draw,above of=s2,yshift=\disply] (s1) {
\begin{minipage}{\statew}
\centering
f(a)$\mapsto$a~
g(a)$\mapsto$a\\
~\\
\hrule
~\\
P(a)~R(a)~Q(a,a)
\end{minipage}
};

\node [draw,below of=s2,yshift=-\disply] (s3) {
\begin{minipage}{\statew}
\centering
f(a)$\mapsto$a~
g(a)$\mapsto$c\\
~\\
\hrule
~\\
P(a)~R(a)~Q(a,c)
\end{minipage}
};


\node [draw, right of=s2,xshift=\displx] (s21) {
\begin{minipage}{\statew}
\centering
f(a)$\mapsto$a~
g(a)$\mapsto$b\\
~\\
\hrule
~\\
P(a)~Q(a,b)
\end{minipage}
};

\node [draw, right of=s3,xshift=\displx] (s31) {
\begin{minipage}{\statew}
\centering
f(a)$\mapsto$a~
g(a)$\mapsto$c\\
~\\
\hrule
~\\
P(a)~Q(a,c)
\end{minipage}
};

\node [below of=s0,xshift=2 cm,yshift=-2 cm, minimum height=0 cm] (dots) {\ldots};


\path
(s0) edge [->,bend left=10] (s1)
(s0) edge [->] (s2) 
(s0) edge [->, bend right=10]  (s3) 
(s0) edge [->, bend right=15, dashed] (dots)
(s1) edge [->,loop right, min distance=1cm] (s1)   
(s2) edge [->] (s21)  
(s21) edge [->,loop right, min distance=1cm] (s21)   
(s3) edge [->] (s31)  
(s31) edge [->,loop right, min distance=1cm] (s31);   
\end{tikzpicture}


%% file: detAS-ec.tex
\renewcommand{\statew}{1.8 cm}
\renewcommand{\stateh}{1 cm}
\renewcommand{\displx}{2.5 cm}
\renewcommand{\disply}{1 cm}
\begin{tikzpicture}[node distance=.1 cm, >=latex, minimum height=\stateh,font=\scriptsize, rounded corners=3mm,thick]

\node at (0,0) [draw] (s0) {
\begin{minipage}{\statew}
\centering
~\\
~\\
\hrule
~\\
P(a)~Q(a,a)
\end{minipage}
};

\node [draw,right of=s0,xshift=\displx,yshift=-.55 cm] (s2) {
\begin{minipage}{\statew}
\centering
f(a)$\mapsto$a~
g(a)$\mapsto$b\\
~\\
\hrule
~\\
P(a)~R(a)~Q(a,b)
\end{minipage}
};

\node [draw,right of=s0,xshift=\displx,yshift=.55 cm] (s1) {
\begin{minipage}{\statew}
\centering
f(a)$\mapsto$a~
g(a)$\mapsto$a\\
~\\
\hrule
~\\
P(a)~R(a)~Q(a,a)
\end{minipage}
};


\node [draw, right of=s2,xshift=\displx] (s21) {
\begin{minipage}{\statew}
\centering
f(a)$\mapsto$a~
g(a)$\mapsto$b\\
~\\
\hrule
~\\
P(a)~Q(a,b)
\end{minipage}
};


\path
(s0) edge [->] node [yshift=.2 cm] {} (s1)
(s0) edge [->] node [yshift=.2 cm] {} (s2) 
(s1) edge [->,loop right, min distance=1cm] (s1)   
(s2) edge [->] (s21)  
(s21) edge [->,loop right, min distance=1cm](s21); 
\end{tikzpicture}


%% file: detCS.tex
\renewcommand{\statew}{1.8 cm}
\renewcommand{\stateh}{1 cm}
\renewcommand{\displx}{2.5 cm}
\renewcommand{\disply}{1 cm}
\begin{tikzpicture}[node distance=.1 cm, >=latex, minimum height=\stateh,font=\scriptsize, rounded corners=3mm,thick]

\node at (0,0) [draw] (s0) {
\begin{minipage}{\statew}
\centering
~\\
~\\
\hrule
~\\
P(a)~Q(a,a)
\end{minipage}
};

\node [draw, right of=s0,xshift=\displx] (s4) {
\begin{minipage}{\statew}
\centering
f(a)$\mapsto$b~
g(a)$\mapsto$b\\
~\\
\hrule
~\\
P(a)~R(a)~Q(b,b)
\end{minipage}
};

\node [draw, above of=s4,yshift=\disply] (s3) {
\begin{minipage}{\statew}
\centering
f(a)$\mapsto$b~g(a)$\mapsto$a\\
~\\
\hrule
~\\
P(a)~R(a)~Q(b,a)
\end{minipage}
};

\node [draw,above of=s3,yshift=\disply] (s2) {
\begin{minipage}{\statew}
\centering
f(a)$\mapsto$a~g(a)$\mapsto$b\\
~\\
\hrule
~\\
P(a)~R(a)~Q(a,b)
\end{minipage}
};

\node [draw,above of=s2,yshift=\disply] (s1) {
\begin{minipage}{\statew}
\centering
f(a)$\mapsto$a~
g(a)$\mapsto$a\\
~\\
\hrule
~\\
P(a)~R(a)~Q(a,a)
\end{minipage}
};

\node [draw,below of=s4,yshift=-\disply] (s5) {
\begin{minipage}{\statew}
\centering
f(a)$\mapsto$c~
g(a)$\mapsto$b\\
~\\
\hrule
~\\
P(a)~R(a)~Q(c,b)
\end{minipage}
};

\node [draw,below of=s5,yshift=-\disply] (s6) {
\begin{minipage}{\statew}
\centering
f(a)$\mapsto$b~
g(a)$\mapsto$c\\
~\\
\hrule
~\\
P(a)~R(a)~Q(b,c)
\end{minipage}
};

\node [draw,below of=s6,yshift=-\disply] (s7) {
\begin{minipage}{\statew}
\centering
f(a)$\mapsto$c~
g(a)$\mapsto$c\\
~\\
\hrule
~\\
P(a)~R(a)~Q(c,c)
\end{minipage}
};


\node [draw, right of=s2,xshift=\displx] (s21) {
\begin{minipage}{\statew}
\centering
f(a)$\mapsto$a~
g(a)$\mapsto$b\\
~\\
\hrule
~\\
P(a)~Q(a,b)
\end{minipage}
};

\node [draw,below of=s21,yshift=-\disply] (s31) {
\begin{minipage}{\statew}
\centering
f(a)$\mapsto$b~
g(a)$\mapsto$a\\
~\\
\hrule
~\\
P(a)~Q(b,a)
\end{minipage}
};

\node [draw,below of=s31,yshift=-\disply] (s41) {
\begin{minipage}{\statew}
\centering
f(a)$\mapsto$b~
g(a)$\mapsto$b\\
~\\
\hrule
~\\
P(a)~Q(b,b)
\end{minipage}
};

\node [draw,below of=s41,yshift=-\disply] (s51) {
\begin{minipage}{\statew}
\centering
f(a)$\mapsto$c~
g(a)$\mapsto$b\\
~\\
\hrule
~\\
P(a)~Q(c,b)
\end{minipage}
};

\node [draw,below of=s51,yshift=-\disply] (s61) {
\begin{minipage}{\statew}
\centering
f(a)$\mapsto$b~
g(a)$\mapsto$c\\
~\\
\hrule
~\\
P(a)~Q(b,c)
\end{minipage}
};

\node [draw,below of=s61,yshift=-\disply] (s71) {
\begin{minipage}{\statew}
\centering
f(a)$\mapsto$c~
g(a)$\mapsto$c\\
~\\
\hrule
~\\
P(a)~Q(c,c)
\end{minipage}
};

\node [below of=s0,xshift=2 cm,yshift=-4.2 cm, minimum height=0 cm] (dots) {\ldots};


\path
(s0) edge [->,bend left=25] (s1)
(s0) edge [->,bend left=20] (s2) 
(s0) edge [->,bend left=15] (s3) 
(s0) edge [->] (s4) 
(s0) edge [->,bend left=-15] (s5) 
(s0) edge [->,bend left=-20] (s6) 
(s0) edge [->,bend left=-25] (s7) 
(s2) edge [->] (s21)  
(s3) edge [->] (s31)  
(s4) edge [->] (s41)  
(s5) edge [->] (s51)
(s6) edge [->] (s61)
(s7) edge [->] (s71)
(s1) edge [->,loop right,min distance=1cm] (s1)   
(s21) edge [->,loop right,min distance=1cm] (s21)   
(s31) edge [->,loop right,min distance=1cm]  (s31)   
(s41) edge [->,loop right,min distance=1cm] (s41)   
(s51) edge [->,loop right,min distance=1cm]  (s51)
(s61) edge [->,loop right,min distance=1cm]  (s61)
(s71) edge [->,loop right,min distance=1cm]  (s71)
(s0) edge [->,bend left=-25, dashed] (dots);
\end{tikzpicture}


%% file: detAS.tex
\renewcommand{\statew}{1.8 cm}
\renewcommand{\stateh}{1 cm}
\renewcommand{\displx}{2.5 cm}
\renewcommand{\disply}{1 cm}
\begin{tikzpicture}[node distance=.1 cm, >=latex, minimum height=\stateh,font=\scriptsize, rounded corners=3mm,thick]

\node at (0,0) [draw] (s0) {
\begin{minipage}{\statew}
\centering
~\\
~\\
\hrule
~\\
P(a)~Q(a,a)
\end{minipage}
};

\node [draw, right of=s0,xshift=\displx] (s3) {
\begin{minipage}{\statew}
\centering
f(a)$\mapsto$b~g(a)$\mapsto$a\\
~\\
\hrule
~\\
P(a)~R(a)~Q(b,a)
\end{minipage}
};

\node [draw,above of=s3,yshift=\disply] (s2) {
\begin{minipage}{\statew}
\centering
f(a)$\mapsto$a~g(a)$\mapsto$b\\
~\\
\hrule
~\\
P(a)~R(a)~Q(a,b)
\end{minipage}
};

\node [draw,above of=s2,yshift=\disply] (s1) {
\begin{minipage}{\statew}
\centering
f(a)$\mapsto$a~
g(a)$\mapsto$a\\
~\\
\hrule
~\\
P(a)~R(a)~Q(a,a)
\end{minipage}
};

\node [draw, below of=s3,yshift=-\disply] (s4) {
\begin{minipage}{\statew}
\centering
f(a)$\mapsto$b~
g(a)$\mapsto$b\\
~\\
\hrule
~\\
P(a)~R(a)~Q(b,b)
\end{minipage}
};

\node [draw,below of=s4,yshift=-\disply] (s5) {
\begin{minipage}{\statew}
\centering
f(a)$\mapsto$b~
g(a)$\mapsto$c\\
~\\
\hrule
~\\
P(a)~R(a)~Q(b,c)
\end{minipage}
};


\node [draw, right of=s2,xshift=\displx] (s21) {
\begin{minipage}{\statew}
\centering
f(a)$\mapsto$a~
g(a)$\mapsto$b\\
~\\
\hrule
~\\
P(a)~Q(a,b)
\end{minipage}
};

\node [draw,below of=s21,yshift=-\disply] (s31) {
\begin{minipage}{\statew}
\centering
f(a)$\mapsto$b~
g(a)$\mapsto$a\\
~\\
\hrule
~\\
P(a)~Q(b,a)
\end{minipage}
};

\node [draw,below of=s31,yshift=-\disply] (s41) {
\begin{minipage}{\statew}
\centering
f(a)$\mapsto$b~
g(a)$\mapsto$b\\
~\\
\hrule
~\\
P(a)~Q(b,b)
\end{minipage}
};

\node [draw,below of=s41,yshift=-\disply] (s51) {
\begin{minipage}{\statew}
\centering
f(a)$\mapsto$b~
g(a)$\mapsto$c\\
~\\
\hrule
~\\
P(a)~Q(b,c)
\end{minipage}
};


\path
(s0) edge [->,bend left=10] node [yshift=.2 cm] {} (s1)
(s0) edge [->,bend left=5] node [yshift=.2 cm] {} (s2) 
(s0) edge [->] node [yshift=.2 cm] {} (s3) 
(s0) edge [->,bend left=-5] node [yshift=-.2 cm] {} (s4) 
(s0) edge [->,bend left=-10] node [yshift=-.2 cm] {} (s5) 
(s1) edge [->,loop right,min distance=1cm] node {} (s1)   
(s2) edge [->] node [yshift=.2 cm] {} (s21)  
(s3) edge [->] node [yshift=.2 cm] {} (s31)  
(s4) edge [->] node [yshift=.2 cm] {} (s41)  
(s5) edge [->] node [yshift=.2 cm] {} (s51)
(s21) edge [->,loop right,min distance=1cm] node {} (s21)   
(s31) edge [->,loop right,min distance=1cm] node {} (s31)   
(s41) edge [->,loop right,min distance=1cm] node {} (s41)   
(s51) edge [->,loop right,min distance=1cm] node {} (s51);
\end{tikzpicture}


%% file: det-bounded.tex
\subsection{Run-Bounded Systems}
\label{sec:det-bounded}
We now study the verification of $\dcds$s with deterministic
services. In particular, we are interested in the following problem:
given a \dcds \sys and a temporal property $\Phi$, check whether $\cts{\sys} \models \Phi$. Not
surprisingly, given the expressive power of \dcds\ as a computation
model, the verification problem is undecidable for all the $\mu$-calculus
variants introduced in Section~\ref{sec:verif}.  In fact, we can show an
even stronger undecidability result, for a very small fragment of
propositional linear temporal logic (LTL)~\cite{Pnueli1977:LTL}, namely
the safety properties of the form $G p$ where $p$ is propositional.

\begin{theorem}
\label{thm:ltl-undecid-det}
There exists a \dcds\ \sys\ with deterministic services, and a
propositional LTL safety property $\Phi$,
such that checking $\cts{\sys} \models \Phi$ is undecidable.
\end{theorem}


 In the following, we isolate a notable class of
$\dcds$ for which verification of $\muladom$ is not
only decidable, but can also be reduced to standard model checking
techniques.

Consider a transition system $\mf{} = \tup{\Delta, \R, \Sigma, s_0, \db,
  \Rightarrow}$. A \emph{run} $\tau$ in $\mf{}$ is a (finite or infinite)
sequence of states $s_0s_1s_2\cdots$ rooted at $s_0$, where
$s_i \Rightarrow s_{i+1}$. We use $\tau(i)$ to
denote $s_i$ and $\tau[i]$ to represent the finite prefix $s_0 \cdots
s_i$ of $\tau$.
A run $\tau=s_0s_1s_2\cdots$ is \emph{(data) bounded} if the number of values mentioned
inside its databases is bounded, i.e., there exists a finite bound
$b$ such that $| \bigcup_{s \textrm{ state of } \tau} \adom{\db(s)}| < b$.
This is equivalent to saying that, for every finite prefix $\tau[i]$ of $\tau$,
$| \bigcup_{j \in \{0,\ldots,i\}} \adom{\db(s_j)}| < b$.
We say that $\mf{}$ is \emph{run-bounded} if there exists a bound $b$ such that every
run in $\mf{}$ is (data) bounded by $b$. A \dcds \sys is
run-bounded if its concrete transition system $\cts{\sys}$ is run-bounded.

Intuitively, a (data) unbounded run represents an execution of the \dcds in
which infinitely many distinct values occur because infinitely many different
service calls are issued. Since we model deterministic services whose number is
finite, this can only happen if some service is repeatedly called with
arguments that are the result of previous service calls. This means that the
values of the run indirectly depend on arbitrarily many states in the past.

Notice that run boundedness does not impose any restriction about the
branching of the transition system; in particular, $\cts{\sys}$ is
typically infinite-branching because new service calls may return any
possible value.
We show that this restriction guarantees decidability
for $\muladom$ verification of run-bounded $\dcds$s with deterministic services.

\begin{theorem}
\label{thm:decidability-det}
Verification of \muladom properties on run-bounded $\dcds$s with
deterministic services is decidable.
\end{theorem}

We get this result by showing that for run-bounded $\dcds$s there always
exists an abstract finite-state transition system that is history preserving
bisimilar to the concrete one, and hence satisfies the same
$\muladom$ formulae as the concrete transition system.

\begin{theorem}
\label{thm:finitestate-det}
For every run-bounded \dcds \sys with deterministic services, given its
concrete transition system $\cts{\sys}$ there exists an (abstract)
\emph{finite-state} transition system $\sts{\sys}$ such that $\sts{\sys}$ is
history preserving bisimilar to $\cts{\sys}$, i.e.,
$\sts{\sys}\hbsim\cts{\sys}$.
\end{theorem}
Let $\Sigma$ be the set of states of $\sts{\sys}$ and $\adom{\sts{\sys}} =
\bigcup_{s_i \in \Sigma} \adom{\db(s_i)}$.  If $\sts{\sys}$ is finite-state,
then there exists a bound $b$ such that $|\adom{\sts{\sys}}| <
b$. Consequently, it is possible to transform a \muladom property $\Phi$ into
an equivalent \emph{finite} propositional $\mu$-calculus formula $\prop{\Phi}$,
where $\prop{\Phi}$ is inductively defined over the structure of $\Phi$ as the
identity, except for the following case: $\prop{\exists x.\inadom(x)\land
 \Psi(x)} = \bigvee_{t_i \in \adom{\sys}} \inadom(t_i) \land \prop{\Psi(t_i)}$.
Clearly, $\sts{\sys} \models \Phi$ if and only if $\sts{\sys} \models
\prop{\Phi}$.

\begin{theorem}
\label{thm:propositional-det}
Verification of \muladom properties for run-bounded $\dcds$s with
deterministic services
can be reduced to conventional model checking of propositional
$\mu$-calculus over a finite transition system.
\end{theorem}
By the above theorem, and recalling that model checking of
propositional $\mu$-calculus formulae over finite transition systems
is decidable \cite{Emerson96}, we get
Theorem~\ref{thm:decidability-det}.\\

We conclude the Section by observing that the approach presented above for $\muladom$ does not extend to full
$\mu\L$.  

\begin{theorem}
\label{thm:faithful-abstraction}
There exists a \dcds \sys for which it is impossible to find a
faithful finite-state abstraction that satisfies the same $\muL$
properties as \sys.
\end{theorem}

The Theorem \ref{thm:faithful-abstraction} is proved by exhibiting,
for every $n$, a $\muL$ property that requires the existence of at least $n$ objects
in the transition system.

Even if this observation does not imply undecidability of model
checking $\mu\L$ properties over run-bounded $\dcds$s, it shows
that there is no hope of reducing this problem to standard,
finite-state model checking.



%
%
%
%


%% file: det-wa.tex
\subsection{Weakly Acyclic $\dcds$s}
\label{sec:det-wa}

The results presented in Section~\ref{sec:det-bounded} rely on the
hypothesis that the \dcds under study is run-bounded, which is a
semantic restriction. A natural question is whether it is
possible to check run-boundedness of a \dcds. We provide a negative
answer to this question.

\begin{theorem}
\label{thm:det-boundedness}
Checking run-boundedness of $\dcds$s with deterministic services is undecidable.
\end{theorem}
To mitigate this issue, we investigate a sufficient syntactic condition that can
be effectively tested over the process layer of the \dcds: if the
condition is met, then the \dcds is guaranteed to be run-bounded,
otherwise nothing can be said. To this end, we recast the approach of
\cite{Bagheri2011:Artifacts} in the more abstract and expressive
framework here presented.
In particular, we first introduce the ``positive approximate'' of a
\dcds, which abstracts away some of its aspects.
We do so for convenience, but we note that the definition of
weak-acyclicity as well as our results can be stated directly over the
original \dcds (in fact, we do so in condensed presentations of this work).
Technically, given a \dcds $\sys = \tup{\dl,\pl}$ with data layer
$\dl=\tup{\CONST, \schema, \EC, \idb}$ and process layer $\pl=\tup{\FUNC,\aset,\rset}$,
its \emph{positive approximate}  $\pos{\sys}$ is a \dcds
$\tup{\pos{\dl},\pos{\pl}}$, where $\pos{\dl}=\tup{\CONST, \schema,
  \emptyset, \idb}$  corresponds to
$\dl$ without equality constraints, while $\pos{\pl} =
\tup{\FUNC,\pos{\aset},\pos{\rset}}$ is a process layer whose actions
$\pos{\aset}$ and process $\pos{\rset}$ are obtained as follows:
\begin{itemize} \parskip=0in\itemsep=0in
\item Each condition-action rule $\carule{Q}{\alpha}$ in $\rset$ becomes
  $\carule{\true}{\pos{\alpha}}$ in $\pos{\rset}$.  Therefore, $\pos{\rset}$ is
  a process that supports the execution of every action in $\pos{\aset}$ at
  each step.
\item Each action $\alpha(p_1,\ldots,p_n) : \{e_1,\ldots,e_m\}$ in $\aset$
  becomes $\pos{\alpha} : \{\pos{e_1},\ldots,\pos{e_m}\}$ in $\pos{\aset}$,
  where each $e_i = \map{q_i^+\land Q_i^-}{E_i}$ becomes in turn $\pos{e_i} =
  \map{q_i^+}{E_i}$. Intuitively, the positive approximate action is obtained
  from the original action by removing all the parameters from its signature,
  and by removing all ``negative'' components from the query used to
  instantiate its effect specifications; note that the variables of $q_i^+$
  that were parameters in $\alpha$ are now free variables in $\pos{\alpha}$.
\end{itemize}
The positive approximate fulfils the following key property.
\begin{lemma}
\label{lemma:approximate-det}
  Given a \dcds \sys, if its positive approximate $\pos{\sys}$ is
  run-bounded, then \sys is run-bounded as well.
\end{lemma}
To derive a sufficient condition for $\pos{\sys}$ to be run-bounded, we can
exploit a strict correspondence between the execution of an action in
$\pos{\pl}$ and a step in the chase of a set of tuple generating dependencies
(TGDs) in data exchange\cite{AbHV95,Kolaitis05:Exchange}.
In particular, we resort to a well-known result in data exchange,  namely chase
  termination for \emph{weakly acyclic} TGDs
  \cite{Kolaitis05:Exchange}.\footnote{Notice that using other
    variants of weak acyclicity is also possible \cite{Meier2010:Optimization}.}

In our setting, the weak acyclicity of
 a process layer is a property over a dataflow graph constructed by
  analyzing the corresponding positive approximate process
  layer. A non-weakly acyclic \dcds contains a service that may
be repeatedly called, every time using fresh values that are
directly or indirectly obtained by manipulating previous results
produced by the same service. \mrem{Is ``manipulating previous results
of the same service'' clear? Mainpulating means copied or used to
generate other values by issuing service calls} This self-dependency can potentially
lead to an infinite number of calls of the same service along an
execution of the system, thus making it impossible to put a bound on
the data used throughout the run (see also Example~\ref{ex:nonwa}).
 Weak acyclicity rules out such self dependencies and is actually a
 sufficient condition for run-boundedness.

Given a \dcds $\sys = \tup{\dl,\pl}$ with positive
  approximate $\pos{\sys} = \tup{\pos{\dl},\pos{\pl}}$, the
  \emph{dependency graph} of $\pos{\pl}$ is an edge-labeled
directed graph $\tup{N,E}$ where: \myi $N \subseteq \schema \times \mathbb{N}^+$ is a set of nodes such
  that $\tup{R,i} \in N$ for every $R/n \in \schema$ and every $i \in
  \{1,\ldots,n\}$; \myii $E \subseteq N \times N \times \{\true,\false\}$ is a set of
  labeled edges where
  \begin{itemize}\parskip=0in\itemsep=0in
    \item an ordinary edge $\tup{\tup{R_1,j},\tup{R_2,k},\false} \in E$ if there exists an action
      $\pos{\alpha} \in \pos{\aset}$, an effect $\map{q_i^+}{E_i}
      \in \effect{\pos{\alpha}}$ and a variable $x$ such that
      $R_1(\ldots,t_{j-1},x,t_{j+1},\ldots)$ occurs in $q_i^+$ and
      $R_2(\ldots,t'_{k-1},x,t'_{k+i},\ldots)$ occurs in
      $E_i$;
    \item a special edge $\tup{\tup{R_1,j},\tup{R_2,k},\true} \in E$ if there exists an action
      $\pos{\alpha} \in \pos{\aset}$, an effect $\map{q_i^+}{E_i}
      \in \effect{\pos{\alpha}}$ and a variable $x$ such that
      $R_1(\ldots,t_{j-1},x,t_{j+1},\ldots)$ occurs in $q_i^+$,
      $R_2(\ldots,t'_{k-1},t,t'_{k+i},\ldots)$ occurs in $E_i$, and
      $t = f(\ldots,x,\ldots)$, with $f \in \FUNC$.
  \end{itemize}
$\pl$ is \emph{weakly acyclic} if the dependency
graph of its approximate $\pos{\pl}$ does not contain any cycle going
through a special edge. We say that a \dcds is weakly acyclic if its
process layer is weakly acyclic (e.g., see Figure~\ref{fig:dgraph-acyclic}).

Intuitively, ordinary edges represent the possible propagation (copy)
of a value across states: $\tup{\tup{R_1,j},\tup{R_2,k},\false} \in E$
reflects the possibility that the value currently stored inside the $j$-th
component of an $R_1$ tuple will be moved to the $k$-th component
of an $R_2$ tuple in the next state. Contrariwise, special edges
represent that a value can be taken as parameter of a service call,
thus contributing to the creation of (possibly new) values across
states: $\tup{\tup{R_1,j},\tup{R_2,k},\true} \in E$ means that the
value currently stored inside the $j$-th component of an $R_1$ tuple
could be used as parameter for a service call, whose result is then
stored inside the $k$-th component of an $R_2$ tuple.

\mrem{Similar to the paragraph I have introduced in the main file}
A cycle going through a special edge, forbidden by the weak acyclicity
condition, represents that a service may be repeatedly called, every
time using fresh values that are indirectly or directly obtained by
manipulating previous results produced by the same service. This
self-dependency can potentially lead to an infinite number of calls of
the same service along an execution of the system, thus making it
impossible to put a bound on the data used throughout the run.

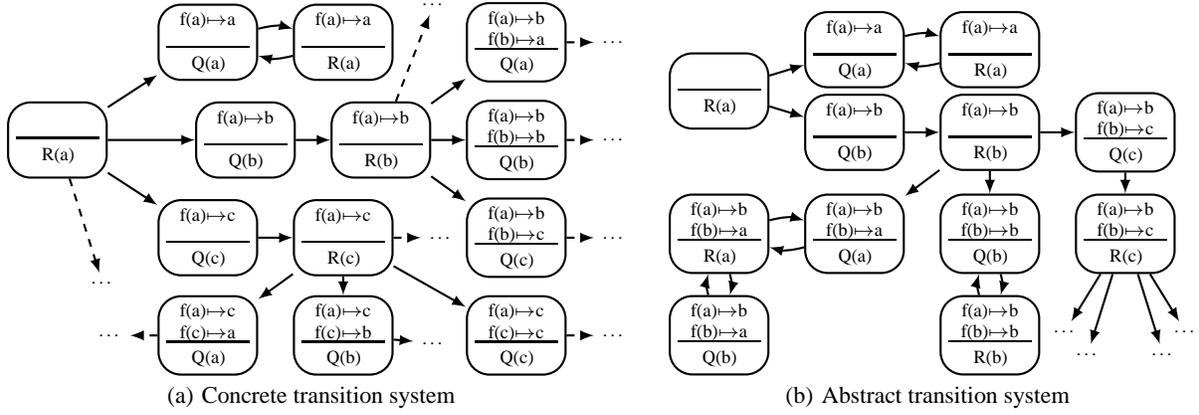
\begin{figure*}[!thb]
\centering
\subfigure[Concrete transition system\label{fig:nonwa-concrete}]{\input{detCS-nonwa}}
\subfigure[Abstract transition system\label{fig:nonwa-abstract}]{\input{detAS-nonwa}}
\caption{Concrete and abstract transition systems of the
  run-unbounded \dcds
  with deterministic services \sys described in Example~\ref{ex:nonwa};
  special relations that store the service calls results are
  represented using a $call\mapsto value$ notation}
\end{figure*}

\begin{example}
\label{ex:nonwa}
Let $\sys = \tup{\dl,\pl}$ be a \dcds with data layer
$\dl=\tup{\CONST, \schema, \emptyset, \idb}$ and process layer $\pl=\tup{\FUNC,\aset,\rset}$,
where $\FUNC = \{f/1\}$, $\schema = \{R/1, Q/1\}$, $\idb =
\{R(a)\}$, $ \rset = \{true
\mapsto \alpha\}$ and $\aset=\{\alpha\}$, where
$\alpha: ~ \{R(x) \rightsquigarrow{Q(f(x))},~Q(x)
\rightsquigarrow{R(x)} \}$. 

\sys is not weakly acyclic, due to the
mutual dependency between $R$ and $Q$ that involves a call to service
$f$. This can be easily seen from the dataflow graph (shown in Figure \ref{fig:dgraph-cyclic}), which contains
a special edge from $\tup{R,1}$ to $\tup{Q,1}$, and a normal edge from
$\tup{Q,1}$ to $\tup{R,1}$.
Notice that, in this case, the positive approximate of \sys coincides with \sys
itself. Starting from the initial state, $\alpha$ calls $f(a)$ and
stores the result inside $Q$. A second
execution of $\alpha$ transfers the result of $f(a)$ into $R$. When
$\alpha$ is executed for the third time, $f$ is called
again, but using as parameter the previously obtained
result. Consequently, $f$ may return a
new, fresh result, because $f(f(a))$ may be
different from $f(a)$. This chain can be repeated forever, leading to possibly generate infinitely many
distinct values along the run. The existence of a run in which $a$,
$f(a)$, $f(f(a))$, $f(f(f(a)))$, \ldots, are all distinct values,
makes it impossible to obtain a finite-state abstraction for \sys (see Figure \ref{fig:nonwa-abstract}).
\end{example}

\begin{theorem}
\label{thm:weak-acyclicity}
Every weakly acyclic \dcds with deterministic services is run-bounded.
\end{theorem}
\arem{
``The notion of weak acyclicity is independent from the initial active
domain and equality constraints of the \dcds, as well as from its process. Therefore, given a
weakly acyclic \dcds \sys, all variations of \sys that maintain the same set of
actions are weakly acyclic as well.''\\
I think this is confusing, as it requires the reader to recall that there is a difference
between the ``process layer'' (which contains the actions) and the ``process'' (which contains
the condition-action rules).  I predict he won't. WA is independent from process, but NOT from process layer.\\
Anyway, I propose we cut this because it is a subordinate point and we need the space for
higher-priority claims.}

Checking weak acyclicity is polynomial in the size of the \dcds.
Thus it gives us an effective way to verify $\dcds$s.
\begin{theorem}
\label{thm:synt-decidability-det}
Verification of \muladom properties for weakly acyclic $\dcds$s with
deterministic services is decidable, and can be reduced to model
checking of propositional $\mu$-calculus over a finite transition
system.
\end{theorem}

\begin{example}
Consider  the $\dcds$s described in
Example~\ref{ex:openruntime} and~\ref{ex:openruntime-ec}. They have the same
dataflow graph, which is weakly acyclic
(see Figure~\ref{fig:dgraph-acyclic}.
\begin{figure*}[!tb]

\subfigure[width=\columnwidth][Weakly acyclic dataflow graph for the $\dcds$s of Example~\ref{ex:openruntime} and~\ref{ex:openruntime-ec}
\label{fig:dgraph-acyclic}]
{
\begin{minipage}{\columnwidth}
\centering
\begin{tikzpicture}[node distance=.8 cm, >=latex,font=\scriptsize,
  rounded corners=3mm,thick]
\node [circle,draw] (P1) {P,1}; 
\node [circle,draw,left of=P1,xshift=-1cm] (R1) {R,1}; 
\node [circle,draw,right of=P1,xshift=1cm] (Q1) {Q,1}; 
\node [circle,draw,below of=Q1] (Q2) {Q,2}; 
\path
(P1) edge [->] (R1)
(P1) edge [->] node[auto] {*} (Q1)
(P1) edge [->] node[auto,yshift=-.5cm] {*} (Q2)
(P1) edge [->,loop below] (P1);
\end{tikzpicture}
\end{minipage}
}
\hfill
\subfigure[Non weakly
 acyclic dataflow graph for the \dcds of Example~\ref{ex:nonwa}\label{fig:dgraph-cyclic}]
{
\begin{minipage}{\columnwidth}
\centering
\begin{tikzpicture}[node distance=1 cm, >=latex,font=\scriptsize,
  rounded corners=3mm,thick]
\node [circle,draw] (R1) {R,1}; 
\node [circle,draw,right of=R1,xshift=1cm] (Q1) {Q,1};
\node [below of=Q1] (phantom) {};  
\path
(R1) edge [->,bend left=15] node[auto] {*} (Q1)
(Q1) edge [->,bend left=15]  (R1);
\end{tikzpicture}
\end{minipage}
}
\caption{Examples of dataflow graphs for $\dcds$s with deterministic
  services; special edges are decorated with *}
\end{figure*}
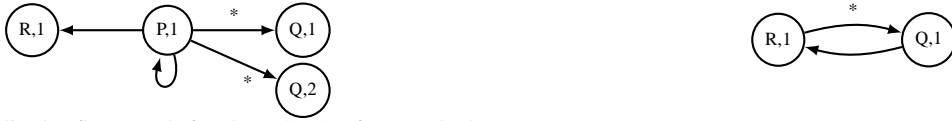
This guarantees that they are run-bounded  and that it is possible to find a faithful finite-state
abstraction from them. Two such abstractions are respectively shown in
Figure~\ref{fig:openRS-abstract} and \ref{fig:openRS-abstract-ec}.
\end{example}



%% file: detCS-nonwa.tex
\renewcommand{\statew}{1.1 cm}
\renewcommand{\stateh}{1 cm}
\renewcommand{\displx}{1.7 cm}
\renewcommand{\disply}{1.2 cm}

\begin{tikzpicture}[node distance=.1 cm, >=latex, minimum height=\stateh,font=\scriptsize, rounded corners=3mm,thick]

\node at (0,0) [draw] (s0) {
\begin{minipage}{\statew}
\centering
~\\
~\\
\hrule
~\\
R(a)
\end{minipage}
};

\node [draw, right of=s0,xshift=2.4cm] (s2) {
\begin{minipage}{\statew}
\centering
f(a)$\mapsto$b\\
~\\
\hrule
~\\
Q(b)
\end{minipage}
};

\node [draw,above of=s2,yshift=\disply,xshift=-.5cm] (s1) {
\begin{minipage}{\statew}
\centering
f(a)$\mapsto$a\\
~\\
\hrule
~\\
Q(a)
\end{minipage}
};

\node [draw, below of=s2,yshift=-\disply,xshift=-.5cm] (s3) {
\begin{minipage}{\statew}
\centering
f(a)$\mapsto$c\\
~\\
\hrule
~\\
Q(c)
\end{minipage}
};


\node [draw, right of=s1,xshift=\displx] (s11) {
\begin{minipage}{\statew}
\centering
f(a)$\mapsto$a\\
~\\
\hrule
~\\
R(a)
\end{minipage}
};

\node [draw,right of=s2,xshift=\displx] (s21) {
\begin{minipage}{\statew}
\centering
f(a)$\mapsto$b\\
~\\
\hrule
~\\
R(b)
\end{minipage}
};


\node [draw,right of=s21,xshift=\displx] (s212) {
\begin{minipage}{\statew}
\centering
f(a)$\mapsto$b\\
f(b)$\mapsto$b\\
\hrule
~\\
Q(b)
\end{minipage}
};

\node [draw,above of=s212,yshift=\disply] (s211) {
\begin{minipage}{\statew}
\centering
f(a)$\mapsto$b\\
f(b)$\mapsto$a\\
\hrule
~\\
Q(a)
\end{minipage}
};

\node [draw,below of=s212,yshift=-\disply] (s213) {
\begin{minipage}{\statew}
\centering
f(a)$\mapsto$b\\
f(b)$\mapsto$c\\
\hrule
~\\
 Q(c)
\end{minipage}
};

\node [draw,right of=s3,xshift=\displx] (s31) {
\begin{minipage}{\statew}
\centering
f(a)$\mapsto$c\\
~\\
\hrule
~\\
 R(c)
\end{minipage}
};

\node [draw,below of=s31,yshift=-\disply] (s312) {
\begin{minipage}{\statew}
\centering
f(a)$\mapsto$c\\
f(c)$\mapsto$b\\
\hrule
~\\
Q(b)
\end{minipage}
};

\node [draw,left of=s312,xshift=-\displx] (s311) {
\begin{minipage}{\statew}
\centering
f(a)$\mapsto$c\\
f(c)$\mapsto$a\\
\hrule
~\\
Q(a)
\end{minipage}
};

\node [draw,below of=s213,yshift=-\disply] (s313) {
\begin{minipage}{\statew}
\centering
f(a)$\mapsto$c\\
f(c)$\mapsto$c\\
\hrule
~\\
 Q(c)
\end{minipage}
};

\node [below of=s0,xshift=.6 cm,yshift=-1.8 cm, minimum height=0 cm] (dots0) {\ldots};

\node [right of=s21,xshift=.6 cm,yshift=1.8cm,minimum height=0 cm] (dots211) {\ldots};

\node [right of=s211,xshift=1.2cm,minimum height=0 cm] (dots2112) {\ldots};

\node [right of=s212,xshift=1.2cm,minimum height=0 cm] (dots2122) {\ldots};

\node [right of=s213,xshift=1.2cm,minimum height=0 cm] (dots2132) {\ldots};

\node [right of=s31,xshift=1.2 cm,yshift=0cm,minimum height=0 cm] (dots311) {\ldots};

\node [left of=s311,xshift=-1.2cm,minimum height=0 cm] (dots3112) {\ldots};

\node [below of=s312,xshift=1.2cm,yshift=-0cm,minimum height=0 cm] (dots3122) {\ldots};

\node [right of=s313,xshift=1.2cm,minimum height=0 cm] (dots3132) {\ldots};

\path
(s0) edge [->] (s1)
(s0) edge [->] (s2)
(s0) edge [->] (s3)
(s0) edge [->,dashed] (dots0)
(s1) edge [->,bend left=15] (s11)
(s11) edge [->,bend left=15] (s1)
(s2) edge [->] (s21)
(s21) edge [->] (s211)
(s21) edge [->] (s212)
(s21) edge [->] (s213)
(s21) edge [->,dashed] (dots211)
(s211) edge [->,dashed] (dots2112)
(s212) edge [->,dashed] (dots2122)
(s213) edge [->,dashed] (dots2132)
(s3) edge [->] (s31)
(s31) edge [->] (s311)
 (s31) edge [->] (s312)
(s31) edge [->] (s313)
(s31) edge [->,dashed] (dots311)
(s311) edge [->,dashed] (dots3112)
(s312) edge [->,dashed] (dots3122)
(s313) edge [->,dashed] (dots3132)
;

\end{tikzpicture}


%% file: detAS-nonwa.tex
\renewcommand{\statew}{1.1 cm}
\renewcommand{\stateh}{1 cm}
\renewcommand{\displx}{1.7 cm}
\renewcommand{\disply}{1.25 cm}

\begin{tikzpicture}[node distance=.1 cm, >=latex, minimum height=\stateh,font=\scriptsize, rounded corners=3mm,thick]

\node at (0,0) [draw] (s0) {
\begin{minipage}{\statew}
\centering
~\\
~\\
\hrule
~\\
R(a)
\end{minipage}
};

\node [draw, right of=s0,xshift=\displx,yshift=-.55cm] (s2) {
\begin{minipage}{\statew}
\centering
f(a)$\mapsto$b\\
~\\
\hrule
~\\
Q(b)
\end{minipage}
};

\node [draw,right of=s0,yshift=.55cm,xshift=\displx] (s1) {
\begin{minipage}{\statew}
\centering
f(a)$\mapsto$a\\
~\\
\hrule
~\\
Q(a)
\end{minipage}
};


\node [draw, right of=s1,xshift=\displx] (s11) {
\begin{minipage}{\statew}
\centering
f(a)$\mapsto$a\\
~\\
\hrule
~\\
R(a)
\end{minipage}
};

\node [draw,right of=s2,xshift=\displx] (s21) {
\begin{minipage}{\statew}
\centering
f(a)$\mapsto$b\\
~\\
\hrule
~\\
R(b)
\end{minipage}
};

\node [draw,below of=s2,yshift=-\disply] (s211) {
\begin{minipage}{\statew}
\centering
f(a)$\mapsto$b\\
 f(b)$\mapsto$a\\
\hrule
~\\
Q(a)
\end{minipage}
};

\node [draw,left of=s211,xshift=-\displx] (s2111) {
\begin{minipage}{\statew}
\centering
f(a)$\mapsto$b\\
f(b)$\mapsto$a\\
\hrule
~\\
R(a)
\end{minipage}
};

\node [draw,below of=s2111,yshift=-\disply] (s21111) {
\begin{minipage}{\statew}
\centering
f(a)$\mapsto$b\\
f(b)$\mapsto$a\\
\hrule
~\\
Q(b)
\end{minipage}
};

\node [draw,below of=s21,yshift=-\disply] (s212) {
\begin{minipage}{\statew}
\centering
f(a)$\mapsto$b\\
f(b)$\mapsto$b\\
\hrule
~\\
Q(b)
\end{minipage}
};

\node [draw,below of=s212,yshift=-\disply] (s2121) {
\begin{minipage}{\statew}
\centering
f(a)$\mapsto$b\\
f(b)$\mapsto$b\\
\hrule
~\\
R(b)
\end{minipage}
};

\node [draw,right of=s21,xshift=\displx] (s213) {
\begin{minipage}{\statew}
\centering
f(a)$\mapsto$b\\
f(b)$\mapsto$c\\
\hrule
~\\
Q(c)
\end{minipage}
};

\node [draw,below of=s213,yshift=-\disply] (s2131) {
\begin{minipage}{\statew}
\centering
f(a)$\mapsto$b\\
f(b)$\mapsto$c\\
\hrule
~\\
R(c)
\end{minipage}
};

\node [below of=s2131,yshift=-1.2cm,xshift=-.8cm,minimum height=0 cm] (dots21311) {\ldots};
\node [below of=s2131,yshift=-1.5cm,xshift=-.5cm, minimum height=0 cm] (dots21312) {\ldots};
\node [below of=s2131,yshift=-1.5cm,xshift=.5cm,minimum height=0 cm] (dots21313) {\ldots};
\node [below of=s2131,yshift=-1.2cm,xshift=.8cm,minimum height=0 cm] (dots21314) {\ldots};

\path
(s0) edge [->] (s1)
(s0) edge [->] (s2)
(s1) edge [->,bend left=15] (s11)
(s11) edge [->,bend left=15] (s1)
(s2) edge [->] (s21)
(s21) edge [->] (s211)
(s21) edge [->] (s212)
(s21) edge [->] (s213)
(s211) edge [->,bend left=15] (s2111)
(s2111) edge [->,bend left=15] (s211)
(s2111) edge [->,bend left=15] (s21111)
(s21111) edge [->,bend left=15] (s2111)
(s212) edge [->,bend left=15] (s2121)
(s2121) edge [->,bend left=15] (s212)
(s213) edge [->] (s2131)
(s2131) edge [->] (dots21311)
(s2131) edge [->] (dots21312)
(s2131) edge [->] (dots21313)
(s2131) edge [->] (dots21314)
;

\end{tikzpicture}


%% file: nondeterministic.tex
\newcommand{\iso}[1]{{\ensuremath \mapsto_{#1}}}
\newcommand{\id}{\mathit{id}\xspace}

\section{Nondeterministic Services}\label{sec:nondet}

We now consider $\dcds$s under the assumption that services behave nondeterministically,
i.e., two calls of a service with the same arguments may return distinct results during the same
run.
This case captures both services that model a truly nondeterministic process
(e.g., human operators, random processes), and services that model stateful servers.
In the remainder of this section, whenever we refer to a \dcds, services are implicitly
assumed nondeterministic.

\subsection{Semantics}
As in the case of deterministic services, we define the semantics of a \dcds\ $\sys$ in terms of a
(possibly infinite) transition system $\cts{\sys}$.

Let $\sys = \tup{\dl,\pl}$ be a \dcds with
data layer $\dl = \tup{\CONST,\schema,\EC,\idb}$ and
process layer $\pl=\tup{\FUNC,\aset,\rset}$.
A \emph{state} is simply a relational instance of $\schema$ over $\CONST$ satisfying each
constraint in $\EC$. We denote the \emph{initial state} with \idb.

Next, we define the semantics of action application.
Let $\alpha$ be an action in $\aset$
of the form $\actex{\alpha}(p_1,\ldots,p_m): \set{e_1,\ldots,e_m}$
with effects $e_i=\map{q_i^+\land Q_i^-}{E_i}$.  The parameters for $\alpha$ are
guarded by the condition-action rule $Q \mapsto \alpha \in \rset$. Let $\sigma$ be a
legal substitution for the input parameters $p_1,\ldots,p_m$ with values
taken from $\CONST$.

We reuse the definition of $\doo{}{}{\I, \alpha\sigma}$ from
Section~\ref{sec:determ-cts},
as the instance obtained by evaluating the effects of $\alpha$ on
instance $\I$.\mrem{In the det semantics we don't refer to it as an
  instance, but as a set of facts}
Recall that $\doo{}{}{}$ generates an instance over values from the
domain $\CONST$ but also over Skolem terms, which model service calls.
For any such instance $\bar \I$, we denote with
$\skolems{\bar \I}$ the set of calls it contains.
For a given set $D\subseteq\CONST$, we denote with
$\groundexec{D}{\I,\alpha,\sigma}$ the set of substitutions that replace all service calls
in $\doo{}{}{\I, \alpha,\sigma}$ with values in $D$,
\begin{tabbing}
$\groundexec{D}{\I,\alpha,\sigma}= \{ \theta\ |\ $ \=
			$\theta \mbox { is a total function }$\+\\
			$\theta: \skolems{\doo{}{}{\I, \alpha,\sigma}} \ra D \}$.
\end{tabbing}
Each substitution in $\groundexec{D}{\I,\alpha,\sigma}$
models the simultaneous evaluation of all service calls, which replaces the calls with results
selected nondeterministically from $D$. In the following, we refer to these substitutions
as \emph{evaluations}.\\

\noindent
{\bf Concrete action execution.\ }
We introduce a transition relation $\nrexec{\sys}$ between states,
called \emph{concrete execution} of $\alpha\sigma\theta$, such that 
$\tup{\I,\alpha\sigma\theta,\I'} \in \nrexec{\sys}$ if the following
holds:
 \begin{enumerate}\itemsep=0in\parskip=0in
 \item $\sigma$ is a legal parameter assignment for $\alpha$ in state $\I$,
 \item $\theta \in \groundexec{\CONST}{\I,\alpha,\sigma}$,
 \item $\I' = \doo{}{}{\I, \alpha,\sigma}\theta$, and
 \item $\I'$ satisfies the constraints $\EC$.
 \end{enumerate}
Notice that, in contrast to the deterministic services case,
the choice of evaluation $\theta$ is not subject to the requirement that it
evaluates a service call to the same result \emph{across} concrete execution steps.
However, notice that \emph{within} a concrete execution step,
all occurrences of the same service call evaluate to the same result
(modeling the fact that a call with given arguments is invoked only once per transition,
and the returned result is copied as needed).
\\

\noindent
{\bf Concrete transition system.\ }
The \emph{concrete transition system}
$\cts{\sys}$ for $\sys$ is a transition system
whose states are labeled by databases. More precisely,\\
$\cts{\sys} = \tup{\CONST, \schema, \Sigma, s_0, \db,\Rightarrow}$
where $s_0 = \idb$ and $\db$ is such that $\db(\I) = \I$.
$\Sigma$ and $\Rightarrow$ are defined by simultaneous induction as the smallest sets
satisfying the following properties: \myi $\idb \in \Sigma$; \myii if $\I \in \Sigma$ ,
then for all $\alpha$, $\sigma$, $\theta$ and $\I'$ such that
$\tup{\I,\alpha\sigma\theta,\I'}\in\nrexec{\sys}$, we have that $\I'\in\Sigma$, and
$\I \Rightarrow \I'$.

\subsection{State-Bounded Systems}
We consider the verification problem for \dcds with nondeterministic services.
As in the deterministic case, restrictions on both the processes and the properties are required,
motivated by the following undecidability result.

\begin{theorem}\label{thm:ltl-undecid}
There exists a \dcds\ \sys\ with nondeterministic services,
and a propositional LTL safety property $\Phi$,
such that checking $\cts{\sys} \models \Phi$ is undecidable.
\end{theorem}

\noindent
{\bf State-bounded \dcds.\ }\mrem{Should we also do this in the
  run-bounded section? Is it ``bounded'' DCDS or system?}
Since we are interested in verifying more expressive temporal properties,
we need to consider restricted classes of \dcds. We observe first that,
with nondeterministic services, the run-boundedness restriction of Section~\ref{sec:det-bounded}
is very limiting on the form of the \dcds, as it boils down to imposing a bound on how many times each service may
be called with the same arguments.
Observe that this was not the case for deterministic services, where the unlimited same-argument
calls are allowed, as they all return the same result. We propose a less restrictive alternative.
We say that \dcds\ \sys\ is \emph{state-bounded} if there is a finite bound $b$ such that
for each state $\I$ of $\cts{\sys}$, $|\adom{\I}| < b$.
Notice that, in contrast to the notion of run-boundedness, state-boundedness
does allow runs in which infinitely many distinct values occur because infinitely many
service calls are issued. The unboundedly many call results are distributed \emph{across}
states of the run, but may not accumulate \emph{within} a single state.
The following result shows that we also need to restrict the logic, as the one used in the
deterministic case is too expressive for decidability.

\begin{theorem}\label{thm:sbounded-undecid}
Verification of \muladom\ properties on state-bounded $\dcds$s with nondeterministic services
is undecidable.
\end{theorem}

We therefore restrict the property class to the logic $\mulpers \subset \muladom$
presented in Section~\ref{sec:mulp}.

\begin{theorem}\label{thm:sbounded-decid}
Verification of $\mulpers$ properties by state-bounded \dcds\ with nondeterministic services is
decidable.
\end{theorem}

\subsection{Abstract Transition System}\label{sec:nondet-abstract}
We relegate the proof of Theorem~\ref{thm:sbounded-decid} to Appendix~\ref{app:nondet-abstract},
but provide the main ideas here.

Given a \dcds \sys, we show that if concrete transition system \ncts{\sys} is state-bounded,
then there is a finite-state abstract transition system $\nsts{\sys}$ whose states and edges are subsets of
those in \ncts{\sys}, such that $\nsts{\sys}$ is persistence-preserving bisimilar to \ncts{\sys} (and hence satisfies the same
\mulpers properties, by Theorem~\ref{theorem:mulpers-bisimulation}).
Since $\nsts{\sys}$ is finite-state, the verification of \mulpers properties on \ncts{\sys}
reduces to finite-state model checking on $\nsts{\sys}$, and hence is decidable.

The existence of $\nsts{\sys}$ follows from the key fact that
if two states of \ncts{\sys} are isomorphic, then they are persistence-preserving bisimilar.
This implies that one can construct a finitely-branching transition system $\nsts{\sys}$
(i.e. with finite number of successors per state), such that $\nsts{\sys}$ is persistence-preserving
bisimilar to \ncts{\sys}, by dropping sibling states from \ncts{\sys} as follows:
instead of listing among the successors of $s$ one state for each possible instantiation of the
service call results, just keep a \emph{representative} state for each isomorphism type.
Since the number of service calls made in each state is finite, the number of distinct isomorphism
types is finite, so the finite branching follows. We call a transition
system $\nsts{\sys}$ obtained as above a
\emph{pruning of \ncts{\sys}}.

Notice that despite being finitely-branching, any pruning $\nsts{\sys}$ can
still have infinitely many states, as it may contain infinitely long simple
runs\footnote{We call a run \emph{simple} if no state appears more than once in
 the run.} $\tau$, along which the service calls return in each state ``fresh''
values, i.e., values distinct from all values appearing in the predecessors of
this state on $\tau$.  This problem is solved by judiciously selecting which
representatives to keep in $\nsts{\sys}$ for the successors of a state $s$.
Namely, whenever the representatives of a given isomorphism type $\T$ include
states generated exclusively by service calls that ``recycle'' values, select
only such states (finitely many thereof, of course). By recycled values we mean
values appearing on a path leading into $s$.

If \ncts{\sys} is state-bounded, then the number of service calls per state is bounded, and due to
the construction's preference for recycling, it follows that all simple runs in $\nsts{\sys}$ must
have finite length. Together with the finite branching, this implies finiteness of $\nsts{\sys}$.

Notice that proving the existence of $\nsts{\sys}$ does not suffice for decidability, as the proof
is non-constructive. We therefore provide an algorithm for constructing $\nsts{\sys}$ (Algorithm \ERP).
One of the technical problems we need to overcome in developing the algorithm is that we
evidently cannot start from the infinite-state concrete transition system, instead exploring a
portion thereof. This means that it is not obvious how to decide whether the successors of a state
are generated by recycling service calls, since these calls may recycle from paths that \ERP\
hasn't explored yet. Therefore, \ERP\ may sometimes select non-recycling service calls
even when a recycling alternative exists. However, we can prove that \ERP\ constructs
what we call an \emph{eventually recycling pruning}, which in essence means it may
fail to detect recycling service calls, but only a bounded number of times.

We formalize the above discussion in Appendix~\ref{app:nondet-abstract}, where we prove the
following result:

\begin{theorem}\label{thm:algo}
If input \dcds $\sys$ is state-bounded, then every possible run of Algorithm \ERP\ terminates,
yielding a finite eventually recycling pruning $\nsts{\sys}$ of $\cts{\sys}$, with $\cts{\sys} \pbsim \nsts{\sys}$.
\end{theorem}

Theorem~\ref{thm:algo} and Theorem~\ref{theorem:mulpers-bisimulation} directly imply
Theorem~\ref{thm:sbounded-decid}.

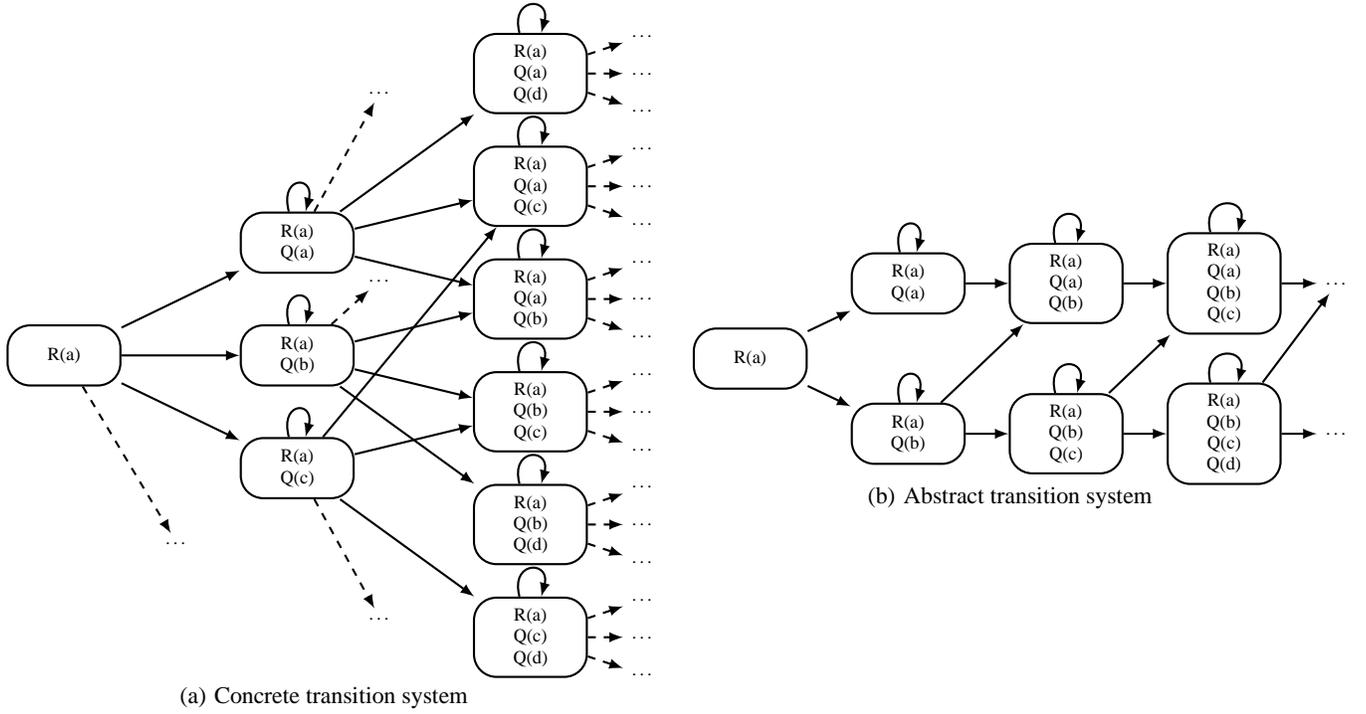
\begin{figure*}[!thb]
\centering
\subfigure[\label{fig:nondetCS-cyclic-copy}Concrete transition system]
{
\begin{minipage}{\columnwidth}
\centering
\input{nondetCS-cyclic-copy}
\end{minipage}
}
\hfill
\subfigure[\label{fig:nondetAS-cyclic-copy}Abstract transition system]
{
\begin{minipage}{\columnwidth}
\centering
\input{nondetAS-cyclic-copy}
\end{minipage}
}
\caption{Concrete and abstract transition systems of the
  state-unbounded \dcds
  with nondeterministic services of
  Example~\ref{ex:nondet-cyclic-copy}. \label{fig:nondet-cyclic-copy}}
\end{figure*}

Figures~\ref{fig:nondet-acyclic} and \ref{fig:nondet-cyclic-copy} illustrate two
concrete transition systems, and possible recycling prunings for them.

\subsection{GR-Acyclic $\dcds$s}

As with run-boundedness in the deterministic services case,
for nondeterministic services the state-boundedness restriction is a semantic property.
We investigate whether it can be effectively checked.
\begin{theorem}\label{thm:sbound-check}
Checking state-boundedness of $\dcds$s is undecidable.
\end{theorem}
Consequently we propose a sufficient syntactic restriction.

Intuitively, for a run to have unbounded states, it must issue unboundedly many service calls.
Since there are only a bounded number of effects in the process layer specification, there must
exist some service-calling effect that ``cyclically generates'' fresh values
(i.e. is invoked infinitely many times during the run).
Notice that unbounded generation of fresh values is insufficient for state-unboundedness:
these values must also accumulate in the states. But by definition of the \dcds\ semantics,
a transition drops (``forgets'') all values that are not explicitly copied (``recalled'')
into the successor.
Therefore, to accumulate, a value must be ``cyclically recalled'' througout the run
(it must be copied infinitely many times from relation to relation).
\mrem{It ``jumps'' here...}

GR-acyclicity is stated in terms of a dataflow graph constructed by analyzing the process layer.
The graph identifies how service calls and
value recalls can chain. In essence, GR-acyclicity requires the absence of a ``generate cycle''
that feeds into a ``recall cycle''.\\

\noindent
{\bf GR-acycliclicity.\ }
Let $\aset$ be a set of actions, and $\aset^+$ its positive approximate
(Section~\ref{sec:det-wa}).
We call \emph{dataflow graph} of $\aset$ the directed edge-labeled graph
$\tup{N,E}$ whose set $N$ of nodes is the set of relation names occurring in
$\aset$, and in which each edge in $E$ is a 4-tuple $(R_1,\id,R_2,b)$, where
$R_1$ and $R_2$ are two nodes in $N$, $\id$ is a (unique) edge identifier,
and $b$ is a boolean flag used to mark \emph{special} edges.  Formally, $E$
is the minimal set satisfing the following condition:
for each effect $e$ of $\aset^+$, each $R(t_1,\ldots,t_m)$ in the body of
$e$, each $Q(t'_1,\ldots,t'_{m'})$ in the head of $e$, and each $i\in\{1,\ldots,m'\}$:
\begin{itemize}\parskip=0in\itemsep=0in
 \item if $t'_i$ is either an element of
   $\adom{\idb}$ or a free variable, then $(R,\id,Q,\false)\in E$, where
   $\id$ is a fresh edge identifier.
 \item if $t'_i$ is a service call, then $(R,\id,Q,\true)\in E$, where
   $\id$ is a fresh edge identifier.
\end{itemize}
We say that $\aset$ is \emph{GR-acyclic} if there is no path
$\pi=\pi_1\pi_2\pi_3$ in the dataflow graph of $\aset$,
such that $\pi_1,\pi_3$ are simple cycles and
$\pi_2$ is a path containing a special edge that is
disjoint from the edges of $\pi_1$.
We say that a process layer $\pl=\tup{\FUNC,\aset,\rset}$ is GR-acyclic,
if $\aset$ is GR-acyclic. We call a \dcds GR-acyclic if its process layer is GR-acyclic.

Notice that GR-acyclicity is a purely syntactic notion. Moreover, it can be checked in
PTIME since the dataflow graph has size polynomial in the size of the process layer specification.

\begin{theorem}\label{thm:gracyclic-sbounded}
Any GR-acyclic \dcds\ is state-bounded.
\end{theorem}

We show the proof in Appendix~\ref{app:gr-acyclic} but provide some intuition
here, noting that the dataflow analysis is significantly more subtle than suggested above.

First, note that ordinary edges correspond to an effect copying a value from a relation
of the current state to a relation of the successor state.
Special edges correspond to feeding a value of the current state to a service call and
storing the result in a relation of the successor state.
Note that the cycles $\pi_1$ and $\pi_3$ allow both kinds of edges, reflecting the
insight that the size of the state is affected in the same way regardless of
whether a value is \emph{copied} to
the successor, or it is \emph{replaced} with a service call result
(see Example~\ref{ex:nondet-cyclic-copy} and Example~\ref{ex:nondet-cyclic-replace}
for illustrations of state-unboundedness arising
from each case).
$\pi_1,\pi_3$ are both ``recall cycles'': the number of values moving around them does not
decrease (this is of course a conservative statement; reality depends on the semantics of
queries in the effects, which is abstracted away).
Note that $\pi_2$ contains a special edge $E$, which means that the values moving
around $\pi_1$ are cyclically fed into the service call $f$ of $E$.
\arem{I have a great name for a cycle around which values are moving: orbit.
Cycle is too static. And instead of saying ``values moving around the cycle $\pi$'', say
``values in orbit $\pi$''. But no time to implement carefully.}
The key insight here is that, even if the set of
values moving around $\pi_1$ does not change (no special edges in $\pi_1$ replace them),
and thus the service call $f$ sees the same bounded set of distinct arguments over time,
it can still generate an unbounded number of fresh values because $f$ is nondeterministic.
$\pi_1\pi_2$ constitute the ``generate cycle'' we mention above.
The generated values are stored in the recall cycle $\pi_3$, where they accumulate
and force the size of the relations of $\pi_3$ to grow unboundedly.

\begin{figure*}[!thb]
\centering
\subfigure[\label{fig:nondetCS-acyclic}Concrete transition system]
{
\begin{minipage}{\columnwidth}
\centering
\input{nondetCS-acyclic}
\end{minipage}
}
\hfill
\subfigure[\label{fig:nondetAS-acyclic}Abstract transition system]
{
\begin{minipage}{\columnwidth}
\centering
\input{nondetAS-acyclic}
\end{minipage}
}
\caption{Concrete and abstract transition systems obtained 
  when the \dcds described in Example~\ref{ex:nonwa} has nondeterministic
  services
\label{fig:nondet-acyclic}}
\end{figure*}
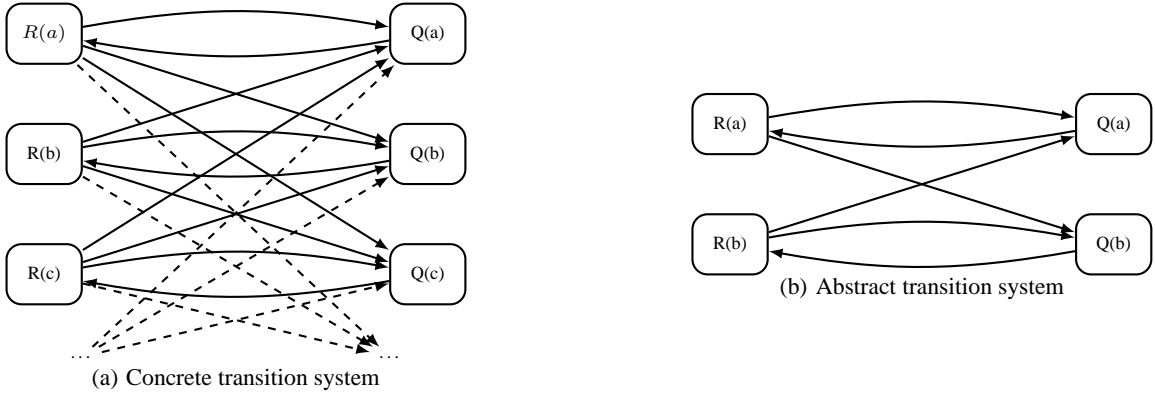

 \begin{example}
\label{ex:nondet-acyclic}
Let us consider again the \dcds \sys described in Example~\ref{ex:nonwa}, this time
considering $f/1$ as a nondeterministic service. The resulting
concrete transition system is shown in Figure~\ref{fig:nondetCS-acyclic}. Even if \sys is not
run-bounded, it is state-bounded, because in every state its database
consists of only one tuple. This is attested by the dataflow graph
shown in Figure~\ref{fig:nondetCS-acyclic}, and guarantees the
existence of a faithful finite-state abstraction. One such finite-state
abstraction is reported in Figure~\ref{fig:nondetAS-acyclic}.
\end{example}

\begin{example}
\label{ex:nondet-cyclic-copy}
Let $\sys = \tup{\dl,\pl}$ be a \dcds with data layer
$\dl=\tup{\CONST, \schema, \emptyset, \idb}$ and process layer $\pl=\tup{\FUNC,\aset,\rset}$,
where $\FUNC = \{f/1\}$, $\schema = \{R/1, Q/1\}$, $\idb =
\{R(a)\}$, $ \rset = \{true
\mapsto \alpha\}$, $\aset=\{\alpha\}$ and
$\alpha: ~ \{R(x) \rightsquigarrow{R(x)},R(x) \rightsquigarrow{Q(f(x))},Q(x) \rightsquigarrow{Q(x)} \}$. 

\sys is not GR-acyclic, because each $R$ tuple is continuously copied,
and at the same time continuously issues a call to service $f$ that is
then stored into a $Q$ tuple, which is continuously copied as
well. This is attested by the dataflow graph of
Figure~\ref{fig:dgraph-GRcyclic-copy}).

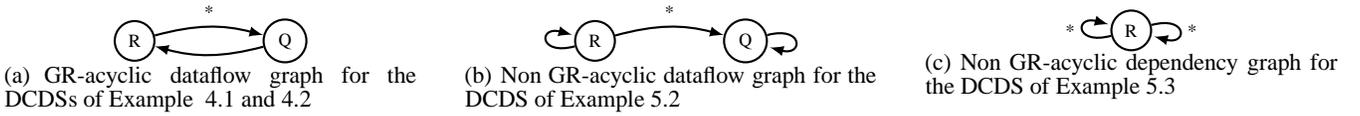
\begin{figure*}[!thb]
\subfigure[GR-acyclic dataflow graph for the $\dcds$s of Example~
  \ref{ex:openruntime} and \ref{ex:openruntime-ec}
\label{fig:dgraph-GRacyclic}]
{
\begin{minipage}{.3\textwidth}
\centering
\begin{tikzpicture}[node distance=1 cm, >=latex,font=\scriptsize,
  rounded corners=3mm,thick]
\node [circle,draw] (R) {R}; 
\node [circle,draw,right of=R,xshift=1cm] (Q) {Q}; 
\path
(R) edge [->,bend left=15] node[auto] {*} (Q)
(Q) edge [->,bend left=15]  (R);
\end{tikzpicture}
\end{minipage}
}
\hfill
\subfigure[Non GR-acyclic dataflow graph for the \dcds of Example~\ref{ex:nondet-cyclic-copy}\label{fig:dgraph-GRcyclic-copy}]
{
\begin{minipage}{.3\textwidth}
\centering
\begin{tikzpicture}[node distance=1 cm, >=latex,font=\scriptsize,
  rounded corners=3mm,thick]
\node [circle,draw] (R) {R}; 
\node [circle,draw,right of=R,xshift=1cm] (Q) {Q}; 
\path
(R) edge [->,loop left]  (R)
(R) edge [->,bend left=15] node[auto] {*} (Q)
(Q) edge [->,loop right]  (Q);
\end{tikzpicture}
\end{minipage}
}
\hfill
\subfigure[Non GR-acyclic dependency graph for the \dcds of Example~\ref{ex:nondet-cyclic-replace}\label{fig:dgraph-GRcyclic-replace}]
{
\begin{minipage}{.3\textwidth}
\centering
\begin{tikzpicture}[node distance=1 cm, >=latex,font=\scriptsize,
  rounded corners=3mm,thick]
\node [circle,draw] (R) {R}; 
\path
(R) edge [->,loop left] node[auto] {*}  (R)
(R) edge [->,loop right] node[auto] {*}  (R);
\end{tikzpicture}
\end{minipage}
}
\caption{Examples of dependency graphs for $\dcds$s with nondeterministic
  services; special edges are decorated with *. }
\end{figure*}

The overall effect caused by the iterated application of $\alpha$ is that fresh values are continuously generated and
accumulated, making \sys state-unbounded. Consider for example the
application of action $\alpha$ in state $\idb$. It leads to an
infinite number of successors, each one of the form $\{R(a), Q(v)\}$
where $v$ is the value returned by $f(a)$. Consider now a second
application of $\alpha$ in one of these states. It again leads to an
infinite number of successors, due to the nondeterminism of
$f(a)$. In particular, each successor has the form $\{R(a), Q(v),
Q(v')\}$, where $v'$ is the result of the second call $f(a)$. When $v'
\neq v$, the number of tuples is increased from 2 to 3. By
executing $\alpha$ over and over again, for some successors the value
returned by a new call $f(a)$ will be distinct
from all the ones already stored in $Q$. This causes an
indefinite increment of the database size due to the continuous insertion of fresh
$Q$ tuples. Such behavior is clearly shown in the concrete transition
system of \sys, depicted in
Figure~\ref{fig:nondetCS-cyclic-copy}. 
Figure~\ref{fig:nondetAS-cyclic-copy} shows instead one possible corresponding
abstraction; even if the abstraction approach ensures that the
generated transition system is finite-branching, some of its runs pass
through an infinite number of distinct, growing states.
\end{example}

\begin{example}
\label{ex:nondet-cyclic-replace}
Let $\sys = \tup{\dl,\pl}$ be a \dcds with data layer
$\dl=\tup{\CONST, \schema, \emptyset, \idb}$ and process layer $\pl=\tup{\FUNC,\aset,\rset}$,
where $\FUNC = \{f/1,g/1\}$, $\schema = \{R/1\}$, $\idb =
\{R(a)\}$, $ \rset = \{true
\mapsto \alpha\}$, $\aset=\{\alpha\}$ and
$\alpha: ~ \{R(x) \rightsquigarrow\{R(f(x)),R(g(x))\}\}$. 

\sys is not GR-acyclic, because its dataflow graph, shown in
Figure~\ref{fig:dgraph-GRcyclic-replace}, contains a unique
node $R$ with two distinct special looping edges from $R$ to $R$
itself.
Indeed, every time $\alpha$ is executed, each $R$ tuple contained in
the current database may generate two $R$ tuples in the next
state, such that each such new tuple contains a value different from
all the other ones. Therefore, even if the newly generated values are
not accumulated, in the ``worst'' case the number of $R$
tuples is doubled every time $\alpha$ is executed. A sample run of the
system could be the following. Starting from $\idb$, $\alpha$ calls
$f(a)$ and $g(a)$, getting $b$ and $c$ as result and obtaining the state
$\{R(b),R(c)\}$. A second execution of $\alpha$ involves now 4 service
calls ($f(b)$, $g(b)$, $f(c)$, $g(c)$), which may return 4 different
new values, e.g.~leading to state $\{R(d),R(e),R(f),R(g)\}$, and so on.
\end{example}

\newcommand{\actions}{\ensuremath{\mathit{actions}}}
\noindent
{\bf GR$^{+}$-Acyclicity.\ }
This relaxation of GR-acyclicity is based on the insight that, for a cycle $\ncts{\sys}$
in the dataflow graph to truly preserve the number of values moving in it,
$\ncts{\sys}$'s edges must not all be simultaneously inactive. 
We say that an edge is {\em active} in a step of the run when some action
corresponding to it executes. By the \dcds semantics, if all edges of $\ncts{\sys}$ 
are simultaneously inactive, then none of the corresponding copy/call operations 
are executed and all relations involved in $\ncts{\sys}$ forget their value in the next state.
$\ncts{\sys}$ is effectively flushed.

GR$^+$-Acyclicity is a relaxation that does allow
path $\pi = \pi_1\pi_2\pi_3$ as in the definition of GR-Acyclicity, 
provided that $\pi_2$ contains an edge $e$ that 
cannot be active at the same time as any of the subsequent edges in $\pi_2\pi_3$.

Semantically this ensures that in order for the generate cycle $\pi_1\pi_2$ to push fresh values
toward recall cycle $\pi_3$, some action corresponding to $e$ must execute,
and in the meantime all actions maintaining the values in cycle $\pi_3$ are disabled,
thus flushing $\pi_3$. $\pi_3$ thus receives an unbounded number of waves of fresh
values from $\pi_1\pi_2$, but it forgets each wave before the next arrives.

Of course, the property of being active at the same time is semantic in nature,
but we give a sufficient syntactic condition. Associate with every edge $e$ in the dataflow
graph the set $\actions(e)$ of actions it corresponds to (this set can be computed
via simple inspection of the process layer). Then edges $e_1,e_2$ are
not simultaneously active if $\actions(e_1) \cap \actions(e_2) = \emptyset$.

The $\dcds$s discussed in Example~\ref{ex:nondet-cyclic-copy}
and~\ref{ex:nondet-cyclic-replace} are not GR$^+$-acyclic. Indeed,
they are not GR-acyclic, and all the edges contained in their dataflow
graphs can be simultaneously active, because they all correspond to a
single action.


We observe that GR-acyclicity is not related to weak
acyclicity. In particular, a \dcds may be GR-acyclic but not weakly acyclic
(see Example~\ref{ex:nondet-acyclic}).

\mrem{Examples follow- please double check whether it is readable or not}
\arem{``It is worth noting that the accumulation of unboundedly many values in
a state could occur for subtle reasons. An apparent source of
state-unboundedness occurs when a service continuously called,
and all its values are stored into a tuple of the database (see
Ex.~\ref{ex:nondet-cyclic-copy} in the appendix). A less apparent source
of state-unboundedness is when the values obtained from the
service calls are not copied into the database, but are instead
continuously used to replace the old tuples with more new tuples,
thus causing a continuous increment of the database size (see
Ex.~\ref{ex:nondet-cyclic-replace} in the appendix).''\\
As written, this is confusing to me. Replacing requires copying into the db, no?
It is unclear how the two are different. I believe that the intention is to capture the
intuition given above, in the paragraph starting with ``First, note that...''.
There, we already talk about replace and copy and their effect on the size of the db.
I added there a parenthesis mentioning the two examples. I don't think we should
dedicate more space to discuss this intuition.}
\arem{
On a different topic: since the figure and example numbers do not include the section,
it is insufficient to say ``look at figure 5'' (which we used to do in other parts of the text)
without mentioning the appendix where it is. It is very frustrating to the reviewer, who has
to linearly scan through a monster appendix. It is also insufficient to say ``Ex. 7''
in the appendix. We must name the appendix. I tried to refine these references everywhere I saw them,
but keep an eye out for more.
}

As with any sufficient syntactic condition for an undecidable semantic property,
an infinite succession of refinements of GR-acyclicity is possible, each
relaxing the condition to allow more \dcds classes. We propose a very powerful relaxation
in Appendix~\ref{app:gr-acyclic}, GR$^+$-acyclicity.
Appendix~\ref{ex:travelReimburseSystem} shows a full-fledged \dcds example that conforms
to GR$^+$-acyclicity, showing that it admits a practically relevant \dcds class.

Theorem~\ref{thm:gracyclic-sbounded} and Theorem~\ref{thm:sbounded-decid} imply:
\begin{theorem}\label{thm:gracyclic-decid}
Verification of \mulpers\ properties for GR$^+$-acyclic \dcds\ with nondeterministic services
is decidable.
\end{theorem}

%% file: nondetCS-cyclic-copy.tex
\renewcommand{\statew}{1.5 cm}
\renewcommand{\stateh}{.8 cm}
\renewcommand{\displx}{3 cm}
\renewcommand{\disply}{1.4 cm}

\begin{tikzpicture}[node distance=.1 cm, >=latex, minimum
  height=\stateh,minimum width=\statew,font=\scriptsize, rounded corners=3mm,thick]

\node at (0,0) [rectangle,draw] (ra) {$\textrm{R(a)}$};
\node [rectangle,draw,right of=ra,xshift=\displx] (raqb) {$\begin{array}{c}\textrm{R(a)}\\\textrm{Q(b)}\end{array}$};
\node [rectangle,draw,above of=raqb,yshift=\disply] (raqa) {$\begin{array}{c}\textrm{R(a)}\\\textrm{Q(a)}\end{array}$};
\node [rectangle,draw,below of=raqb,yshift=-\disply] (raqc) {$\begin{array}{c}\textrm{R(a)}\\\textrm{Q(c)}\end{array}$};
\node [rectangle,draw,right of=raqb,xshift=\displx,yshift=.75cm] (raqaqb) {$\begin{array}{c}\textrm{R(a)}\\\textrm{Q(a)}\\\textrm{Q(b)}\\\end{array}$};
\node [rectangle,draw,above of=raqaqb,yshift=\disply] (raqaqc) {$\begin{array}{c}\textrm{R(a)}\\\textrm{Q(a)}\\\textrm{Q(c)}\\\end{array}$};
\node [rectangle,draw,above of=raqaqc,yshift=\disply] (raqaqd) {$\begin{array}{c}\textrm{R(a)}\\\textrm{Q(a)}\\\textrm{Q(d)}\\\end{array}$};
\node [rectangle,draw,below of=raqaqb,yshift=-\disply] (raqbqc) {$\begin{array}{c}\textrm{R(a)}\\\textrm{Q(b)}\\\textrm{Q(c)}\\\end{array}$};
\node [rectangle,draw,below of=raqbqc,yshift=-\disply] (raqbqd) {$\begin{array}{c}\textrm{R(a)}\\\textrm{Q(b)}\\\textrm{Q(d)}\\\end{array}$};
\node [rectangle,draw,below of=raqbqd,yshift=-\disply] (raqcqd) {$\begin{array}{c}\textrm{R(a)}\\\textrm{Q(c)}\\\textrm{Q(d)}\\\end{array}$};

\node [right of=ra,xshift=1.4cm,yshift=-2.5cm,minimum height=0 cm,minimum width=0 cm] (dotsra) {\ldots};

\node [right of=raqa,xshift=1cm,yshift=2cm,minimum height=0 cm,minimum width=0 cm] (dotsraqa) {\ldots};

\node [right of=raqb,xshift=1cm,yshift=1cm,minimum height=0 cm,minimum width=0 cm] (dotsraqb) {\ldots};

\node [right of=raqc,xshift=1cm,yshift=-2cm,minimum height=0 cm,minimum width=0 cm] (dotsraqc) {\ldots};

\node [right of=raqaqb,xshift=1.4cm,yshift=.5cm,minimum height=0 cm,minimum width=0 cm] (dots1raqaqb) {\ldots};
\node [right of=raqaqb,xshift=1.4cm,minimum height=0 cm,minimum width=0 cm] (dots2raqaqb) {\ldots};
\node [right of=raqaqb,xshift=1.4cm,yshift=-.5cm,minimum height=0 cm,minimum width=0 cm] (dots3raqaqb) {\ldots};

\node [right of=raqaqc,xshift=1.4cm,yshift=.5cm,minimum height=0 cm,minimum width=0 cm] (dots1raqaqc) {\ldots};
\node [right of=raqaqc,xshift=1.4cm,minimum height=0 cm,minimum width=0 cm] (dots2raqaqc) {\ldots};
\node [right of=raqaqc,xshift=1.4cm,yshift=-.5cm,minimum height=0 cm,minimum width=0 cm] (dots3raqaqc) {\ldots};

\node [right of=raqaqd,xshift=1.4cm,yshift=.5cm,minimum height=0 cm,minimum width=0 cm] (dots1raqaqd) {\ldots};
\node [right of=raqaqd,xshift=1.4cm,minimum height=0 cm,minimum width=0 cm] (dots2raqaqd) {\ldots};
\node [right of=raqaqd,xshift=1.4cm,yshift=-.5cm,minimum height=0 cm,minimum width=0 cm] (dots3raqaqd) {\ldots};

\node [right of=raqbqc,xshift=1.4cm,yshift=.5cm,minimum height=0 cm,minimum width=0 cm] (dots1raqbqc) {\ldots};
\node [right of=raqbqc,xshift=1.4cm,minimum height=0 cm,minimum width=0 cm] (dots2raqbqc) {\ldots};
\node [right of=raqbqc,xshift=1.4cm,yshift=-.5cm,minimum height=0 cm,minimum width=0 cm] (dots3raqbqc) {\ldots};

\node [right of=raqbqd,xshift=1.4cm,yshift=.5cm,minimum height=0 cm,minimum width=0 cm] (dots1raqbqd) {\ldots};
\node [right of=raqbqd,xshift=1.4cm,minimum height=0 cm,minimum width=0 cm] (dots2raqbqd) {\ldots};
\node [right of=raqbqd,xshift=1.4cm,yshift=-.5cm,minimum height=0 cm,minimum width=0 cm] (dots3raqbqd) {\ldots};

\node [right of=raqcqd,xshift=1.4cm,yshift=.5cm,minimum height=0 cm,minimum width=0 cm] (dots1raqcqd) {\ldots};
\node [right of=raqcqd,xshift=1.4cm,minimum height=0 cm,minimum width=0 cm] (dots2raqcqd) {\ldots};
\node [right of=raqcqd,xshift=1.4cm,yshift=-.5cm,minimum height=0 cm,minimum width=0 cm] (dots3raqcqd) {\ldots};

\path
(ra) edge [->] (raqa)
(ra) edge [->] (raqb)
(ra) edge [->] (raqc)
(ra) edge [->,dashed] (dotsra)
(raqa) edge [->,loop above,min distance=.5cm] (raqa) 
(raqa) edge [->] (raqaqb)
(raqa) edge [->] (raqaqc)
(raqa) edge [->] (raqaqd)
(raqa) edge [->,dashed] (dotsraqa)
(raqb) edge [->,loop above,min distance=.5cm] (raqb)
(raqb) edge [->] (raqaqb)
(raqb) edge [->] (raqbqc)
(raqb) edge [->] (raqbqd) 
(raqb) edge [->,dashed] (dotsraqb)
(raqc) edge [->,loop above,min distance=.5cm] (raqc)
(raqc) edge [->] (raqaqc)
(raqc) edge [->] (raqbqc)
(raqc) edge [->] (raqcqd) 
(raqc) edge [->,dashed] (dotsraqc)
(raqaqb) edge [->,loop above,min distance=.5cm] (raqaqb)
(raqaqb) edge [->,dashed] (dots1raqaqb)
(raqaqb) edge [->,dashed] (dots2raqaqb)
(raqaqb) edge [->,dashed] (dots3raqaqb)
(raqaqc) edge [->,loop above,min distance=.5cm] (raqaqc)
(raqaqc) edge [->,dashed] (dots1raqaqc)
(raqaqc) edge [->,dashed] (dots2raqaqc)
(raqaqc) edge [->,dashed] (dots3raqaqc)
(raqaqd) edge [->,loop above,min distance=.5cm] (raqaqd)
(raqaqd) edge [->,dashed] (dots1raqaqd)
(raqaqd) edge [->,dashed] (dots2raqaqd)
(raqaqd) edge [->,dashed] (dots3raqaqd)
(raqbqc) edge [->,loop above,min distance=.5cm] (raqbqc)
(raqbqc) edge [->,dashed] (dots1raqbqc)
(raqbqc) edge [->,dashed] (dots2raqbqc)
(raqbqc) edge [->,dashed] (dots3raqbqc)
(raqbqd) edge [->,loop above,min distance=.5cm] (raqbqd)
(raqbqd) edge [->,dashed] (dots1raqbqd)
(raqbqd) edge [->,dashed] (dots2raqbqd)
(raqbqd) edge [->,dashed] (dots3raqbqd)
(raqcqd) edge [->,loop above,min distance=.5cm] (raqcqd)
(raqcqd) edge [->,dashed] (dots1raqcqd)
(raqcqd) edge [->,dashed] (dots2raqcqd)
(raqcqd) edge [->,dashed] (dots3raqcqd)
;
\end{tikzpicture}

%% file: nondetAS-cyclic-copy.tex
\renewcommand{\statew}{1.5 cm}
\renewcommand{\stateh}{.8 cm}
\renewcommand{\displx}{2 cm}
\renewcommand{\disply}{2 cm}

\begin{tikzpicture}[node distance=.1 cm, >=latex, minimum
  height=\stateh,minimum width=\statew,font=\scriptsize, rounded corners=3mm,thick]

\node at (0,0) [rectangle,draw] (ra) {$\textrm{R(a)}$};
\node [rectangle,draw,right of=ra,xshift=\displx,yshift=1cm] (raqa) {$\begin{array}{c}\textrm{R(a)}\\\textrm{Q(a)}\end{array}$};
\node [rectangle,draw,right of=ra,xshift=\displx,yshift=-1cm] (raqb) {$\begin{array}{c}\textrm{R(a)}\\\textrm{Q(b)}\end{array}$};
\node [rectangle,draw,right of=raqa,xshift=\displx] (raqaqb) {$\begin{array}{c}\textrm{R(a)}\\\textrm{Q(a)}\\\textrm{Q(b)}\\\end{array}$};
\node [rectangle,draw,right of=raqb,xshift=\displx] (raqbqc) {$\begin{array}{c}\textrm{R(a)}\\\textrm{Q(b)}\\\textrm{Q(c)}\\\end{array}$};
\node [rectangle,draw,right of=raqaqb,xshift=\displx] (raqaqbqc) {$\begin{array}{c}\textrm{R(a)}\\\textrm{Q(a)}\\\textrm{Q(b)}\\\textrm{Q(c)}\\\end{array}$};
\node [rectangle,draw,right of=raqbqc,xshift=\displx] (raqbqcqd) {$\begin{array}{c}\textrm{R(a)}\\\textrm{Q(b)}\\\textrm{Q(c)}\\\textrm{Q(d)}\\\end{array}$};

\node [right of=raqaqbqc,xshift=1.4cm,minimum height=0 cm,minimum width=0 cm] (dots1) {\ldots};
\node [right of=raqbqcqd,xshift=1.4cm,minimum height=0 cm,minimum width=0 cm] (dots2) {\ldots};

\path
(ra) edge [->] (raqa)
(ra) edge [->] (raqb)
(raqa) edge [->,loop above,min distance=.5cm] (raqa) 
(raqa) edge [->] (raqaqb)
(raqb) edge [->,loop above,min distance=.5cm] (raqb)
(raqb) edge [->] (raqaqb)
(raqb) edge [->] (raqbqc)
(raqaqb) edge [->,loop above,min distance=.5cm] (raqaqb) 
(raqaqb) edge [->] (raqaqbqc)
(raqbqc) edge [->,loop above,min distance=.5cm] (raqbqc)
(raqbqc) edge [->] (raqaqbqc)
(raqbqc) edge [->] (raqbqcqd)
(raqaqbqc) edge [->,loop above,min distance=.5cm] (raqaqbqc)
(raqbqcqd) edge [->,loop above,min distance=.5cm] (raqbqcqd)
(raqaqbqc) edge [->] (dots1)
 (raqbqcqd) edge [->] (dots1)
(raqbqcqd) edge [->] (dots2)
;
\end{tikzpicture}


%% file: nondetCS-acyclic.tex
\renewcommand{\statew}{1 cm}
\renewcommand{\stateh}{.8 cm}
\renewcommand{\displx}{5 cm}
\renewcommand{\disply}{1.5 cm}

\begin{tikzpicture}[node distance=.1 cm, >=latex, minimum
  height=\stateh,minimum width=\statew,font=\scriptsize, rounded corners=2mm,thick]

\node at (0,0) [rectangle,draw] (ra) {$R(a)$};
\node [rectangle,draw,right of=ra,xshift=\displx] (qa) {Q(a)};
\node  [rectangle,draw, below of=ra,yshift=-\disply] (rb) {R(b)};
\node [rectangle,draw,right of=rb,xshift=\displx] (qb) {Q(b)};
\node  [rectangle,draw, below of=rb,yshift=-\disply] (rc) {R(c)};
\node [rectangle,draw,right of=rc,xshift=\displx] (qc) {Q(c)};
\node [below of=qc,yshift=-1cm,xshift=-.5cm,minimum width=0cm,minimum height=0cm] (dots) {\ldots};
\node [below of=rc,yshift=-1cm,xshift=.5cm,minimum width=0cm,minimum height=0cm] (dots2) {\ldots};

\path
(ra) edge [->,bend left=10] (qa)
(ra) edge [->] (qb)
(ra) edge [->] (qc)
(ra) edge [->,dashed] (dots)
(rb) edge [->] (qa)
(rb) edge [->,bend left=10] (qb)
(rb) edge [->] (qc)
(rb) edge [->,dashed] (dots)
(rc) edge [->] (qa)
(rc) edge [->] (qb)
(rc) edge [->,bend left=10] (qc)
(rc) edge [->,dashed] (dots)
(qa) edge [->,bend left=10] (ra)
(qb) edge [->,bend left=10] (rb)
(qc) edge [->,bend left=10] (rc)
(dots2) edge [->,dashed] (qa)
(dots2) edge [->,dashed] (qb)
(dots2) edge [->,dashed] (qc)
;
\end{tikzpicture}


%% file: nondetAS-acyclic.tex
\renewcommand{\statew}{1 cm}
\renewcommand{\stateh}{.8 cm}
\renewcommand{\displx}{5 cm}
\renewcommand{\disply}{1.5 cm}

\begin{tikzpicture}[node distance=.1 cm, >=latex, minimum
  height=\stateh,minimum width=\statew,font=\scriptsize, rounded corners=2mm,thick]

\node at (0,0) [rectangle,draw] (ra) {R(a)};
\node [rectangle,draw,right of=ra,xshift=\displx] (qa) {Q(a)};
\node  [rectangle,draw, below of=ra,yshift=-\disply] (rb) {R(b)};
\node [rectangle,draw,right of=rb,xshift=\displx] (qb) {Q(b)};

\path
(ra) edge [->,bend left=10] (qa)
(ra) edge [->] (qb)
(rb) edge [->] (qa)
(rb) edge [->,bend left=10] (qb)
(qa) edge [->,bend left=10] (ra)
(qb) edge [->,bend left=10] (rb)
;
\end{tikzpicture}


%% file: discussion.tex
\section{Discussion}\label{sec:discussion}

\noindent
{\bf Complexity.\ }
Both in the case of weakly acyclic $\dcds$s with deterministic
services and of GR$^+$-acyclic $\dcds$s with non-deterministic
services, our construction generates a finite transition system whose
number of states is exponential in the size of the $\dcds$. Let $\Phi$
be a \muladom or \mulpers formula of size $\ell$ with $k$ alternating
nested fixpoints. Then, considering the complexity or propositional
$\mu$-calculus model checking on finite transition systems
\cite{Emerson96}, the  complexity of verification of $\Phi$ over a
\dcds of size $n$ is $O(2^n \cdot n^\ell)^k$, hence in
\textsc{ExpTime}.\mrem{Shall we say that the
  translation from det to nondet is irrelevant?}\\

\noindent
{\bf Comparison of the two semantics.\ }
It is natural to ask how the expressivities of the two $\dcds$ flavors compare.
Interestingly, we can show that for unrestricted $\dcds$s, the two semantics are equivalent from
the point of view of expressive power, i.e. any \dcds with deterministic services can be simulated
by a \dcds\ with nondeterministic services, and conversely. However, we show
below that the two semantics are not equivalent with respect to decidability of verification.

Consider first the reduction from deterministic to non-deterministic services.

\begin{theorem}\label{thm:det-2-nondet}
Let $D$ be a \dcds with deterministic services and schema $\schema_D$.
Then one can rewrite $D$ in linear time to a \dcds $N$ with nondeterministic services and schema
$\schema_N$, such that
(i) $\schema_N$ includes $\schema_D$ and
(ii) the projection of $\cts{N}$ to $\schema_D$ coincides with $\cts{D}$, and
(iii) if $D$ is run-bounded, then $N$ is state-bounded.
\end{theorem}

We turn next to the converse reduction.

\begin{theorem}\label{thm:nondet-2-det}
Let $N$ be a \dcds with nondeterministic services and schema $\schema_N$.
Then one can rewrite $N$ in linear time to a \dcds $D$ with deterministic services and schema
$\schema_D$, such that
(i) $\schema_D$ includes $\schema_N$ and
(ii) the projection of $\cts{D}$ to $\schema_N$ coincides with $\cts{N}$.
\end{theorem}

The above reductions show that for unrestricted \dcds, deterministic and nondeterministic
services are equivalent with respect to expressive power. However, they are not
equivalent with respect to decidability of verification.
This is because state-boundedness of the \dcds with nondeterministic services does not
imply run-boundedness of the rewritten \dcds with deterministic services.
In fact, one can prove that there exists no reduction from state-bounded \dcds
with nondeterministic services to run-bounded \dcds with deterministic services:
recall that for properties from $\muladom - \mulpers$,
verification is decidable for the latter
(by Theorem~\ref{thm:finitestate-det}), and undecidable for the former
(Theorem~\ref{thm:sbounded-undecid}).
In particular, the reduction we use to prove Theorem~\ref{thm:nondet-2-det}
yields a non-weakly-acyclic \dcds and is therefore not pertinent to verification
decidability.

In contrast, for the converse reduction of Theorem~\ref{thm:det-2-nondet},
observe that whenever $D$ is run-bounded, $N$ is state-bounded.
Therefore, if we restrict the property language to $\mulpers$,
decidability of verification for run-bounded $\dcds$s with deterministic services
follows as a corollary of the reduction. Recall however that
decidability holds even for the larger logic \muladom (by Theorem~\ref{thm:decidability-det}).
Our proof of Theorem~\ref{thm:decidability-det} exploits the reduction
as well, though additional technical contribution is needed to handle \muladom.\\

\noindent
{\bf Mixed semantics.\ }
The reduction in Theorem~\ref{thm:det-2-nondet} allows us to verify \mulpers properties
for $\dcds$s with a \emph{mix} of deterministic and nondeterministic services,
by first rewriting to a \dcds with exclusively nondeterministic services
(as long as the rewritten \dcds is GR-acyclic). We give an example of a \dcds\ with mixed
service semantics in Appendix~\ref{ex:travelReimburseSystem}.\\

\noindent
{\bf Support for arbitrary integrity constraints.\ }
We remark that, by exploiting the equality constraints, we can extend
our decidability results to support integrity constraints on the database expressed
as arbitrary FO sentences under the active domain semantics.
First, note that the definition of \dcds semantics is independent of the type of constraints used,
as it simply requires their satisfaction by each state of the concrete transition system.
Now consider a \dcds \sys with an FO integrity constraint \emph{IC} defined on its schema.
We can rewrite \sys to enforce \emph{IC} using equality constraints.
To this end, we add a binary auxiliary relation \emph{aux} to the schema, initialized in the initial
state to contain the tuple $\tup{a,b}$ of distinct constants. We add to each action an effect
that simply copies \emph{aux} between states, ensuring the persistence of fact $\mathit{aux}(a,b)$
throughout the run.
Finally, we add an equality constraint
$ec := \neg \mathit{IC} \land \mathit{aux}(x,y) \rightarrow x = y$. Notice now that \sys' will never
execute an action that violates \emph{IC}, because that would violate $ec$.
\arem{guys, what am I missing? why isn't Marco's trick applicable to arbitrary FO
integrity constraints, rather than just denials? See above. Anyway, if I made a mistake,
then be aware that the original discussion of ICs is still in the main text, only commented out
using command ``eat''. You can easily restore it. I fyou agree with above, then the appendix
on denials can go, and the one reference to it from main text should point instead to the Discussion
section.}
\eat{
We remark that our decidability results to support without change arbitrary FO
integrity constraints on the database.
First, note that the definition of \dcds semantics is independent of the type of constraints used,
as it simply requires their satisfaction by each state of the concrete transition system.
Regarding decidability
under arbitrary constraints, we observe that these can be incroporated into the
property such that it is checked only on those runs whose states
satisfy the constraints.
In more detail, consider a constraint-free \dcds \sys ($\EC = \emptyset$)
and an FO sentence $IC$. Let $i\sys$ be the \dcds obtained by setting $\EC = \{ IC \}$.
Then we claim that $\cts{\sys} \models IC \land i\Phi$ if and only if
$\cts{{i\sys}} \models \Phi$.
Here, $i\Phi$ is obtained by rewriting $\Phi$ as follows:
each maximal FO sub-formula $Q$ is replaced with $Q \land IC$, and each formula
$\DIAM{\Phi}$ is replaced with $\DIAM{IC \land \Phi}$. Notice that if
$\Phi \in \mulpers (\muladom)$, then $i\Phi \in \mulpers (\muladom)$.
The above shows that, from the point of view of verification, it does not matter
whether one incorporates the constraints into the \dcds data layer or into the property.
A reason to pick the data layer is to allow specification of processes for purposes
that are independent of their verification.
}
Equality constraints also prove
instrumental in modeling artifact systems, described next.\\

\noindent
{\bf Connection with the artifact model.\ }
In terms of expressive capabilities, our \dcds model is equivalent to a business process model
known in the literature as the \emph{artifact model} (see Section~\ref{sec:related}).
While variations thereof abound, they are virtually all special cases of the following
general model.
In it, given a relational schema $\T$, an \emph{artifact of type $\T$} (or $\T$-artifact)
is a tuple of schema $\T$. The attributes of the tuple are known as \emph{artifact variables},
and they must include an \emph{id} attribute that uniquely identifies each artifact.
An \emph{artifact system} has a schema comprising a collection of types
$\{T_i\}_{ i \in \{1,\ldots,n\}}$,
and the schema $\R_{DB}$ of an underlying relational database. The \emph{instance} of an artifact
system consists of a relation $I_i$ for each type $\T_i$ and a database of schema $\R_{DB}$.
The artifact system also has a collection of actions (usually called ``services'',
a term we avoid here to rule out confusion with the external services of the \dcds model).
The exection of an action evolves the current instance into its successor.
Each action has a \emph{pre-condition} which is a FO sentence over the
artifact schema, evaluated over the current artifact instance under the active domain
semantics. The pre-condition must hold for an action to be eligible to execute.
Actions are also equipped with a \emph{post-condition}
which is usually an $\exists$FO formula relating the current and the successor instances
(if $R$ is a relation in the schema,
the post-condition's $R$-atoms refer to the current instance, while $R'$ atoms refer to the
successor). By $\exists$FO we mean existential FO logic, in which only existential
quantifiers are allowed, and they must appear in the scope of an even number of negations.
Existentially quantified variables are not interpreted over the
active domain, but over the possible infinite domain. They model external inputs from the
environment the artifact system evolves in.

While we do not show a formal reduction between the two models, we sketch here
how a \dcds process can simulate an artifact-based one.
The \dcds can model the sets $I_i$ of $\T_i$-artifacts using an
integrity constraint to enforce the uniqueness of the \emph{id} attribute.
The pre-conditions of artifact actions correspond to the conditions
in the \dcds condition-action rules. Artifact post-conditions $\psi$ can be simulated by
\dcds effects, after rewriting $\psi$ to Skolem normal form and introducing for each resulting
Skolem term a nondeterministic service call. The fact that post-conditions can contain
disjunction while effects are conjunctive and positive
is no impediment: the additional expressivity needed can be transferred to the
\dcds condition-action rules, if necessary modeling one artifact transition step with
several \dcds transition steps.

%% file: relatedwork.tex
\section{Related Work}
\label{sec:related}

As discussed in Section~\ref{sec:discussion}, the unrestricted  
artifact-centric and \dcds models have equivalent expressive capabilities. 
Our work is therefore most closely related to prior work on verification of artifact-centric 
business processes. The difference lies in how each work trades off between restricting the 
class of business processes versus the class of properties to verify.\\

\noindent{\bf Artifact-centric processes with no database.\ }
Work on formal analysis of artifact-based business processes
in restricted contexts has been reported in
\cite{GBS:SOCA07,GeredeSu:ICSOC2007,BGHLS-2007:artifacts-analysis}.
Properties investigated include reachability
\cite{GBS:SOCA07,GeredeSu:ICSOC2007}, general temporal
constraints \cite{GeredeSu:ICSOC2007}, and the existence of complete
execution or dead end \cite{BGHLS-2007:artifacts-analysis}.
For the variants considered in each paper, verification is
generally undecidable; decidability results were obtained only
under rather severe restrictions, e.g., restricting all pre-conditions to be
"true" \cite{GBS:SOCA07}, restricting to bounded
domains \cite{GeredeSu:ICSOC2007,BGHLS-2007:artifacts-analysis},
or restricting the pre- and post-conditions to be propositional,
and thus not referring to data values \cite{GeredeSu:ICSOC2007}.
\cite{CGHS:ICSOC:09} adopts an artifact model variation
with arithmetic operations but no database.
It proposes a criterion for comparing the expressiveness of specifications using the notion of
{\em dominance}, based on the input/output pairs of business processes.
Decidability relies on restricting runs to bounded length.
\cite{ZSYQ:TASE:09} addresses the problem of the existence of a run
that satisfies a temporal property, for a restricted case
with no database and only propositional LTL properties.
All of these works model no underlying database (and hence no
integrity constraints).\\

\noindent{\bf Artifact-centric processes with underlying database.\ }
More recently, two lines of work have considered artifact-centric
processes that also model an underlying relational database.
One considers branching time, one only linear time.\\
{\bf Branching time.\ }
Our approach stems from a line of research that has started with
\cite{ICSOC10} and continued with \cite{Bagheri2011:Artifacts} and \cite{ICSOC11} in
the context of artifact-centric processes.
The connection between evolution of data-centric dynamic systems and
data exchange that we exploit in this paper was first devised in
\cite{ICSOC10}. There the dynamic system transition relation itself is
described in terms of TGDs mapping the current state to the next, and
the evolution of the system is essentially a form of chase. Under
suitable weak acyclicity conditions such a chase terminates, 
thus making the DCDS transition system finite. 
A first-order $\mu$-calculus without first-order quantification across
states is used as the verification formalism for which decidability is
shown. 
Notice the role of getting new objects/values from the external
environment, played here by service calls, is played there by nulls.
These ideas where further developed in \cite{Bagheri2011:Artifacts}, where TGDs where
replaced by action rules with the same syntax as here. Semantically
however the dynamic system formalism there is deeply different: what
we call here service calls are treated there as uninterpreted Skolem
terms. This results in an ad-hoc interpretation of equality which sees
every Skolem term as equal only to itself (as in the case of 
nulls~\cite{ICSOC10}).  
The same first-order $\mu$-calculus without first-order quantification across states of
\cite{ICSOC10} is used as the verification formalism, and a form of
weak acyclicity is used as a sufficient condition for getting finite-state transition systems and decidability. 

In the case of deterministic services, our framework is directly inspired by~\cite{Bagheri2011:Artifacts}, 
though here we do interpret service calls. 
This decision is motivated by
our goal of modeling real-life external services, for which two distinct service
calls may very well return equal results, even under the deterministic semantics 
(for instance if the same service is called with different arguments, or
if distinct services are invoked). 
Interpreting service calls raises a major challenge: even under the run-bounded restriction, 
the concrete transition system is infinite, because it is infinitely branching.
(a service call can be interpreted with any of the constants from the 
infinite domain). 
In contrast to~\cite{Bagheri2011:Artifacts}, what we show in this case 
is not that the concrete transition system is finite (it never is), 
but that it is {\em bisimilar} to a finite abstract transition system.
This leads to a proof technique that is interesting in its own right,
being based on novel notions of bisimilarity for the considered $\mu$-calculus variants.
The reason standard bisimilarity is insufficient is that our logics  $\mulpers$ and $\muladom$
allow first-order quantification across states, so bisimilarity must respect the connection between
values appearing both in the current and successor state. Our decision to include first-order quantification across
states was motivated by the need to express liveness properties that refer to 
the same data at various points in time (e.g. 
``if student $x$  is enrolled now and continues to be enrolled in the future, then $x$ will eventually graduate'').

\eat{
\arem{no longer a difference between the works, in light of the new insight.}

Moreover, the proof technique we adopt allows us to also take into
account integrity constraints over the
data layer. 
In the technical development, we have dealt only with
equality constraints, but our results can be directly extended to the
case of arbitrary denial constraints \cite{} (see Ex.~\ref{ex:denials}).
}

Inspired by \cite{Bagheri2011:Artifacts}, \cite{ICSOC11} builds a similar framework
where actions are specified via pre- and post-conditions given as FO formulae
interpreted over active domains. The verification logic considered is
a first-order variant of CTL with no quantification across states. Thus, it inherits the
limitations discussed above on expressibility of liveness properties. 
In addition, the limited temporal expressivity of CTL precludes
expressing certain desirable properties such as fairness.
\cite{ICSOC11} shows that under the assumption that each state has a bounded active domain, 
one can construct an abstract finite transition system that can be checked 
instead of the original concrete transition system, which is infinite-state in general.  
The approach is similar to the one we developed independently for nondeterministic services,
however without quantification across states, standard bisimilarity suffices. 
As opposed to our work, the decidability of checking state-boundedness is not investigated 
in \cite{ICSOC11}, and no sufficient syntactic conditions are proposed.\\
{\bf Linear time.\ }
Publication \cite{DHPV:ICDT:09} considers an artifact model that has the same expressive 
capabilities as an unrestricted class of
\dcds in which the infinite domain is equipped with a dense linear order, which can be mentioned in pre-, post-conditions, 
and properties. Runs can receive unbounded external input from an infinite domain, and this input corresponds to 
nondeterministic services in a \dcds. Verification is decidable even if the input accumulates in states, and runs are 
neither run-bounded, nor state-bounded. However, this expressive power requires restrictions that render the result 
incomparable to ours. First, the property language is a first-order extension of LTL,
and it is shown that extension to branching time (CTL$^*$) leads to undecidability. 
Second, the formulae in pre-, post-conditions and properties access read-only and read-write database relations differently, 
querying the latter only in limited fashion. In essence, data can be arbitrarily accumulated in read-write relations, but 
these can be queried only by checking that they contain a given tuple of constants. 
It is shown that this restriction is tight, as even the ability to check emptiness of a read-write relation leads to 
undecidability. In addition, no integrity constraints are supported as it is shown that allowing a single functional 
dependency leads to undecidability.
\cite{Damaggio2001:Artifact} disallows read-write relations entirely (only the artifact variables are writable), but this allows
the extension of the decidability result to integrity constraints expressed as embedded dependencies with terminating
chase, and to any decidable arithmetic. 
Again the result is incomparable to ours, as our modeling needs include
read-write relations and their unrestricted querying.\\

\eat{
{\bf Beyond artifact-centric processes.\ }
Static analysis for semantic web services is considered in
\cite{McIlraith-www2002}, but in a context restricted to finite domains.
More recently, \cite{ASV:TODS:09} has
studied automatic verification in the context of business processes based on Active XML documents.

The works~\cite{jcss,S00,AVFY98} are ancestors of~\cite{DHPV:ICDT:09}
from the context of verification of electronic commerce applications.
Their models could conceptually (if not naturally) be encoded as
artifact systems (and thus as \dcds), but they correspond only to particular cases
of the model in~\cite{DHPV:ICDT:09}.
Also, they limit external inputs  to essentially
come from the active domain of the database, thus ruling out fresh values introduced
during the run.
}

\noindent
{\bf Infinite-state systems.\ }
$\dcds$s are a particular case of infinite-state systems.
Research on automatic verification of infinite-state systems has
also focused on extending classical model checking techniques
(e.g., see \cite{Burkartetal01} for a survey).
However, in much of this work the emphasis is on studying
recursive control rather than data,
which is either ignored or finitely abstracted.
More recent work has been focusing specifically on data as a source of
infinity. This includes augmenting recursive procedures with integer
parameters \cite{Bouajjani&Habermehl&Mayr03}, rewriting systems with data
\cite{Bouajjani&Habermehl&Jurski&Sighireanu07},
Petri nets with data associated to tokens \cite{Lazicetal07},
automata and logics over infinite alphabets
\cite{Bouyer&Petit&Therien03,Bouyer02,Neven&Schwentick&Vianu04,Demri:2009:LFQ,JL07,BMSSD06,Bouajjani&Habermehl&Jurski&Sighireanu07},
and temporal logics manipulating data \cite{Demri:2009:LFQ}.
However, the restricted use of data and the particular properties verified
have limited applicability to the business process setting we target with the \dcds\ model.

%% file: conclusions.tex
\section{Conclusions}\label{sec:conclusions}
\newcommand{\ddec}{\textbf{D}\xspace}
\newcommand{\uundec}{\textbf{U}\xspace}
\newcommand{\dec}{\textbf{D}\xspace}
\newcommand{\undec}{\textbf{U}\xspace}
\newcommand{\golu}{\begin{rotate}{90}$\Rsh$\end{rotate}}
\newcommand{\gold}{\begin{rotate}{90}$\Lsh$\end{rotate}}
\newcommand{\gour}{\begin{rotate}{0}$\Rsh$\end{rotate}}
\newcommand{\goul}{\begin{rotate}{90}$\Lsh$\end{rotate}}

\newcolumntype{C}{>{\centering\arraybackslash}X}
\newcolumntype{R}{>{\raggedleft\arraybackslash}X} 
\newcolumntype{L}{>{\raggedright\arraybackslash}X}

\begin{table}
\begin{small}
\centering
\begin{threeparttable}
\renewcommand{\arraystretch}{1.3}
\begin{tabularx}{\columnwidth}{|r@{~}|l@{}c@{}c@{}c@{}r@{}|r@{~}|l@{}c@{}c@{}c@{}r@{\,}|}
\multicolumn{6}{c}{\textsc{deterministic services}}&
\multicolumn{6}{c}{\textsc{nondeterministic services}}\\
\hline
& $\mu\L$ && \muladom && \mulpers&& $\mu\L$ && \muladom && \mulpers\\
\hline
{\em unrestricted}
& \undec\phantom{\tnote{2}} & $\leftarrow$ 
& \phantom{\tnote{3}}\undec\phantom{\tnote{3}} & $\leftarrow$
& \uundec\tnote{1} &
{\em unrestricted}
& \undec & $\leftarrow$
& \undec & $\leftarrow$
& \uundec\tnote{1} \\
&  && && &&$\uparrow$&&$\uparrow$&&\\
{\em bounded-run}
& \textbf{?} \tnote{2} & 
& \phantom{\tnote{3}}\ddec\tnote{3} & $\rightarrow$
& \ddec\phantom{1} &
{\em bounded-state}
& \undec & $\leftarrow$ 
& \uundec & 
&\ddec\tnote{3}\\
\hline
\end{tabularx}
\begin{tablenotes}
\item [1] The result is even stronger: it holds for propositional LTL.
\item [2] Decidability cannot be established via a faithful finite-state abstraction.
\item [3] Decidability is obtained via reduction to finite-state model checking.
\end{tablenotes}
\end{threeparttable}
\end{small}
\caption{Summary of our (un)decidability results\label{tab:summary}.}
\end{table}

We summarize our results in Table~\ref{tab:summary} (arrows denote implications between results).
We note that exhibiting a finite faithful abstraction of a concrete transition system
is more than a means towards showing decidability, being a desirable goal in its own right
as the most promising avenue towards practical implementation.
Notice that we list as open the verification of
$\mu\L$ properties on bounded-run $\dcds$s with deterministic services, but recall from 
Section~\ref{sec:det-bounded} that in this case there exists no faithful 
finite-state abstract transition system. 

We believe that $\dcds$s provide a natural and expressive model for business processes powered by an 
underlying database, and thus are an ideal vehicle for foundational research with potential to
transfer to alternative models.

Note that the design space for FO extensions of propositional $\mu$-calculus is broad, and 
notoriously contains bounded-state settings for which satisfiability of even modest extensions
of propositional LTL is highly undecidable 
(e.g. LTL with the freeze quantifier over infinite data words~\cite{Demri:2009:LFQ}).
In light of this, our decidability results come as a pleasant surprise, and the two $\mu\L$ variants 
studied here, paired with the respective \dcds classes,
strike a fortuitous\mrem{Is it the case that ``fortuituous'' could be
  interpreted negatively?} balance between expressivity and verification feasibility.

%% file: appendix-mula.tex
\section{Verification}

In this appendix we give the  bisimulation invariance results for 
$\muladom$ and $\mulpers$.

\subsection{History Preserving Mu-Calculus}

We prove \emph{history preserving bisimulation invariance} for 
$\muladom$.  We adopt a two-step approach. We first prove the
result for the logic $\hmadom$, obtained from $\muladom$ dropping the
predicate variables and the fixpoint constructs. Such a logic
corresponds to a first-order variant of the Hennessy Milner logic;
note that the semantics of this logic is completely independent from
the second-order valuation. We then extend the result to the whole
\muladom by dealing with fixpoints.

\begin{lemma}
\label{lemma:hmadom-bisimulation}
Consider two transition systems $\mf{}_1 = \langle$$\Delta_1$,$\R$, $\Sigma_1$,$s_{01}$,$\db_1$,$\Rightarrow_1 \rangle$ and $\mf{}_2 =
\tup{\Delta_2,\R,\Sigma_2,s_{02},\db_2,\Rightarrow_2}$, a partial
bijection $h$ between $\Delta_1$ and $\Delta_2$ and
two states $s_1 \in \Sigma_1$ and $s_2 \in \Sigma_2$ such that $s_1
\hbsim_h s_2$. Then for every (open) formula $\Phi$ of $\hmadom$, and every valuations $\vfo_1$ and $\vfo_2$ that
 assign to each of its free variables a value $d_1 \in \adom{\db_1(s_1)}$ and
 $d_2 \in \adom{\db_2(s_2)}$, such that $d_2 = h(d_1)$, we have that
\[\mf{}_1,s_1 \models \Phi \vfo_1 \textrm{ if and only if } \mf{}_2,s_2 \models \Phi \vfo_2. \]
\end{lemma}
\begin{proof}
We proceed by induction on the structure of $\Phi$.

\begin{description}
 \item[Local first-order queries (base case)]  Consider $\Phi=Q$,
   where $Q$ is an (open) FO query. Since $h$ induces
   an isomorphism between $\db(s_1)$ and $\db(s_2)$, for every
   valuations $\vfo_1$ and $\vfo_2$ that assign to each free variable
   of $Q$ a value $d_1 \in  \adom{\db_1(s_1)}$ and
 $d_2 \in  \adom{\db_2(s_2)}$, such that $d_2 = h(d_1)$, we have that
 $\ans{Q\vfo_1,\db(s_1)}\equiv\ans{Q\vfo_2,\db(s_2)}$.
\item[Negation] By induction hypothesis, for every (open) formula
  $\Phi$ and every valuations $\vfo_1$ and $\vfo_2$ that assign to each of its free variables a value $d_1 \in  \adom{\db_1(s_1)}$ and $d_2 \in  \adom{\db_2(s_2)}$, such that $d_2 = h(d_1)$, we have that
$\mf{}_1,s_1 \models \Phi\vfo_1$ if and
  only if  $\mf{}_2,s_2 \models \Phi\vfo_2$. By definition, $\mf{}_1,s_1 \models (\neg\Phi)\vfo_1$ if and only if
  $\mf{}_1,s_1 \not\models \Phi\vfo_1$, and, by induction hypothesis,
  $\mf{}_1,s_1 \not\models \Phi\vfo_1$ if and only if $\mf{}_2,s_2
  \not\models \Phi\vfo_2$, which corresponds to $\mf{}_2,s_2 \models (\neg\Phi)\vfo_2$.
\item[Conjunction] By induction hypothesis,  for every (open) formula
  $\Phi$ and every valuations $\vfo_1$ and $\vfo_2$ that assign to
  each of its free variables a value $d_1 \in  \adom{\db_1(s_1)}$ and $d_2 \in
  \adom{\db_2(s_2)}$, such that $d_2 = h(d_1)$, we have that $\mf{}_1,s_1 \models \Phi_i\vfo_1$ if and
  only if $\mf{}_2,s_2 \models \Phi_i\vfo_2$, with $i \in
  \{1,2\}$. Hence, $\mf{}_1,s_1 \models \Phi_1\vfo_1$ and
   $\mf{}_1,s_1 \models \Phi_2\vfo_1$ if and only if
  $\mf{}_2,s_2 \models \Phi_1\vfo_2$ and
   $\mf{}_2,s_2 \models \Phi_2\vfo_2$. By definition, we therefore have  $\mf{}_1,s_1 \models (\Phi_1 \land \Phi_2) \vfo_1$ if and
  only if  $\mf{}_2,s_2 \models (\Phi_1 \land \Phi_2) \vfo_2$.
\item[Modal operator]
Consider two states $s_1 \in \Sigma_1$ and $s_2 \in \Sigma_2$ such
that $s_1 \hbsim_h s_2$.
By definition, given a valuation $\vfo_1$ that assigns to each free
variable of $\Phi$ a value $d_1 \in  \adom{\db_1(s_1)}$, we have that
  $\mf{}_1,s_1 \models (\DIAM{\Phi})\vfo_1$ if there exists a transition
  $s_1 \Rightarrow_1 s'_1$ such that  $\mf{}_1,s'_1 \models
  \Phi\vfo_1$. Since $s_1 \hbsim_h s_2$,
  there exists a transition $s_2 \Rightarrow_2 s'_2$ such that $s'_1
  \hbsim_{h'} s'_2$, where $h'$ extends $h$. By
  induction hypothesis, for every valuation $\vfo_2$
  that assigns to each free variable $x$ of $\Phi$ a value $d_2 \in
 \adom{\db_2(s'_2)}$, such that $d_2 = h'(d_1)$ with $x/d_1 \in \vfo_1$, we
  have that  $\mf{}_1,s'_1 \models
  \Phi\vfo_1$ if and only if  $\mf{}_2,s'_2 \models
  \Phi\vfo_2$. Since $h'$ is an extension of $h$, and $\vfo_1$ assigns to each free
variable of $\Phi$ a value $d_1 \in  \adom{\db_1(s_1)} \subseteq \domain{h}$, we observe that for
every pair of assignments $x/d_1 \in \vfo_1$ and $x/d_2 \in \vfo_2$, it holds that
$d_2 = h'(d_1) = h(d_1)$. Furthermore, since $h$ induces an
isomorphism between $db_1(s_1)$ and $db_2(s_2)$, for each assignment
$x/d_2 \in \vfo_2$, we have that $d_2 \in \adom{\db_2(s_2)}$. Considering that $s_2
\Rightarrow_2 s'_2$, by definition we therefore get $\mf{}_2,s_2 \models
  (\DIAM{\Phi})\vfo_2$.

The other direction can be proven in a symmetric way.
\item[Quantification]
Consider two states $s_1 \in \Sigma_1$ and $s_2 \in \Sigma_2$ such
that $s_1 \hbsim_h s_2$.
By definition, given a formula $\Phi$ and a valuation $\vfo'_1$ that assigns to each free
variable of $\Phi$ a value $d_1 \in \domain{h}$, we have that $\mf{}_1,s_1 \models (\exists
  x.\inadom(x) \land \Phi)\vfo'_1$ if and only if there exists $d \in
  \adom{\db_1(s_1)}$ such that  $\mf{}_1,s_1 \models
  \Phi\vfo_1$, where $\vfo_1 = \vfo'_1[x/d]$. By
  induction hypothesis, for every valuation $\vfo_2$
  that assigns to each free variable $y$ of $\Phi$ a value $d_2 \in
   \adom{\db_2(s_2)}$, such that $d_2 = h(d_1)$ with $y/d_1 \in \vfo_1$, we have that $\mf{}_1,s_1 \models
  \Phi\vfo_1$ if and only if $\mf{}_2,s_2 \models \Phi
  \vfo_2$. More specifically, the structure of $\vfo_2$ is
  $\vfo_2 = \vfo'_2[x/d']$, where $d' = h(d) \in  \adom{\db_2(s_2)}$
  because $h$ induces an isomorphism between $db_1(s_1)$ and $db_2(s_2)$.
Hence, we get $\mf{}_2,s_2 \models (\exists x. \inadom(x)
  \land \Phi)\vfo'_2$.

The other direction can be proven in a symmetric way.\qed
\end{description} 
\end{proof}

\eat{
\begin{theorem}
\label{theorem:muladom-bisimulation}
Consider two transition systems $\mf{}_1 = \langle
\Delta_1,\R$, $\Sigma_1,s_{01},\db_1,\Rightarrow_1 \rangle$ and $\mf{}_2 =
\tup{\Delta_2,\R,\Sigma_2,s_{02},\db_2,\Rightarrow_2}$ such that
$\mf{}_1 \hbsim \mf{}_2$.
Then for every closed formula $\Phi$ of $\muladom$, we have that
\[\mf{}_1 \models \Phi \textrm{ if and only if } \mf{}_2 \models \Phi. \]
\end{theorem}
}

\begin{proof}[of Theorem~\ref{theorem:muladom-bisimulation}]
We prove the theorem in two steps. First, we show that Lemma
\ref{lemma:hmadom-bisimulation} can be extended to the
infinitary version of $\hmadom$ that supports arbitrary
countable disjunction. Then, we recall that fixpoints can be
translated into this infinitary logic, thus guaranteeing
invariance for the whole $\muladom$ logic.

Let $\Psi$ be a countable ordered set of open $\hmadom$
formulae. Given a transition system $\mf{} =  \tup{\Delta,\R,\Sigma,s_0,\db,\Rightarrow}$, the semantics of
$\bigvee \Psi$ is $\MODAfirst{\bigvee \Psi} = \bigcup_{\psi \in
  \Psi} \MODAfirst{\psi}$. Therefore, given a state $s$ of $\mf{}$ and
a variable valuation $\vfo$ that assigns to each free variable of
$\Psi$ a value $d \in \adom{\db(s)}$ , we
have $\mf{},s \models \Psi\vfo$ if and only if
$\mf{},s \models \psi\vfo$ for some $\psi \in
\Psi$. Arbitrary countable conjunction is obtained for free because
of negation.

We show that the invariance result proven in Lemma
\ref{lemma:hmadom-bisimulation} trivially extends to this
arbitrary countable disjunction. Lemma
\ref{lemma:hmadom-bisimulation} guarantees that invariance is
preserved for any finite disjunction. Formally, let
$\{\Phi_1,\ldots,\Phi_n\}$ be a finite set of open
$\hmadom$ formulae. Consider two states $s_1 \in \Sigma_1$
and $s_2 \in \Sigma_2$ such that $s_1 \hbsim_h s_2$. Then, for every valuations $\vfo_1$ and $\vfo_2$ that
 assign to each free variable of $\{\Phi_1,\ldots,\Phi_n\}$ a value $d_1 \in \adom{\db_1(s_1)}$ and
 $d_2 \in \adom{\db_2(s_2)}$, such that $d_2 = h(d_1)$, we have that
 $\mf{}_1,s_1 \models (\bigvee_{i \in
    \{1,\ldots,n\}}\Phi_i)\vfo_1$ if and only if $\mf{}_2,s_2 \models (\bigvee_{i \in
    \{1,\ldots,n\}}\Phi_i)\vfo_2$.

Now consider two valuations $\vfo'_1$ and $\vfo'_2$ that
 assign to each free variable of $\bigvee\Psi$ a value $d_1 \in \adom{\db_1(s_1)}$ and
 $d_2 \in \adom{\db_1(s_2)}$, such that $d_2 = h(d_1)$. By definition,
 $\mf{}_1,s_1 \models (\bigvee
  \Psi)\vfo'_1$ if and only if there exists $\psi_k \in \Psi$ such that $\mf{}_1,s_1 \models \psi_k\vfo'_1$.
 The proof of invariance for the infinitary $\hmadom$ logic is then
obtained by observing that $\mf{}_1,s_1 \models (\bigvee
  \Psi)\vfo'_1$ if and only if
$\mf{}_1,s_1 \models (\bigvee_{i \in \{1,\ldots,k\}}
  \psi_i)\vfo'_1$ if and only if
$\mf{}_2,s_2 \models (\bigvee_{i \in \{1,\ldots,k\}}
  \psi_i)\vfo'_2$ if and only if $\mf{}_2,s_2 \models (\bigvee
  \Psi)\vfo'_2$.

In order to extend the result to the whole \muladom, we resort to
the well-known result stating that fixpoints of the $\mu$-calculus can be
translated into the infinitary Hennessy Milner logic by iterating over
\emph{approximants}, where the approximant of index $\alpha$ is
denoted by $\mu^\alpha Z.\Phi$ ($\nu^\alpha Z.\Phi$). This is a
standard result that also holds for \muladom. In particular, approximants are
built as follows:
\begin{align*}
  \mu^0 Z.\Phi & = \false
  &  \nu^0 Z.\Phi & = \true\\
  \mu^{\beta+1} Z.\Phi & = \Phi[Z/\mu^\beta Z.\Phi]
  & \nu^{\beta+1} Z.\Phi & = \Phi[Z/\nu^\beta Z.\Phi]\\
  \mu^\lambda Z.\Phi & = \bigvee_{\beta < \lambda} \mu^\beta Z. \Phi &
  \nu^\lambda Z.\Phi & = \bigwedge_{\beta < \lambda} \nu^\beta Z. \Phi
\end{align*}
where $\lambda$ is a limit ordinal, and where fixpoints and their
approximants are connected by the following properties: given a transition
system $\Upsilon$ and a state $s$ of $\Upsilon$
\begin{itemize}
\item $s \in \MODA{\mu Z.\Phi}$ if and only if there exists an
  ordinal $\alpha$ such that $s \in \MODA{\mu^\alpha Z.\Phi}$ and, for
  every $\beta < \alpha$,  it holds that $s \not\in \MODA{\mu^\beta Z.\Phi}$;
\item $s \not\in \MODA{\nu Z.\Phi}$ if and only if there exists an ordinal $\alpha$ such that $s \not\in \MODA{\nu^\alpha
    Z.\Phi}$ and, for every $\beta < \alpha$, it holds that $s \in \MODA{\nu^\beta
    Z.\Phi}$.\qed
\end{itemize}
\end{proof}

%% file: appendix-mulp.tex
\subsection{Persistence Preserving Mu-Calculus}

We prove \emph{persistence preserving bisimulation invariance} for $\mulpers$.
To prove the invariance result, we adopt a two-step approach. We first
prove the result for the logic $\hmpers$, obtained from $\mulpers$
dropping the predicate variables and the fixpoint constructs. Such a
logic corresponds to a first-order variant of the Hennessy Milner logic; note that the
semantics of this logic is completely independent from the
second-order valuation. We then extend the result to the whole
$\mulpers$ by dealing with fixpoints.

\begin{lemma}\label{lemma:hmpers-bisimulation}
Consider two transition systems $\mf{}_1 = \tup{\Delta_1,\R,\Sigma_1,s_{01},\db_1,\Rightarrow_1}$ and $\mf{}_2 =
\tup{\Delta_2,\R,\Sigma_2,s_{02},\db_2,\Rightarrow_2}$, a partial
bijection $h$ between $\Delta_1$ and $\Delta_2$ and
two states $s_1 \in \Sigma_1$ and $s_2 \in \Sigma_2$ such that $s_1
\pbsim_h s_2$. Then for every (open) formula $\Phi$ of $\hmpers$, and every valuations $\vfo_1$ and $\vfo_2$ that
 assign to each of its free variables a value $d_1 \in \adom{\db_1(s_1)}$ and
 $d_2 \in \adom{\db_2(s_2)}$, such that $d_2 = h(d_1)$, we have that
\[\mf{}_1,s_1 \models \Phi \vfo_1 \textrm{ if and only if } \mf{}_2,s_2 \models \Phi \vfo_2. \]
\end{lemma}
\begin{proof}
We proceed by induction on the structure of $\Phi$. In particular, we
discuss the two base cases of $\DIAM{(\inadom(x) \land \Phi)}$ and
$\BOX{(\inadom(x) \land \Phi')}$ with one variable. For convenience,
we rewrite the latter case to $\DIAM{(\inadom(x) \limp
  \Phi)}$, where $\Phi = \neg \Phi'$. The other cases
are derived, or proven in the same way as done for Lemma \ref{lemma:hmadom-bisimulation}.

\begin{description}
\item[Modal operator (conjunction)]
Consider two states $s_1 \in \Sigma_1$ and $s_2 \in \Sigma_2$ such
that $s_1 \pbsim_h s_2$. Let $x$ be the only free variable of $\Phi$,
and $x/d$ a valuation such that $d \in  \adom{\db_1(s_1)}$. Then, by
definition we have that
  $\mf{}_1,s_1 \models (\DIAM{(\inadom(x) \land \Phi})[x/d]$ if there exists a transition
  $s_1 \Rightarrow_1 s'_1$ such that $d \in \adom{\db_1(s'_1)}$ and $\mf{}_1,s'_1 \models
  \Phi[x/d]$.
 Since $s_1 \pbsim_h s_2$,
  there exists a transition $s_2 \Rightarrow_2 s'_2$ such that
$s'_1
  \pbsim_{h'} s'_2$, where $h'$ is compatible with $h$.
By induction hypothesis and by considering that $h'$ is an isomorpshim
between $\db_1(s'_1)$ and $\db_2(s'_2)$, we
  have that  $\mf{}_1,s'_1 \models
  \Phi[x/d]$ if and only if $h'(d) \in
 \adom{\db_2(s'_2)}$ and $\mf{}_2,s'_2 \models
  \Phi[x/h'(d)]$. Now we observe that  $d \in \adom{\db_1(s_1)} \cap
\adom{\db_1(s'_1)}$ and $h'$ is an extension of $\restrict{h}{\adom{\db_1(s_1)} \cap
\adom{\db_1(s'_1)}}$. This implies that $h'(d) = h(d) \in
\adom{\db_2(s_2)}$, because $h$ is an isomorphism between $\db_1(s_1)$
and $\db_2(s_2)$.
Considering that $s_2
\Rightarrow_2 s'_2$, by definition we therefore
get $\mf{}_2,s_2 \models
  (\DIAM{(\inadom(x) \land \Phi)})[x/h(d)]$.

The other direction can be proven in a symmetric way.
\item[Modal operator (implication)]
Consider two states $s_1 \in \Sigma_1$ and $s_2 \in \Sigma_2$ such
that $s_1 \pbsim_h s_2$.  Let $x$ be the only free variable of $\Phi$,
and $x/d$ a valuation such that $d \in  \adom{\db_1(s_1)}$. Then, by definition we have that
  $\mf{}_1,s_1 \models (\DIAM{(\inadom(x) \limp \Phi})[x/d]$ if there exists a transition
  $s_1 \Rightarrow_1 s'_1$ such that $d \not\in \adom{\db_1(s'_1)}$ or $\mf{}_1,s'_1 \models
  \Phi[x/d]$.
 Since $s_1 \pbsim_h s_2$,
  there exists a transition $s_2 \Rightarrow_2 s'_2$ such that
$s'_1
  \pbsim_{h'} s'_2$, where $h'$ is an extension of $\restrict{h}{\adom{\db_1(s_1)}
  \cap \adom{\db_1(s'_1)}}$
Now we discuss the two cases in which  $d \not\in \adom{\db_1(s'_1)}$
and  $d \in \adom{\db_1(s'_1)}$.
\begin{itemize}
\item Assume that $d \not\in \adom{\db_1(s'_1)}$. Since $s_1 \pbsim_h
  s_2$, we have that $h(d) \in \adom{\db_2(s_2)}$. Now, towards
  contradiction, let us assume that
  $h(d) \in \adom{\db_2(s'_2)}$. Hence, we have $h(d) \in \adom{\db_2(s_2)}
  \cap \adom{\db_2(s'_2)}$. Observe that $h'$ is an extension of $\restrict{h}{\adom{\db_1(s_1)}
  \cap \adom{\db_1(s'_1)}}$, which is equivalent to state that
$h'^{-1}$ is an extension of $\restrict{h^{-1}}{\adom{\db_2(s_2)}
  \cap \adom{\db_2(s'_2)}}$. This implies that  $h^{-1}(d) = h'^{-1}(d) = d$. Since $h'$
  is an isomorphism between $\db_1(s'_1)$ and $\db_2(s'_2)$, then $d
  \in \adom{\db_1(s'_1)}$, and this contradicts the hypothesis.
\item Assume that $d \in \adom{\db_1(s'_1)}$. Then we can proceed
  following the line of reasoning used for the case of $\DIAM{(\inadom(x) \land \Phi)}$.
\end{itemize}
The other direction can be proven in a symmetric way. \qed

\end{description}
\end{proof}

\eat{
\begin{theorem}
\label{theorem:mulpers-bisimulation}
Consider two transition systems $\mf{}_1$ and $\mf{}_2$ such that
$\mf{}_1 \pbsim \mf{}_2$.
Then for every $\mulpers$ closed formula $\Phi$, we have:
\[\mf{}_1 \models \Phi \textrm{ if and only if } \mf{}_2 \models \Phi. \]
\end{theorem}
}

\begin{proof}[of Theorem~\ref{theorem:mulpers-bisimulation}]
The proof is analogous to that of Theorem~\ref{theorem:muladom-bisimulation}, but now using Lemma~\ref{lemma:hmpers-bisimulation}.
\end{proof}

%% file: appendix-det-verification-results.tex
\section{Deterministic Services}\label{app:det}



\addtocounter{subsection}{1} 
\subsection{Run-Bounded Systems}

\newcommand{\succS}{\ensuremath{\mathit{right}}}
\newcommand{\succT}{\ensuremath{\mathit{succ}_t}}
\newcommand{\lab}{\ensuremath{\mathit{sym}}}
\newcommand{\head}{\ensuremath{\mathit{head}}}
\newcommand{\st}{\ensuremath{\mathit{state}}}
\newcommand{\halted}{\ensuremath{\mathit{halted}}}
\newcommand{\new}{\ensuremath{\mathit{newCell}}}

\begin{proof}[of Theorem \ref{thm:ltl-undecid-det}]
\arem{guys, I am reusing this construction in 3 other undecidability proofs, so if you want
to edit it, please talk to me before so I can synch up.}

The proof is by reduction from the halting problem.
Given a deterministic Turing Machine {\em TM}, we define
\dcds\ \sys with deterministic services and propositional safety property $\Phi$,
such that {\em TM} halts if and only if $\cts{\sys} \models \Phi$.

Intuitively, every run of $\cts{\sys}$ simulates a run of {\em TM}.
Each state $s$ of $\cts{\sys}$ models a configuration of {\em TM}.
A transition in $\cts{\sys}$ models a transition in {\em TM}.
We give the construction next.

{\bf The $\dcds$.\ }
To model a configuration of {\em TM} in a relation of the \dcds state, 
we model the visited tape segment as a graph whose nodes are cell identifiers, 
and whose edges form a linear path. 
The edge relation is called \succS,
with the intended meaning that $\succS(x,y)$ declares cell $y$ to be the right neighbor of cell $x$
on the tape.
We also introduce a relation $\lab$, with $\lab(c,s)$ intended to model that cell $c$ holds symbol $s$.
Unary relation $\head$ models the head position: $\head(c)$ means that the head points to cell $c$.
Finally, unary relation $\st$ keeps the state of {\em TM}, and a boolean predicate \halted\
is meant to detect that {\em TM} has halted.
In summary, the data layer $\dl = \tup{\CONST,\schema,\EC,\idb}$ 
of \sys contains the schema $\schema  = \{ \succS/2, \lab/2, \head/1, \halted/0 \}$.
We detail $\EC$ and $\idb$ after sketching the process layer.

There is a single action $\alpha$, in charge of simulating the transitions of {\em TM}.
It has no parameters, and its guard is always true: $\true \mapsto \alpha$.
$\alpha$ contains the following effects.

$e_{\mathit{copy}}$ simply copies the part of the tape that stays unchanged in the transition
because the head doesn't point to it:
\begin{tabbing}
$e_{\mathit{copy}}:$  \= $\succS(X,Y) \land \succS(Y,Z) \land$\+\\ 
         $\lab(X,SX) \land \lab(Y,SY) \land \lab(Z,SZ) \land$\\
         $\neg (\head(X) \land \head(Y) \land \head(Z))$\\
         $\rightsquigarrow$\\ 
         $\{\succS(X,Y) , \succS(Y,Z) ,$\\ 
         $\phantom{\{}\lab(X,SX),\lab(Y,SY), \lab(Z,SZ)\}$
\end{tabbing}

In addition, we add effects for each entry of {\em TM}'s transition relation $\delta$.

For instance, if $(p,b,\ra) \in \delta(s,a)$ 
(i.e. $\delta$ prescribes that in state $s$, if the head points to a cell containing
symbol $a$, {\em TM} changes state to $p$, the cell's symbol is overwritten with $b$,
and the head moves to the right), we introduce two effects. 
One for the case when the tape needs no extension to the right,

\begin{tabbing}
$e_{s,a,p,b,\ra}^{noext}:$  \= $\succS(X,Y) \land \lab(X,a) \land \lab(Y,SY) \land SY \not= \omega \land$\+\\         $\head(X) \land \st(s)$\\
         $\rightsquigarrow$\\ 
         $\{\succS(X,Y) , \lab(X,b) , \lab(Y,SY) ,$\\
         $\phantom{\{}\head(Y) , \st(p)\}$,
\end{tabbing}
and one when it does:
\begin{tabbing}
$e_{s,a,p,b,\ra}^{ext}:$  \= $\succS(X,Y) \land \lab(X,a) \land \lab(Y,\omega) \land$\+\\ 
         $\head(X) \land \st(s)$\\
         $\rightsquigarrow$\\ 
         $\{\succS(X,Y) , \succS(Y,\new(Y)) ,$\\ 
         $\phantom{\{}\lab(X,b) , \lab(Y,\bot) , \lab(\new(Y),\omega) ,$\\
         $\phantom{\{}\head(Y) , \st(p)\}$.
\end{tabbing} 
To distinguish among constants and variables in the above effect specifications, 
we use capital letters for the latter and lower-case letters for the former.
Notice that the extension is performed by calling service \new, which is meant to return
a fresh cell id (we show below how to ensure this). 
Also notice the use of special symbol $\omega$, which is reserved for labeling the
end of the tape segment. Finally, special symbol $\bot$ is by convention used
to initialize the tape prior to starting the run.

We are not quite done, as we still need to ensure that \succS\ induces a linear order
on the collection of cell identifiers generated during the run. Notice that this cannot
be achieved exclusively by declaring FO constraints in \succS, 
as linear orders are not FO-axiomatizable. The solution must exploit the interplay between
constraints on \succS\ and the way $\cts{\sys}$ transitions.

Observe that, by definition of the effects that extend \succS\ (e.g. $e_{s,a,p,b,\ra}^{ext}$ above),
at each step the current right end of the tape segment obtains at most one new successor.
However, if the call of service $\new$ returns a cell id that already appears in the tape segment, 
then there can be some cell with several predecessors  according to \succS. 
We rule out this case by declaring the second component of \succS\ to be a key.
It follows that \succS\ must be either 
(i) a linear path (possibly starting from a source node that has a self-loop), 
or 
(ii) it must contain a simple cycle involving more than one cell id. 
The simple cycle is created at the step when \new\ returns the id of the leftmost cell.

We wish to force case (i). To rule out case (ii), we proceed as follows: 
we initialize \succS\ to contain a source node $0$ that can never be a cell id
because it cannot be returned by \new\ without violating the key constraint on \succS. 
To this end, we initialize \idb to
\begin{itemize}\parskip=0in\itemsep=0in
\item 
$\succS^{\idb} = \{(0,0),(0,1),(1,2)\}$,
\item
$\lab^{\idb} = \{ (1,\$), (2,\omega)\}$,
\item
$\head^{\idb} = \{2\}$, 
\item
$\st^{\idb} = \{ s_0 \}$,
\item
$\halted^{\idb} = \{ \}$,
\end{itemize}
where $s_0$ is the initial state of {\em TM}.

Notice that, if we disregard cell $0$, \idb contains the representation of an
empty tape (symbol $\$$ labels the left end, symbol $\omega$ the right end).
Also notice that $\succS^{\idb}$ has type (i).
An easy induction shows that every run prefix must also construct a \succS\ relation of 
type (i), since any attempt to extend \succS\ with an edge
back to one of its existing nodes violates the key constraint. 

Because symbol $\$$ denotes the left end of the tape, 
it also follows easily from the behavior of {\em TM} that during the run, 
the head will never reach the special cell $0$, so \head\ can only take values from the suffix of \succS\ 
starting at cell $1$, which is a true linear path.

Now assume without loss of generality that the {\em TM} is normalized to enter a particular sink state $h$ 
when it halts. We add effect $e_h$, which detects the halting state and sets the boolean predicate \halted.

Observe that for the cases when the head stays in place or moves left, no tape extension is required,
so each such entry in the transition relation corresponds to a single
effect.

{\bf The property.\ }
We define the propositional safety property $\Phi$ as
$$
\Phi:  {\bf G} \neg \halted.
$$

It is easy to see that the runs of $\cts{\sys}$ correspond ono-to-one to the runs of {\em TM}.
Since $\Phi$ is a linear-time property, this run correspondence suffices to
guarantee that $\cts{\sys} \models \Phi$ if and only if {\em TM} does not halt.
\end{proof}

\begin{proof}[of Theorem \ref{thm:decidability-det}]
The proof is directly obtained from Theorem~\ref{thm:propositional-det}, 
noticing that model checking of propositional
$\mu$-calculus formulae over finite transition systems is decidable \cite{Emerson96}.
\end{proof}

\noindent
{\bf Proof of Theorem~\ref{thm:finitestate-det}.\ }
In view of proving this result, we first introduce a key lemma. 
We say that a transition system is {\em adom-inflationary} if the active domain of every
state is included in its successor's active domain.
We say that a \dcds \sys is adom-inflationary if $\cts{\sys}$ is adom-inflationary.
We can show that, for adom-inflationary transition systems, persistence-preserving
bisimilarity coincides with history-preserving bisimilarity.
 
\begin{lemma}\label{lem:adom-infl}
Consider two adom-inflationary $\dcds$s with non-deterministic services, 
$\sys_1,\sys_2$. Then $\cts{{\sys_1}} \pbsim \cts{{\sys_2}}$ if and only if 
$\cts{{\sys_1}} \hbsim \cts{{\sys_2}}$. 
\end{lemma}
\begin{proof}[of Lemma~\ref{lem:adom-infl}]
A comparison the two notions of bisimilarity reveals that the difference is
in the local condition, as follows.

Notice first that what both bisimilarity notions have in common is that they
mention bisimilar states $s_1$ and $s_2$ and witness isomorphism $h$, and their successors
$s_1 \Longrightarrow s_1'$, $s_1 \Longrightarrow s_2'$ such that $s_1'$ and $s_2'$
are bisimilar as witnessed by isomorphism $h'$.
The key difference lies in how $h'$ and $h$ are related. 
In the history-preserving flavor, $h'$ must extend $h$, while 
in the persistence-preserving flavor $h'$ must only extend $h\mid_{\adom{s_1} \cap \adom{s_1'}}$.

Clearly, history-preserving bisimilarity implies persistence-preserving bisimilarity.
However, notice that if the transition systems are adom-inflationary, then the
converse also holds. Indeed, assume $s_1 \pbsim_h s_2$. By definition, $h'$ extends 
$h\mid_{\adom{s_1} \cap \adom{s_1'}}$.
But because of adom-inflation, $\adom{s_1} \subseteq \adom{s_1'}$ and hence 
$\adom{s_1} \cap \adom{s_1'} = \adom{s_1}$, yielding 
$h\mid_{\adom{s_1} \cap \adom{s_1'}} = h$. \mrem{Is this true? In
  history preserving bisimulation, the domain of h could be greater
  than the active domain!} Hence, $h'$ extends $h$, which is the
condition for history-preserving bisimilarity.
\end{proof}

\begin{proof}[of Theorem~\ref{thm:finitestate-det}]
We prove the result by exploiting the reduction postulated by 
Theorem~\ref{thm:det-2-nondet}.

Starting from run-bounded \dcds $D$ with deterministic services, 
the reduction gives us state-bounded \dcds $N$ with non-deterministic services.
Moreover, the two transition systems have the same domains, and
the projection of $\cts{N}$ on the schema of $D$ coincides with $\cts{D}$ 
(Theorem~\ref{thm:det-2-nondet}(ii)).
In more detail, denoting the schema of $D$ with $\schema_D$ and the schema of $N$ with $\schema_N$,
there is a bijection $\beta$ between the states of $\cts{D}$ and the states of $\cts{N}$,
such that $s = \beta(s)\mid_{{\schema_D}}$. Clearly, this implies that 
$\cts{D}$ and $\cts{N}$ satisfy the same $\mu\L$ formulae.

However, a weaker statement suffices for our purpose.
By definition of history-preserving bisimilarity, Theorem~\ref{thm:det-2-nondet}(ii)
implies that 
\begin{quote}
(1)\ 
$\cts{D} \hbsim \cts{N}$.
\end{quote}

We recall that on the way to proving
Theorem~\ref{thm:sbounded-decid}, it is shown in Theorem~\ref{thm:algo} that since $N$ is state-bounded,
we can construct using algorithm \ERP\ a finite-state abstract transition system $F$ such
that $\cts{N} \pbsim \cts{F}$ ($F$ is an eventually recycling pruning of $\cts{N}$).

An inspection of the reduction in Theorem~\ref{thm:det-2-nondet} reveals
that $\cts{N}$ is adom-inflationary. But since $\cts{N} \pbsim \cts{F}$, it follows
that $\cts{F}$ is adom-inflationary as well 
(by the local condition of persistence-preserving bisimilarity).
Thus Lemma~\ref{lem:adom-infl} applies, yielding

\begin{quote} 
(2)\  
$\cts{N} \hbsim \cts{F}$.
\end{quote}

By (1) and (2), and by transitivity of $\hbsim$, we obtain that $\cts{D} \hbsim \cts{F}$. 
\end{proof}

\begin{proof}[of Theorem \ref{thm:propositional-det}]

Theorem \ref{thm:finitestate-det} implies that, given a \dcds \sys,
there exists a finite-state transition system $\sts{\sys} =
\tup{\HERBRAND,\schema,\Sigma_a,s^a_0,\db_a,\Longrightarrow_a}$ that is
history preserving bisimilar to the concrete transition system
$\cts{\sys} =
\tup{\HERBRAND,\schema,\Sigma,s_0,\db,\Longrightarrow}$. Thus, it is possible to use $\sts{\sys}$ in place of $\cts{\sys}$ for
verification. In particular, given a \muladom property $\Phi$, the
verification problem is reduced to $\sts{\sys} \models \Phi$.
Let $\adom{\sts{\sys}} = \bigcup_{s_i \in \Sigma} \adom{\db(s_i)}$.
If $\sts{\sys}$ is finite-state, then there exists a bound $b$ such
that $|\adom{\sts{\sys}}| < b$. Consequently, it is possible to
transform $\Phi$ into an equivalent \emph{finite} propositional
$\mu$-calculus formula $\prop{\Phi}$ as follows:
\begin{align*}
\prop{Q} & = Q\\
\prop{\neg \Psi} & = \neg \prop{\Psi}\\
\prop{\Psi_1 \land \Psi_2} & = \prop{\Psi_1} \land \prop{\Psi_2}\\
\prop{\DIAM{\Psi}} & = \DIAM{\prop{\Psi}}\\
\prop{Z} & = Z\\
\prop{\mu Z.\Psi} & = \mu Z. \prop{\Psi}\\
\prop{\exists x.\inadom(x)\land \Psi(x)} & = \bigvee_{t_i \in
  \adom{\sys}} \inadom(t_i) \land \prop{\Psi(t_i)}
\end{align*}
Clearly, $\sts{\sys} \models \Phi$ if and only if $\sts{\sys} \models \prop{\Phi}$.
The proof is then obtained by observing that verification of
$\mu$-calculus formulae over finite transition systems is decidable \cite{Emerson96}.
\end{proof}

\begin{proof}[of Theorem~\ref{thm:faithful-abstraction}]
The Theorem is proved by exhibiting,
for every $n$, a $\muL$ property that requires the existence of at least $n$ objects
in the transition system.

Let $\sys = \tup{\dl,\pl}$ be a \dcds with data layer
$\dl=\tup{\CONST, \schema, \emptyset, \idb}$ and process layer $\pl=\tup{\FUNC,\aset,\rset}$,
where $\FUNC = \{f/1\}$, $\schema = \{R/1, Q/1\}$, $\idb =
\{R(a)\}$, $ \rset = \{R(x)
\mapsto \alpha(x)\}$ and $\aset = \{\alpha(p)\}$, where $\alpha(p): \{\true
\rightsquigarrow \{Q(f(p))\}\}$.
The concrete transition system $\cts{\sys}$ has the following shape:
\begin{compactitem}
\item The initial state is $s_0 = \tup{\{R(a)\},\emptyset}$;
\item $s_0$ is connected to infinitely many successor states, each one storing
  into $Q$ a distinct value $d$ resulting from the service call $f(a)$; each
  such state has then the form $s_d = \tup{\{Q(d)\},\{f(a) \mapsto d\}}$;
\item each $s_d$ has no outgoing edge, because there is no applicable
  action in $s_d$.
\end{compactitem}
$\sys$ is clearly run-bounded, in particular by a bound $b=3$.

Let us now consider the following $\mu\L$ property
without fixpoints:
\[
  \Phi_n = \exists x_1,\ldots,x_n.\bigwedge_{i\neq j} x_i \neq x_j \land
  \bigwedge_{i \in \{1,\ldots,n\}}\DIAM{Q(x_i)}
\]
The property states that there are $n$ distinct values, each of which
is stored into relation $Q$ in one of the successors of the initial
state. It is easy to see that $\cts{\sys} \models \Phi_n$ for every
$n$. 
On the other hand, for every finite state abstraction $\sts{\sys}$
with $k$ successors of the initial state, we have  that $\sts{\sys} \not \models \Phi_{k+1}$.
\end{proof}

\subsection{Weakly Acyclic $\dcds$s}

\begin{proof}[of Theorem \ref{thm:det-boundedness}]
The proof is by reduction from the halting problem.
We reuse without change the reduction in the proof of Theorem~\ref{thm:ltl-undecid-det}.
This reduction yields for any Turing Machine {\em TM} a \dcds with deterministic services \sys,
such that \sys simulates {\em TM}'s computation. That is, the runs of {\em TM} correspond
one-to-one to the runs of $\cts{\sys}$.
It follows immediately that {\em TM} halts if and only if \sys is run-bounded.
\end{proof}

\begin{proof}[of Lemma \ref{lemma:approximate-det}]
Let $\sys = \tup{\dl,\pl}$ be a \dcds with data layer
$\dl=\tup{\CONST, \schema, \EC, \idb}$ and process layer
$\pl=\tup{\FUNC,\aset,\rset}$. Consider now $\cts{\sys} =
\tup{\CONST,\schema,\Sigma,s_0,\db,\Longrightarrow}$ and $\cts{{\pos{\sys}}} =
\tup{\CONST,\schema,\pos{\Sigma},s_0,\db,\pos{\Longrightarrow}}$.
Since $\pos{\sys}$ is weakly acyclic by hypothesis, to prove that run boundedness of $\cts{{\pos{\sys}}}$
implies run boundedness of $\cts{\sys}$, we show
the following stronger result: for every run $\tau$ in $\cts{\sys}$,
there exists a run $\pos{\tau}$ in $\pos{{\cts{\sys}}}$
such that, for all pairs of states $\tau(i) =  \tup{\I_i,\rmap_i}$ and
$\pos{\tau}(i) =  \tup{\pos{\I_i},\pos{\rmap_i}}$, we have
\begin{enumerate}
\item $\pos{\rmap_i}$
extends $\rmap_i$;
\item $\I_i \subseteq \pos{\I_i}$;
\item for the mappings mentioned in $\pos{\rmap_i}$ but not in
  $\rmap_i$, $\pos{\rmap_i}$ ``agrees'' with the maps contained in the
  suffix of $\tau[i]$, i.e., 
\[\restrict{\pos{\rmap_i}}{C_i} =\restrict{(\bigcup_{j>i} \rmap_j)}{C_i}\]
where $C_i = \domain{\pos{\rmap_i}} \cap
    \bigcup_{j>i} \domain{\rmap_j}$.
\end{enumerate}
We prove this by induction on the length of $\tau$:
\begin{description}
\item[(base case)] The initial state of both runs is $\tau(0)
  =\pos{\tau}(0)=\tup{\idb,\emptyset}$, and therefore all the three
  conditions are trivially satisfied.
\item[(inductive step)] Consider a pair of
  corresponding states $\tau(i)$
and $\pos{\tau}(i)$, with $i>0$. By definition, $\tau(i)
\Longrightarrow \tau(i+1)$ means that there exists an action $\alpha \in \aset$ and a
substitution $\sigma$ for the parameters of $\alpha$ such that
$\tup{\tau(i),\alpha\sigma,\tau(i+1)} \in \rexec{\sys}{}{}$. We first
observe that $\pos{\alpha}$ can be executed in $\pos{\tau}(i)$, since $\pos{\pl}$ does not
impose any restriction on the executability of actions. Let
$\pos{Next} = \{ \pos{s} \in \pos{\Sigma} \mid
\tup{\pos{\tau}(i),\alpha,\pos{s}} \in \rexec{\pos{\sys}}{}{}
\}$ be the set of successor states of $\pos{\tau}(i)$ that are obtained from
the application of $\pos{\alpha}$. 

We now show that there exists $\underline{s} \in \pos{Next}$ that satisfies
the three claims above. The proof is then obtained by simply imposing
$\pos{\tau}(i+1) = \underline{s}$.
\begin{enumerate}
\item 
By definition, $\domain{\rmap_{i+1}} =
\domain{\rmap_i} \cup \skolems{\doo{}{}{\I_i,
    \alpha\sigma}}$, and, for every $s_k =
\tup{\pos{\rmap}_k,\pos{\I}_k} \in \pos{Next}$, we have $\domain{\pos{\rmap}_k} =
\domain{\pos{\rmap}_i} \cup \skolems{\doo{}{}{\pos{\I}_i,
    \pos{\alpha}\sigma}}$.
Consider each effect specification $\map{q^+_j\land Q^−_j}{E_j} \in
\effect{\alpha}$. By definition of $q^+_j$ and $Q^-_j$, $\theta \in \ans{(q^+_j\land Q^+_j)\sigma,\I_i}$
implies $\theta \in \ans{q^+_j\sigma,\I_i}$, which in turn implies
$\theta \in \ans{q^+_j\sigma,\pos{\I}_i}$, because $\I_i
\subseteq \pos{\I_i}$ by induction hypothesis. Consequently, we have
$\doo{}{}{\I_i, \alpha\sigma} \subseteq \doo{}{}{\pos{\I}_i,
  \pos{\alpha}\sigma}$,
and hence $\skolems{\doo{}{}{\I_i,
   \alpha\sigma}} \subseteq \skolems{\doo{}{}{\pos{\I}_i,
  \pos{\alpha}\sigma}}$. Since $\domain{\rmap_i} \subseteq
\domain{\pos{\rmap}_i}$ by induction hypothesis, then we obtain
$\domain{\rmap_{i+1}} \subseteq \domain{\pos{\rmap_k}}$. Since
$\pos{\sys}$ has no equality constraint, the states in $\pos{Next}$ cover
every possible result obtained by calling the service call in
$\pos{\rmap_k}\setminus \pos{\rmap_i}$, including those states for
which $\pos{\rmap}_k$ is an extension of $\rmap_{i+1}$. We use
$\pos{\underline{Next}}$ to denote such states.
\item By definition, for each state $s_k =
\tup{\pos{\rmap}_k,\pos{\I}_k} \in \pos{\underline{Next}}$, we have
that $\pos{\rmap}_k$ extends $\rmap_{i+1}$. Therefore, since 
$\doo{}{}{\I_i, \alpha\sigma} \subseteq \doo{}{}{\pos{\I}_i,
  \pos{\alpha}\sigma}$, we have
$\I_{i+1} = \rmap_{i+1}(\doo{}{}{\I_i, \alpha\sigma} ) \subseteq \pos{\I}_{k} = \pos{\rmap}_{k}(\doo{}{}{\pos{\I}_i,
  \pos{\alpha}\sigma})$.
\item Since $\pos{\sys}$ has no equality constraints, we observe that
  the states in $\pos{\underline{Next}}$ cover all possible values for the service
  calls that are not mentioned in $\rmap_{i+1}$. Therefore, there must
  exist at least one state $\underline{s} \in \pos{\underline{Next}}$
  that satisfies the third claim. In other words, by imposing
  $\pos{\tau}(i+1) = \underline{s}$, we have
\[\restrict{\pos{\rmap_{i+1}}}{C_{i+1}} =\restrict{(\bigcup_{j>{i+1}}
  \rmap_j)}{C_{i+1}} \qed
\]
\end{enumerate}
 \end{description}
\end{proof}

\begin{proof}[of Theorem \ref{thm:weak-acyclicity}]
%
%
%

Let $\sys = \tup{\dl,\pl}$ be a \dcds with data layer $\dl=\tup{\CONST, \schema, \EC, \idb}$ and process layer
$\pl=\tup{\FUNC,\aset,\rset}$. 
We consider the positive approximate $\pos{\sys}$, showing that if the
the dependency graph $G = \tup{N,E}$ of $\sys$ (which corresponds by definition to the
one of $\pos{\sys}$) is weakly acyclic, then $\pos{\sys}$ is
run-bounded. 
The complete proof is then directly obtain by appealing to Lemma
\ref{lemma:approximate-det}, which states that if $\pos{\sys}$ is run-bounded, then $\pos{\sys}$ is
run-bounded as well. 

To prove that weak acyclicity of $\sys$ implies that $\pos{\sys}$ is
run-bounded, we exploit the connection with the chase of a set of
tuple generating dependencies (TGDs) in data exchange.
In particular, we resort to the proof given in
\cite{Kolaitis05:Exchange}, Theorem 3.9.
  For every node $p \in N$, we consider
  an incoming path to be any (finite or infinite) path ending in
  $p$. For simplicity, we say that a value appears in position
$p = \tup{R_k,j} \in N$ if it appears in the $j$-th component of an
$R_k$ tuple.
We define the rank of $p$, denoted \underline{rank}$(p)$, as
  the maximum number of special edges on any such incoming path.
  Since $\pos{\sys}$ is weakly acyclic by hypothesis, $G$ does not contain cycles going
  through special edges, and therefore \underline{rank}$(p)$ is finite. Let $r$
  be the maximum among \underline{rank}$(p_i)$ over all nodes. We
  observe that $r \leq |N|$; indeed no path can lead to the same node twice using special edges,
  otherwise $G$ would contain a cycle going through special
  edges, thus breaking the weak acyclicity hypothesis. Notice also
  that $|N|$ is a constant value, because it is obtained from $\R$,
  which is fixed.
We now partition the nodes in $N$ according to their rank, obtaining a
set of sets $\set{N_0, N_1, \ldots, N_r}$, where $N_i$ is the set of all nodes
  with rank $i$. 
 The proof is then a natural consequence of the following claim:

\begin{quote}
{\bf Claim.}  Consider a trace $\tau$ in $\cts{{\pos{\sys}}}$. For
every $i \in \{1,\ldots,r\}$, the total number
of distinct values occurring in the databases of $\tau$ inside
position $p \in N_i$ is bounded by a polynomial $P_i(|\adom{\idb}|)$. 
\end{quote}
%
 We prove the claim by induction on $i$:
\begin{description}
  \item[(Base case)] Consider $p \in N_0$. By
    definition, $p$ has no
  incoming path containing special edges. Therefore, no new values are
  stored in $p$ along the
  run: $p$ can just store values that are part of the initial database
  $\idb$.  This holds for all nodes in $N_0$, and hence we can fix $P_0(|\adom{\idb}|) = |\adom{\idb}|$.
  \item[(Inductive step)] Consider $p \in N_i$, with $i \in \{1,\ldots,r\}$. The first kind of values that may be stored 
  inside $p$ are those values that
  were stored inside the component itself in $\idb$. The number of such values is at most
  $|\adom{\idb}|$. In addition, a value may be stored in $p$ for two reasons: either it is copied from some
  other position $p' \in N_j$ with $i\neq j$, or it is generated by
  means of a service call. 

We first determine how many fresh values can be generated by service
calls. 
The possibility of generating and storing a new value in $p$ as a
  result of an action is reflected by the presence of special
  edges. By definition, any special edge entering $p$ must start from
  a node $p' \in N_0 \cup \ldots \cup
  N_{i-1}$. By induction hypothesis, the number of distinct
  values that can exist in $p'$ is bounded by $H(|\adom{\idb}|) =
  \sum_{j \in \{0,\ldots,i-1\}}P_j(|\adom{\idb}|)$. Let $b_a$
  be the maximum number of special edges that enter a position, over
  all positions in the schema; $b_a$ bounds the arity taken by
  service calls in $\FUNC$.
Then for every choice of $b_a$
  values in $N_0 \cup \ldots \cup N_{i-1}$ (one for each
  special edge that can enter a position) and for every action in
  $\pos{\aset}$, the number of new values generated at position $p$ is
  bounded by $t_{f}\cdot H(n)^{b_a}$, where $t_f$
  is the total number of facts mentioned in the effects of actions
  that belong to  $\pos{\aset}$. 
Notice that this number does not depend on the data in $\idb$. 
By considering all positions in $N_i$, the total number of values that
  can be generated is then bounded by $G(|\adom{\idb}|) = |N_i| \cdot t_f\cdot
  H(|\adom{\idb}|)^{b_a}$. Obviously, $G(\cdot)$
  is a polynomial, because $t_f$ and $b_a$ are values extracted from
  the schema $\schema$ of the \dcds, which is fixed. 

We count next the number of distinct values that
  can be copied to positions of $N_i$ from positions of $N_j$, with
  $j\neq i$. A copy is represented in the graph as a normal edge going
  from a node in $N_j$ to a node in
  $N_i$, with $j \not= i$. We observe first that such normal edges can start
  only from nodes in $N_0 \cup \ldots \cup N_{i-1}$, that is, they
  cannot start from nodes in $N_j$ with $j > i$. We prove this by
  contradiction. Assume that there exists $\tup{p',p,\false}\in E$,
  such that $p \in N_i$ and $p' \in N_j$ with $j > i$. In this case, the rank of
  $p$ would be $j > i$, which contradicts the fact that $p \in
  N_i$. As a consequence, the number of distinct values that can be
  copied to positions in $N_i$ is bounded by the total number of
  values in $N_0 \cup \ldots \cup N_{i-1}$, which corresponds to $H(|\adom{\idb}|)$ from our
  previous consideration. Putting it all together, we define $P_i(|\adom{\idb}|)
  = |\adom{\idb}| + G(|\adom{\idb}|) + H(|\adom{\idb}|)$. $P_i(\cdot)$ is a
  polynomial, and therefore the claim is proven.  
\end{description}
In the above claim, $i$ is bounded by the maximum rank $r$,
  which is a constant. Hence, there exists a fixed polynomial $P(\cdot)$ such
  that the number of distinct values that can exist in the active
  domains of the run $\tau$ is bounded by
  $P(|\adom{\idb}|)$. Technically, given $\cts{{\pos{\sys}}} = \tup{\CONST, \R, \Sigma, s_0, \db,
  \Longrightarrow}$, we have: \[|\bigcup_{s \textrm{ state of }
    \tau} \db(s)| < P(|\adom{\idb}|)\]
which attests that $\tau$ is (data) bounded, and consequently that
$\sys$ is run-bounded.
\end{proof}


%% file: appendix-nondet-verification-results.tex
\section{Nondeterministic Services}

\addtocounter{subsection}{1}

\subsection{State-bounded Systems}\label{sec:nondet-results}

\begin{proof}[of Theorem~\ref{thm:ltl-undecid}]
We reuse the proof of Theorem~\ref{thm:ltl-undecid-det}. 
Recall that the reduction in this proof constructs for every Turing Machine {\em TM}
a \dcds\ {\em with deterministic services} \sys that simluates the computation of {\em TM}.
It also constructs a propositional safety property $\Phi$ such that $\cts{\sys} \models \Phi$
if and only if {\em TM} halts.

What we need here is a reduction to a \dcds with {\em nondeterministic services}.
However, we recall from the proof of Theorem~\ref{thm:ltl-undecid-det} that the only service 
in the process layer, service \new, is guaranteed to be called only with distinct arguments 
across distinct transitions, and so its behavior is unaffected by the choice of deterministic 
versus nondeterministic semantics. Therefore, the reduction applies unchanged to \dcds with 
nondeterministic services.
\end{proof}

\begin{proof}[of Theorem~\ref{thm:sbounded-undecid}]
We prove a stronger result, namely for {\em linear-time} \muladom sentences.
Such sentences can be written using LTL syntax.

We reduce from the problem of satisfiability of LTL with freeze quantifier over infinite data
words, known to be highly undecidable ($\Sigma_1^1$-hard) \cite{Demri:2009:LFQ}.\\

\noindent
{\bf Infinite data words~\cite{Demri:2009:LFQ}.\ }
Let $\Sigma$ be a finite alphabet of labels and $D$ an infinite set
of data values. An infinite \emph{data word} $w = \set{w_i}$
is an infinite sequence over $\Sigma\times D$, i.e., each
$w_i$ is of the form $(a_i, d_i)$ with $a_i\in\Sigma$ and $d_i\in D$.\\

\newcommand{\ltlfreeze}{\ensuremath{\mbox{LTL}^\downarrow}}
\noindent
{\bf LTL with freeze quantifier (\ltlfreeze).\ } 
This logic operates over infinite data words, seen
as runs. It extends propositional LTL with a finite 
number of {\em registers}, which can record the data value at the current step of the run
(position in the data word), and recall it at subsequent steps. 
The operation of recording the data value at the current position into register $i$ is denoted 
with $\downarrow_i$. $\uparrow_i$ denotes the boolean comparison of the data value at the current 
position with the value stored in register $i$. 

As an example, consider the \ltlfreeze\ sentence 
$$
\varphi_{ex}=\downarrow_1 X (G(a\implies \neg \uparrow_1))
$$ 
over alphabet $\set{a,b}$, which states that the data value assigned to each
label $a$ at positions greater than one is different from the data value at the first position
of the data word. Notice that the data value at the first position is recorded in register
$1$ by operation $\downarrow_1$, and it is compared to subsequent data values by $\uparrow_1$.\\

\noindent
{\bf The \dcds construction.\ }  
Given a finite alphabet $\Sigma = \{\sigma_i\}_{i \in \{1,\ldots,n\}}$, we build a DCDS
$\sys =\tup{\D_\Sigma, \P_\Sigma}$ with nondeterministic services, such that each run of $\cts{\sys}$
represents an infinite data word over $\Sigma$. In particular, each state in the run
holds the label and data value for a single position in the data word.
Moreover, given an \ltlfreeze\ sentence $\varphi$ over $\Sigma$, we
construct a \muladom formula $\Phi$, such $\cts{\sys} \models \Phi$ if and only
if $\varphi$ is {\em unsatisfiable}.

The idea is to model the registers with existentially quantified variables, which
\muladom allows us to introduce at any given point in the run and use subsequently,
even if in between their binding does not persist in the run.

More precisely, we define the data layer $\D_\Sigma$ of \sys as
$\D_\Sigma=\tup{\C, \R, \emptyset, \idb}$, where
$\C= \Sigma \cup \set{0}$,
$\R=\set{\textsc{label}/1, \textsc{datum}/1}$, and $\idb=\emptyset$.
Intuitively, $\textsc{label}$ stores the label and 
$\textsc{datum}$ the data value. We then define the process layer
$\P_\Sigma$ of \sys as $\P_\Sigma = \tup{\F,\aset_\Sigma, \rset_\Sigma}$,
where:
\begin{compactitem}
\item $\F=\set{f/0}$. 
\item For each $1 \in \{1,\ldots,n\}$, $\rset_\Sigma$ contains an action $\alpha_i$ with
no parameters and no guard ($\true \mapsto \alpha_i$). 
\item Each $\alpha_i \in \aset_\Sigma$ contains a single
effect $e_i$, which
creates the position of a data word corresponding to label $\sigma_i \in \Sigma$:
\begin{equation*}
\begin{array}{rrl}
  e_i: &\true & \rightsquigarrow  \textsc{label}(\sigma_i) \land \textsc{datum}(f())
\end{array}
\end{equation*}
\end{compactitem}
The service call $f()$ is used to get an arbitrary data value from the
domain during the action execution. It is nondeterministic, and will
therefore return possibly distinct values across the run.

Since actions are always executable, at each step of a run all of them 
qualify, and one is nondeterministically chosen. In this way, the collection of all runs
corresponds to all possible infinite data words. Observe that \sys\ is state-bounded,
as each state contains just one label and one data value.\\

\noindent
{\bf The property.\ }
We now define the property. For simplicity of presentation, we show it using an LTL-based
syntax (branching is irrelevant here), though it is clearly expressible in \muladom.

We obtain $\varphi'$ from $\varphi$ by:
\begin{enumerate}
\item replacing each freeze quantifier $\downarrow_n$ with
  $\exists x_n. \textsc{datum}(x_n)$, and 
\item replacing each occurrence of $\uparrow_n$ with $\textsc{datum}(x_n)$, and
\item replacing each proposition $\sigma \in \Sigma$ with $\textsc{label}(\sigma)$.
\end{enumerate}
Now let $\Phi := \neg \varphi'$.

We illustrate the rewrite on property $\varphi_{ex}$ above, obtaining
$$
\varphi_{ex}' := \exists x_1 \textsc{datum}(x_1) \land {\bf X\ G}
                 (\textsc{label}(a) \implies ~\neg \textsc{datum}(x_1)).
$$

It is easy to see that $\varphi$ is unsatisfiable over infinite data words using 
alphabet $\Sigma$  if and only if $\cts{\sys} \models \Phi$.

As a result, \muladom verification by state-bounded $\dcds$s with nondeterministic
services is undecidable.
\eat{
  The correscponding transition system is shown in the figure
  \ref{fig:openNonDetUndecidability}.

\begin{figure}[t]
\begin{center}
\include{fig-openNonDetUndecidability}
\end{center}
\caption{Transition system for Ex.\
  \ref{ex:openNonDetUndecidability} \label{fig:openNonDetUndecidability}}
\end{figure}

The corresponding $\textsc{CTL}^*_{G}$ property is as follows:
\[ \exists ~ X ~ \exists_x \textsc{datum}(x) \land G
(\textsc{label}(a) \implies ~\neg \textsc{datum}(x))\]
}
\end{proof}

\begin{proof}[of Theorem~\ref{thm:sbounded-decid}]
See Section~\ref{app:nondet-abstract}.
\end{proof}

\subsection{Abstract Transition System}\label{app:nondet-abstract}
We formalize the discussion from Section~\ref{sec:nondet-abstract}.
Since $\dcds$s with nondeterministic services are modeled by means of
transition systems whose states are constituted by database instances, with a slight
abuse of notation we will directly use the state to refer to its database instance.
\\ 

\noindent
{\bf  Equality commitments.\ }
Consider a set $D$ comprised of constants and of Skolem terms built by applying a Skolem function
to constant arguments. An {\em equality commitment} $\H$ on $D$ is a partition
of $D$, i.e. a set of disjoint subsets of $D$, called {\em cells}, such that the union
of the cells in $\H$ is $D$. Moreover, each cell contains at most one
constant (but arbitrarily many Skolem terms). For any $e \in D$, $[e]_\H$ denotes
the cell $e$ belongs to. The intention of the partition is to model equality and non-equality
commitments on the members of $D$ as follows:
for every $e_1,e_2 \in D$, $e_1 = e_2$ if and only if $[e_1]_\H = [e_2]_\H$.\\

\noindent
{\bf Service call evaluations that respect equality commitments.\ }
It is convenient to view the concrete transition system $\ncts{\sys}$ in the following equivalent formulation,
which emphasizes equality commitments on the service calls:
successor states are built by picking an equality commitment $\H$, 
and then picking a service call evaluation that respects $\H$. More specifically,
\begin{itemize}
\item
for each state $\I$, 
\item
for each action $\alpha$, 
\item
for each parameter choice $\sigma$, and 
\item
for each equality commitment $\H$ involving the service calls in 
$\skolems{\doo{}{}{\I, \alpha,\sigma}}$ and the values in $\adom{\I} \cup \adom{\idb}$,
\end{itemize}
$\ncts{\sys}$ contains possibly infinitely many successor states $\I_{next}$, each obtained 
from $\doo{}{}{\I, \alpha,\sigma}$ by picking a service call evaluation that respects $\H$. 
We say that evaluation $\theta$ {\em respects} $\H$ if for every two terms 
$t_1,t_2 \in \skolems{\doo{}{}{\I, \alpha,\sigma}} \cup \adom{\I} \cup
\adom{\idb}$, we have
$[t_1]_\H = [t_2]_\H$ if and only if $t_1\theta = t_2\theta$.

Given $\I,\alpha,\sigma$ and $\H$, 
we denote the set of all legal evaluations with
\begin{tabbing}
$\returns^\H(\I,\alpha,\sigma) := \{\theta\ $ \= $|\ \theta \in \groundexec{\CONST}{\I, \alpha,\sigma}, 
             \theta \mbox{ respects } \H,$\+\\
             $\doo{}{}{\I, \alpha,\sigma}\theta \models \EC \}$.
\end{tabbing}
Notice that we consider legal only those evaluations that respect the equality commitment $\H$
and that, conforming to the semantics of the concrete transition system, generate successors  
which satisfy the constraints $\EC$.
Finally, notice that $\H$ determines an isomorphism type, as all successors of $\I$
generated by the evaluations in $\returns^\H(\I,\alpha,\sigma)$ are isomorphic to each other.
\\

\noindent
{\bf Prunings.\ }
We observe that for each state $\I$ of the concrete transition system $\ncts{\sys}$, 
the number of possible choices of $\alpha, \sigma$ and $\H$ are finite.
The sole reason for infinite branching in $\ncts{\sys}$ are the infinitely many distinct evaluations 
that respect $\H$, whenever $\H$ states that at least one service call result is distinct from 
$\adom{\I} \cup \adom{\idb}$: in that case, the service call can be substituted with any value in 
$\CONST \setminus (\adom{\I} \cup \adom{\idb})$.

In contrast, we obtain a finitely-branching transition system if instead of keeping the successors 
generated by {\em all} evaluations in $\returns^\H(\I,\alpha,\sigma,\H)$, 
we keep the successors generated by a {\em finite subset} of these evaluations
(if $\returns^\H(\I,\alpha,\sigma)$ is non-empty, we pick a non-empty subset, to ensure that
if $\H$ is represented among the successors of $\I$ in $\ncts{\sys}$,
it is also represented among the successors of $\I$ in $\nsts{\sys}$).
We call any transition system obtained in this way a {\em pruning} of $\ncts{\sys}$, and we denote
with $\prunings(\ncts{\sys})$ the set of all such prunings.
By construction, every pruning of $\ncts{\sys}$ is finitely branching.

\arem{mention minimal prunings, which keep singleton subsets}

Formally, let $\sys$ be a \dcds and $\ncts{\sys}$ its concrete transition system, 
with states $\Sigma_C$ and initial state $\idb$.
A {\em pruning of $\ncts{\sys}$} is the restriction of $\ncts{\sys}$ to a subset of states 
$\Sigma_P \subseteq \Sigma_C$, where $\Sigma_P$ satisfies the following properties:
\begin{itemize}
\item[(i)] $\idb \in \Sigma_P$, and
\item[(ii)] 
for each $\I \in \Sigma_C$ and each equality commitment $\H$, if $\H$ is represented by
some successor of $\I$ in $\ncts{\sys}$, it is also represented by a successor of $I$ in $\nsts{\sys}$.
We say that $\H$ is {\em represented} by successor $\I'$ of $\I$ if there exist
$\alpha,\sigma$ and $\theta \in \returns^\H(\I,\alpha,\sigma)$ 
such that $\tup{\I,\alpha\sigma\theta,\I'} \in \nrexec{\sys}$.
\item[(iii)]
for each $\I \in \Sigma_C$, the number of successors of $\I$ that are also in $\Sigma_P$ is finite.
\end{itemize}

Clearly, a concrete transition system $\ncts{\sys}$ admits (potentially infinitely) many prunings,
but we show next that they all are persistence-preserving bisimilar to $\ncts{\sys}$ 
(and therefore to each other, due to transitivity of the $\pbsim$ relation):
\begin{lemma}\label{lem:pruning}
For every concrete transition system $\ncts{\sys}$ and pruning $\nsts{\sys} \in \prunings(\ncts{\sys} )$, we have that $\nsts{\sys}  \pbsim \ncts{\sys} $.
\end{lemma}

The result follows from the fact that state isomorphism implies persistence-preserving bisimilarity.
In the following, we denote with $s \iso{h} s'$ the fact that $h$ is an isomorphism from state $s$ 
to $s'$.
\begin{lemma}\label{lem:iso-is-bisim}
Consider a concrete transition system $\ncts{\sys}$ with initial state $s_0$ and one of its prunings 
$\nsts{\sys}$.
Let $s_C$ be a state of $\ncts{\sys}$ and $s_P$ a state of $\nsts{\sys}$. If there exists function $h$ such that $h$ fixes
$\adom{s_0}$ and $s_P \iso{h} s_C$, then $s_P \pbsim_h s_C$.
\end{lemma}

\begin{proof}[of Lemma~\ref{lem:iso-is-bisim}]
Let $\ncts{\sys} = \tup{\CONST, \R, \Sigma_C,
    s_0, \db,  \Longrightarrow_C}$ and $\nsts{\sys} = \tup{\CONST, \R, \Sigma_P,
    s_0, \db,  \Longrightarrow_P}$. 
The proof follows from the following claim:
\begin{quotation}
\noindent {\bf Claim 1.} Given $s_C \in \Sigma_C$ and $s_P \in
\Sigma_P$, if $s_P \iso{h} s_C$ and $h$ is the identity on $\adom{\idb}$, then 
for each $s_C'$ such that $s_C \Longrightarrow_C s_C'$ 
there exist $s_P'$ and $h'$ such that 
\begin{itemize}\parskip=0in\itemsep=0in
\item[(i)] $s_P \Longrightarrow_P s_P'$; 
\item[(ii)]  $h'$ is an extension of $h\mid_{\adom{s_P} \cap \adom{s_P'}}$; 
\item[(iii)] $h'$ is the identity on $\adom{\idb}$; 
\item[(iv)]  $s_P' \iso{h'} s_C'$.
\end{itemize}
\end{quotation}
Indeed, this claim allows us to exhibit the bisimilarity relation 
$$R = \{ (x, i, y)\ |\ x \in \Sigma_P, y \in  \Sigma_C, x \iso{i} y \}.$$ 
$R$ is a bisimilarity relation because it satisfies the forth condition 
in the definition of persistence-preserving bisimilarity by Claim 1.
It trivially satisfies the back condition because $P$ is constructed by picking a subset
of the states of $\ncts{\sys}$. Since by construction 
$(s_P, h, s_C) \in R$, we have $s_P \pbsim_h s_C$.

To prove Claim 1, we observe that the successor $s_C'$ of $s_C$ is generated by a
particular choice of the 
action $\alpha$ (with condition-action rule $Q \mapsto \alpha$), 
the parameter instantiation $\sigma_C$ (such that $s_C \models Q\sigma_C$),
the equality commitment $\H_C$ on $\skolems{\doo{}{}{s_C, \alpha,\sigma_C}} \cup \adom{s_C} \cup \adom{\idb}$,
and the service call evaluation $\theta_C \in \returns^{\H_C}(s_C,\alpha,\sigma_C)$: 
$s_C' = {\doo{}{}{s_C, \alpha,\sigma_C}}\theta_C$.
We show how to construct $\sigma_P$, $\H_P$ and $\theta_P \in \returns^{\H_P}(s_P,\alpha,\sigma_P)$ 
such that $s_P \models Q\sigma_P$ and $s_P'={\doo{}{}{s_P, \alpha,\sigma_P}}\theta_P$ satisfies the claim.

We let $\sigma_P = h^{-1}(\sigma_C)$, observing that since $Q$ is a first-order query,
it is preserved under isomorphism, so $s_C \models Q\sigma_C$ implies $s_P \models Q\sigma_P$.
Thus, $\sigma_P$ is a legal parameter instantiation.

To construct $\H_P,\theta_P$, we first show that 
$\bar s_C=\doo{}{}{s_C, \alpha,\sigma_C}$ and $\bar s_P = \doo{}{}{s_P, \alpha,\sigma_P}$ 
are isomorphic, as witnessed by the function 
$\bar h : \adom{\bar s_P} \rightarrow \adom{\bar s_C}$ defined as follows:
\begin{eqnarray*}
\bar h &  :=   & \{c \mapsto h(c)\ |\ c \in \adom{s_P} \cup \adom{\idb}\}\\
        & \cup & \{ f(m_P, \dots, m_n) \mapsto f(h(m_P), \dots, h(m_n))\ |\\\
        &      & \hspace{3mm}f(m_P, \dots, m_n) \in \skolems{\bar s_P} \}.
\end{eqnarray*}
From the definition of $\bar h$ and the fact that the service calls are generated by queries 
preserved under isomorphism, it follows immediately that $\bar s_P \iso{\bar h} \bar s_C$.
It is easy to see that $\bar h$ is also an isomorphism between
$
\skolems{\doo{}{}{s_C, \alpha,\sigma_C}}\cup \adom{s_C} \cup \adom{\idb}
$ 
and
$
\skolems{\doo{}{}{s_P, \alpha,\sigma_P}} \cup \adom{s_P} \cup \adom{\idb},
$
(i.e. $\bar h$ preserves the structure of Skolem terms), 
and therefore between the sets of corresponding equality commitments.

We therefore pick $\H_P = \bar h^{-1}(\H_C)$. By construction of $P$, each equality type is represented
among a state's successors in $P$, i.e. there exists $\theta_P$ that respects $\H_P$, and there exists $s_P' \in \Sigma_P$, 
such that $s_P' = \bar s_P \theta_P$.
  
The existence of legal choices for $\alpha,\sigma_P,\H_P$ and $\theta_P$ proves item (i) of Claim 1, 
namely that $s_P \Longrightarrow_P s_P'$.

To prove the remaining items, we exhibit $h'$ defined as follows:
$$
h'(t) := \bar h(\bar t)\theta_C, 
$$
for some choice of $\bar t$ such that $\bar t\theta_P = t$.

To see why $h'$ is well-defined,
observe that, by construction of the successor states in $\ncts{\sys}$, for each $t \in \adom{s_P'}$ there must exist 
$\bar t \in \adom{\bar s_P}$ such that $\theta_P$ evaluates $\bar t$ to $t$ ($\bar t\theta_P = t$). 
Moreover, observe that if there are distinct $\bar t, \bar u \in \adom{\bar s_P}$ 
such that $t = \bar t\theta_P = \bar u\theta_P$, it does not matter which one 
we pick in the definition of $h'$, since $\bar h(\bar t)\theta_C = \bar h(\bar u)\theta_C$. This is because $\theta_P$
respects $\H_P$,
and therefore $[\bar t]_{\H_P} = [\bar u]_{\H_P}$. Since $\H_P \iso{\bar h} \H_C$, it follows that
$[\bar h(\bar t)]_{\H_C} = [\bar h(\bar u)]_{\H_C}$, and since $\theta_C$ respects $\H_C$, we have that
$\bar h(\bar t)\theta_C = \bar h(\bar u)\theta_C$.

Items (ii), (iii) and (iv) of Claim 1 follow by similar reasoning 
from the fact that service call evaluations respect the equality commitments, which are isomorphic.
\end{proof}

\begin{proof}[of Lemma~\ref{lem:pruning}]
This is a corollary of Lemma~\ref{lem:iso-is-bisim}.

Indeed, by definition, $\nsts{\sys} \pbsim \ncts{\sys}$ holds if and only if the initial state $s_0^P$ of $\nsts{\sys}$ 
is bisimilar to the initial state $s_0^C$ of $\ncts{\sys}$, i.e. there exists isomorphism $h$ such that 
$s_0^P \pbsim_h s_0^C$.

By definition, a concrete transition system shares the initial state with all its prunings, 
so $s_0^P = s_0^C$. The identity mapping $id$ witnesses isomorphism: $s_0^P \iso{id} s_0^C$. 
By Lemma~\ref{lem:iso-is-bisim}, we have $s_0^p \pbsim_{id} s_0^C$.
\end{proof}

\noindent
{\bf Eventually Recycling Prunings.\ }
While all prunings of a concrete transition system are finitely-branching, they are not guaranteed
to be finite. The reason is that they don't necessarily rule out infinitely long 
simple runs $\tau$, along which the service calls return in each state $\I$ ``fresh'' values, 
i.e. values distinct from all values appearing in $\I$ and its predecessors on $\tau$. 
Towards addressing this problem, we focus on prunings in which the evaluations are not chosen
arbitrarily.

Given a finite run $\tau$ ending in state $\I$ of $\ncts{\sys}$, 
an action $\alpha$, a parameter choice $\sigma$ and an
equality commitment $\H$ on $\skolems{\doo{}{}{\I, \alpha,\sigma}}$, we say that 
evaluation $\theta \in \returns^\H(\I,\alpha,\sigma)$ {\em recycles from $\tau$} if each
value in the range of $\theta$ occurs in $\tau$. 
We say that pruning $\nsts{\sys}$ is {\em eventually recycling} if every (finite or infinite) path 
$\tau$ in $\nsts{\sys}$ contains only finitely many states generated by non-recycling evaluations.
Formally, if $\tau = s_0s_1\cdots$ and the service call evaluation 
used in $s_i \Longrightarrow_C s_{i+1}$ is denoted as $\theta_i$, 
then there are only finitely many indexes $j$ such that $\theta_j$ does not recycle from $\tau[j]$.

\begin{lemma}\label{lem:recycling}
Let $\ncts{\sys}$ be a concrete transition system.
\begin{itemize}
\item[(i)]
All eventually recycling prunings of $\ncts{\sys}$ are finite.
\item[(ii)]
If $\ncts{\sys}$ is state-bounded, then it has at least one eventually recycling pruning.
\end{itemize}
\end{lemma}

\begin{proof}[of Lemma~\ref{lem:recycling}]
{\em (i):} All eventually recycling prunings are finite.\\

Let $\nsts{\sys}$ be an eventually recycling pruning of concrete transition system $\ncts{\sys}$.
By virtue of being a pruning, $\nsts{\sys}$ is finitely branching. We show next that every simple path in $\nsts{\sys}$
has finite length, which together with finite branching implies finiteness by K\"{o}nig's Lemma.

Towards a contradiction, assume that there exists infinite simple run $\tau$ in $\nsts{\sys}$.
Since $\nsts{\sys}$ is eventually recycling, there is a finite prefix of $\tau$ such that all values occurring 
in $\tau$ occur also in this prefix. Therefore, $\tau$ contains only finitely many distinct values,
and hence only finitely many distinct states (databases of given schema over these values).
If $\tau$ has infinite length, then a pigeonhole argument contradicts the assumption that
$\tau$ is simple.\\

{\em (ii):} If $\ncts{\sys}$ is state-bounded, then it has an eventually recycling pruning.\\

Let $\nsts{\sys}$ be a pruning obtained from $\ncts{\sys}$ by picking the finite subset of evaluations
$SE \subseteq \returns^\H(s,\alpha,\sigma)$  as follows: if there is a run $\tau$ in $\nsts{\sys}$ from
$s_0$ to $s$ such that $\returns^\H(s,\alpha,\sigma )$ includes
at least one evaluation that recycles from $\tau$, then $SE$ contains exclusively recycling evaluations
(i.e. for each evaluation $\theta \in SE$, there is a run $\tau$ from $s_0$ to $s$ in $\nsts{\sys}$ such that
$\theta$ recycles from $\tau$). Otherwise, $SE$ is an arbitrary finite subset of 
$\returns^\H(s,\alpha,\sigma)$).\\

We prove that pruning $\nsts{\sys}$ is eventually recycling.
By definition, if $\ncts{\sys}$ is state-bounded then $|\adom{s}| \leq b$ for each state $s$, where $b$ is 
the size bound on the state. Assume towards a contradiction that 
$\nsts{\sys}$ contains a run  $\tau = s_0s_1s_2 \cdots$
that includes infinitely many states generated by evaluations that do not recycle from $\tau$.
It follows that there must exist a finite 
$k \geq 0$ such that $|\bigcup_{i=0}^{k}\adom{s_i}| > 3b$ and such that $\adom{s_{k+1}}$ contains at 
least one fresh value, i.e. $\adom{s_{k+1}} - \bigcup_{i=0}^{k}\adom{s_i} \not= \emptyset$. 
Let $\theta_{k+1} \in \returns^\H(s_k,\alpha,\sigma)$ be the
service call evaluation that generates $s_{k+1}$. Clearly $\theta_{k+1}$ does not recycle from $\tau$,
since it contains at least one fresh value in its range. 
However observe that, since the $k$-length prefix of $\tau$ contains at least $3b$ distinct values, 
this prefix contains at least $b$ values that are distinct from the values in 
$\adom{\idb} \cup \adom{s_k}$ (since by state-boundedness, $|\adom{\idb} \cup \adom{s_k}| \leq 2b$). 
Call the set of these values $\fresh$.

Also by state-boundedness, $\theta_{k+1}$ introduces at most $b$ fresh values.
Any one of the values in $\fresh$ can be used instead of the
fresh values introduced by $\theta_{k+1}$, to obtain another evaluation
$\theta_{k+1}^r$ that respects $\H$.
Hence $\theta_r$ witnesses an evaluation in $\returns^\H(s_k,\alpha,\sigma)$ that 
{\em does} recycle from $\tau$. 
But this contradicts the definition of $\nsts{\sys}$, which
mandates that $\theta_{k+1}$ be dropped in favor of $\theta_{k+1}^r$. 
\end{proof}

This result implies that if $\ncts{\sys}$ is state-bounded, then there exists a finite-state 
abstract transition system $\nsts{\sys}$ that is persistence-preserving bisimilar to $\ncts{\sys}$. 
Indeed, any eventually recycling pruning of $\ncts{\sys}$ can play the role of $\nsts{\sys}$ 
(it is finite by Lemma~\ref{lem:recycling}(i), it is bisimilar to $\ncts{\sys}$ by Lemma~\ref{lem:pruning}, 
and one is guaranteed to exist by Lemma~\ref{lem:recycling}(ii)).\\

\noindent
{\bf Construction of Eventually Recycling Pruning.\ }
The existence result in Lemma~\ref{lem:recycling}
is non-constructive and therefore does not yet yield decidability
of verification even if the concrete transition system $\ncts{\sys}$ is state-bounded.
We next present Algorithm \ERP, which is guaranteed to construct an eventually recycling pruning 
when its input \dcds\ is state-bounded, but which may diverge otherwise.\\

\begin{tabbing}
{\bf Algorithm} \ERP\\
{\bf Input:} \= $\sys = \tup{\dl,\pl}$, a \dcds with 
                data layer $\dl = \tup{\CONST,\schema,\EC,\idb}$\+\\ and 
                process layer $\pl=\tup{\FUNC,\aset,\rset}$.\-\\
\\
$\Sigma := \{ \idb \}$, $\trans := \emptyset$, $\used := \adom{\idb}$, $\visited := \emptyset$\\
rep\=\kill
{\bf repeat}\+\\
pick \= state $\I \in \Sigma$, action $\alpha$ and legal parameters $\sigma$\+\\
        such that $(\I,\alpha,\sigma) \notin \visited$\-\\
$\recyclable := \used - (\adom{\idb} \cup \adom{\I})$\\
pick    set $\fresh$ of $n$ service call results such that:\+\\ 
        $|\fresh| = n =  |\skolems{\doo{}{}{\I, \alpha,\sigma}}|$ and\\
        {\bf if} \= $|\recyclable| \geq n$\+\\ 
                    {\bf then} $\fresh \subseteq \recyclable$  \% recycled values\\ 
                    {\bf else} $\fresh \subset \CONST - \used$  \% fresh values\-\-\\
\\
$F := \adom{\idb} \cup \adom{\I} \cup \fresh$
\\
{\bf for} \= {\bf each} \= $\theta \in \groundexec{F}{\I, \alpha,\sigma}$ 
                        such that $\I_{next} \models \EC$\+\+\\
                        where $\I_{next} := \doo{}{}{\I, \alpha,\sigma}\theta$ {\bf do}\-\\ 
\begin{tabular}{rcl}
$\Sigma$   & $:=$ & $\Sigma \cup \{\I_{next}\}$\\
$\trans$   & $:=$ & $\trans \cup \{(\I, \I_{next})\}$\\
$\used$    & $:=$ & $\used \cup \adom{\I_{next}}$\\
$\visited$ & $:=$ & $\visited \cup \{(\I,\alpha\,\sigma)\}$ 
\end{tabular}
\-\\
{\bf end}\-\\
{\bf until} $\Sigma$ and $\trans$ no longer change.
\\
{\bf return} $\tup{\CONST, \schema, \Sigma, \idb, \Longrightarrow}$
\end{tabbing}

Observe that algorithm \ERP\ performs several nondeterministic choices in each iteration.
The particular choices (and their order) do not matter, by Theorem~\ref{thm:algo}.

\begin{proof}[of Theorem~\ref{thm:algo}]{\em (Sketch)}

First, we show that algorithm \ERP\ builds a pruning.
Items (i) and (iii) in the definition of pruning are trivially satisfied in every run of \ERP.
Item (ii) follows from the following claim:

\begin{quote}
{\bf Claim:} for any choice of $\fresh$ such that 
$|\fresh| \geq |\skolems{\doo{}{}{\I, \alpha,\sigma}}|$,
the set of equality commitments represented by the successors of $\I$ 
generated by the evaluations in $\groundexec{F}{I,\alpha,\sigma}$
coincides with the set of commitments represented by the successors of $\I$ in $\cts{\sys}$.
\end{quote}

Next, we show that if $\sys$ is state-bounded, every run of \ERP\ terminates. Indeed, state-boundedness
guarantees that in each iteration, only at most $b$ service call values are needed, where $b$
is the state size bound. But after running ``sufficiently'' long, \ERP\ variable $\used$
accumulates at least $3b$ distinct values. At each subsequent step of the algorithm,
there will therefore exist at least $b$ values distinct from the active domains of $\idb$ and
$\I$, so the pick of $\fresh$ will always recycle values
(observe that \ERP\ only picks evaluations from set $\nsts{\sys}$). $\used$ will no longer change, 
and therefore $\Sigma$ and $\Longrightarrow$ must eventually saturate (a key reason for this
is the bookkeeping of variable $\visited$, which avoids repeating the nondeterministic pick
for any combination of state, action and parameter instantiation $(\I,\alpha,\sigma)$).

Finally, since \ERP\ terminates, then it outputs a finite-state pruning, which is trivially
eventually recycling.
\end{proof}

Theorem~\ref{thm:algo} and Theorem~\ref{theorem:mulpers-bisimulation} directly imply 
Theorem~\ref{thm:sbounded-decid}.

\subsection{GR-Acyclic $\dcds$s}\label{app:gr-acyclic}
\label{sec:gr-appendix}
\begin{proof}[of Theorem~\ref{thm:sbound-check}]
For the proof, we reduce from the undecidable problem of checking if the run of a deterministic
Turing Machine is confined to a bounded-length segment of the tape (we say that the TM is tape-bounded). 
This in turn is undecidable by reduction from the halting problem: 
Given deterministic TM T, build TM T' such that T' is tape-bounded if and only 
if T halts. T' simulates T but also records on the tape the historical configurations of T. 
At each step, T' checks if the most recent configuration of T was seen in the history. 
If so, T' stops simulating T and enters a loop in which it keeps extending
the right end of its tape. It is easy to see that T' is tape-bounded if and only if T halts.

We reuse without change the reduction exhibited in the proof of 
Theorem~\ref{thm:ltl-undecid-det}. Recall that the reduction constructs for every 
Turing Machine {\em TM} a \dcds with deterministic services \sys that simluates the 
computation of {\em TM}.
We recall from the proof that the only service in the process layer,
service \new, is guaranteed to be called only with distinct arguments across distinct 
transitions, 
and so its behavior is unaffected by the choice of deterministic versus nondeterministic 
semantics. 
We also note that the state of the \dcds has size linear in the length of the tape segment visited
by {\em TM}, so tape-boundedness reduces to state-boundedness.
\end{proof}

\begin{proof}[of Theorem~\ref{thm:gracyclic-sbounded} (Sketch)] 
We prove the result by counting the maximum number of different values 
in a state of the transition system.

Since this task is undecidable (by Theorem~\ref{thm:sbound-check}), we necessarily have to 
approximate this value. The approximation is performed by analysing a different, much more abstract transition
system we call {\em dataflow} transition system (to distinguish from the abstract system that is bisimilar to the
concrete system).

The dataflow system is a \dcds obtained as follows from the dataflow graph and $\idb$:
For each node of the dataflow graph, there is a unary relation in the dataflow system, 
and for each normal (special) edge in the dataflow graph, there is a normal (special) transition in the
dataflow system between the corresponding relations. The schema of the dataflow system is a set of relation 
names with arity one, in correspondence to the nodes of the dataflow graph. A state of the dataflow system is an 
instantiation of its schema using values from the domain $\CONST$. 

For each term $t$ appearing in a relation in the initial
state of the concrete system, there is a term $t$ in the corresponding relation of the initial state of the dataflow 
system. Being in one state of the dataflow system, the next state is constructed as follows:

  \begin{itemize}
  \item for each normal transition from a relation $A$ to a relation
    $B$, for each term $t$ in the relation $A$ of the current state,
    there is a term $t$ in the relation $B$ of the next state.
  \item for each special edge from a relation $A$ to a relation $B$,
    for each term $t$ in the node $A$ of the current state, there is a
    fresh term $t'$ in node $B$ of the next state.
   \end{itemize}

   It is easy to see the following claim:

\begin{quote} {\bf Claim 1.\ } For any run $\tau$ of length $m \geq 0$ in the concrete system,
there is a run $\tau^d$ of length $m$ in the dataflow system, such that the size of the active domain
of state $\tau(i)$ is at most the size of the active domain of state $\tau^d(i)$.
\end{quote}
As a result, any state bound for the dataflow system also bounds the state of the concrete system.
We compute such a bound next.

Consider the dataflow graph of $\aset$. GR-acyclicity forces cycles with special edges to not be connected to any 
other cycles in the dataflow graph. More specifically, each connected component of the dataflow graph 
must have one of the following types:
  \begin{itemize}\parskip=0in\itemsep=0in
  \item[A:] A simple cycle $C$ (possibly with special edges),
    possibly connected with several directed acyclic graphs (DAG)s, such that 
    the component contains no additional cycle beyond $C$. 
  \item[B:] Several cycles $C_1,\ldots,C_m$ containing only normal edges, 
    each $C_i$ possibly connected to several DAGs, such that the component contains no 
    other cycle beyond the $C_i$'s,  and there is no path with special
    edges connecting two cycles $C_i,C_j$.
    \item[C:] A DAG, possibly containing normal and special edges.
  \end{itemize}
  Denote with
\begin{itemize}\parskip=0in\itemsep=0in
\item[d:] the longest path of the dataflow graph after deleting
  the cycles,
\item[b:] the maximum number of special edges going out
  of a node of the dataflow graph plus one, and 
\item[n:] the number of nodes of the dataflow graph. 
\end{itemize}
It is easy to see that in each transition of the dataflow system, 
for each term in the current state, there can be at most $n\cdot b$ 
distinct terms in the next state.

First, consider the components of type A. Call the DAGs $D$ connected to
the unique cycle $C$ via edges from $D$ to $C$, {\em input DAGs}.
Call {\em output DAGs} the DAGs connected via edges from $C$ to $D$.
It is easy to see that after $d$ transition steps, in any run of the dataflow system
there is no term in any relation of an input DAG (all have been forgotten), 
and at most 
$$m := |\adom{\idb}| + n\cdot b \cdot |\adom{\idb}| + \dots + n^d\cdot b^d \cdot |\adom{\idb}|$$ 
distinct terms may co-exist within the relations of the cycle. 
Moreover, after $d$ steps, the total number of distinct terms in the cycle will no 
longer increase in any run suffix starting from step $d+1$. 
Consider now how the $m$ terms can be copied into the output DAGs.
It is again easy to see that there can be at most $n^d\cdot b^d \cdot m$
distinct terms in any relation of an outgoing DAG. As a result, in any step at most 
$n^{d+1}\cdot b^d \cdot m$ distinct terms may co-exist within a type A component.

Second, consider the components of type B. 
A similar argument yields at most $n^{d+1}\cdot b^d \cdot m$ different terms that may co-exist
within a type B component.

Third, it is easy to see that there can be at most $n^d\cdot b^d \cdot |\adom{\idb}|$ different terms 
within a type C component.

All in all, at most $|\adom{\idb}|\cdot n^{2d+1}\cdot b^{2d}$ distinct terms may co-exist 
in a state of the concrete transition system.
\end{proof}

%% file: appendix-discussion.tex
\section{Discussion}

\begin{proof}[of theorem~\ref{thm:det-2-nondet} (Sketch)]
The technical problem here is to force the results of nondeterministic service
calls to conform to historic evaluations. 

Let $D$ be a deterministic \dcds. We rewrite $D$ to obtain a new \dcds $N$ whose semantics
under nondeterministic services coincides with that of $D$ under deterministic services.
For each term $f(a_1,\ldots,a_n)$ appearing in some effect of $D$ 
$e := \map{q^+\land Q^-}{E}$, we rewrite $D$ as follows. 
We extend the schema with a new $n+1$-ary relation $R_f$.
Intuitively, $R_f(a_1,\ldots,a_n,r)$ states that the call $f(a_1,\ldots,a_n)$ evaluates to $r$. 
We extend the effect to record this fact, replacing $e$ with :
$e' := \map{q^+\land Q^-}{E \land R_f(a_1,\ldots,a_n,f(a_1,\ldots,a_n))}$.
To ensure that $R_f$ records all past calls of $f$, we add to each action an effect
that simply copies $R_f$.
We also add the functional dependency $a_1,\ldots,a_n \ra r$ on $R_f$. Notice that any attempt to record a service 
call with a result distinct from a past invocation violates the
functional dependency and the transition does not occur. 
It is easy to see that, if we project the states of $\cts{N}$ on the schema of $D$, we obtain $\cts{D}$.
\end{proof}

\begin{proof}[of Theorem~\ref{thm:nondet-2-det} (Sketch)]
The challenge here lies in forcing a deterministic service $f_d$ to return possibly distinct 
results for same-argument calls of the nondeterministic service $f_n$ it corresponds to.

The trick is to call $f_d$ with one additional argument, which plays the
role of a timestamp, where each state in the run has its own unique timestamp. 
This way same-argument calls of $f_n$ at distinct steps in the run correspond to distinct-argument 
calls of $f_d$, which therefore simulates the desired nondeterministic behavior. 

An additional trick is used to get the run to generate a sequence of unique values to act as 
timestamps. We use a deterministic service $\mathit{new}(x)$ to generate a new timestamp, 
we record the successor relation over timestamps in binary relation $\mathit{succ}$, and the 
most recent timestamp in unary relation $\mathit{now}$.
We add to each action
\begin{itemize}
\item
the effect 
$$\map{\mathit{now}(x)}{\mathit{now}(\mathit{new}(x)) \land \mathit{succ}(x,\mathit{new(x)})}$$
which extends the successor relation by one timestamp and sets the new timestamp as
most recent; and
\item
the effect 
$$\map{\mathit{succ}(x,y)}{\mathit{succ}(x,y)}$$ 
which accumulates the historical {\em succ} entries.
\end{itemize}
We are not quite done, as we still need to ensure that {\em succ} induces a linear order
on the collection of timestamps generated during the run. For this purpose we employ the same trick
as the proof of Theorem~\ref{thm:ltl-undecid-det}. We describe it below for the sake of 
proof self-containment.

Observe that, by definition of the effect extending {\em succ}, at each step,
the generated timestamp has at most one successor.

However, if the call $\mathit{new}(x)$ returns a previously seen timestamp, 
then there can be some timestamp with several predecessors  in {\em succ}. 
We rule out this case by
declaring the second component of {\em succ} to be a key.
It follows that {\em succ} must be either 
(i) a linear path over timestamps (possibly
starting from a source node that has a self-loop), or 
(ii) must contain a simple cycle involving more than one timestamp. 
The simple cycle is created when {\em new} returns the minimal element of the {\em succ} relation.

We wish to force case (i). To rule out case (ii), we proceed as follows: 
we initialize {\em succ} to contain a source node $0$ that can never be a timestamp
because it cannot be returned by 
{\em new} without violating the key constraint on {\em succ}. 
To this end, we initialize {\em succ} in \idb to 
$\mathit{succ}^{\idb} = \{(0,0),(0,1)\}$
and $\mathit{now}^{\idb} = \{1\}$. Notice that $\mathit{succ}^{\idb}$ has type (i).
An easy induction shows that every run prefix must also construct a {\em succ} relation of 
type (i), since any attempt to extend {\em succ} with an edge
back to one of its existing nodes violates the key constraint. 
It also follows easily that during the run, {\em now} takes values from the linear 
path starting at $1$, and never includes $0$.
\end{proof}

%% file: appendix-example.tex
\section{Example \dcds:\\ Travel Reimbursement System}
\label{ex:travelReimburseSystem} 

We model the process of reimbursing travel expenses in a university,
and the corresponding audit system, in two different subsystems. In
particular, the first subsystem, called the \emph{request system}
manges the submission of reimbursement requests by an employee, and
preliminary inspection and approval of the request by an employee in
the accounting department (we shall call her the \emph{monitor}).  A
log of accepted requests will be submitted to the second subsystem,
the \emph{audit system}, in which requests can be accumulated, and they
can be checked for accuracy by calling external web services (for
instance to obtain the exchange rate from foreign currency to USD on a
past date, or to check that the employee actually was on the declared
flight).

\medskip
\noindent {\bf Request system.} \label{ex:RequestManagement} To keep
the example simple we model a travel reimbursement request as being
associated to the name of the requester, and infor- mation related to
the corresponding flight and hotel costs. 
  After a request is submitted, a monitor will check the request and will decide to
accept or reject the request. If a request is rejected, the employee
needs to modify the information regarding hotel and flight, while employee
name will not be changed while updating. After the update by the
employee, the monitor will again check the request, and the
reject-check loop continues until the monitor accepts the
request. After a request is accepted a log of the request will be sent
to the audit system, and the request system will be ready to process
the next travel request.

We model the request system by a DCDS $\S_R=\tup{\dl,\pl}$, in which 
$\dl = \tup{\CONST,\schema,\EC,\idb}$ such that $\CONST$ is a countably
infinite set of \emph{constants},
$\idb=\set{\eSysStatus(\eRFR),\true}$, and $\schema$ is a database
schema as follows:
   \begin{itemize}
\item $\eSysStatus=\tup{\eStatus}$, a unary relation that
keeps the
      state of the request subsystem, and can take three different
      values: \eRFR, \eRTV, and \eRTU,
    \item $\eTravel=\tup{\eClientName}$, holding the name of
      the employee;
        \item $\eHotel=\tup{\eHotelName, \eDate, \ePrice, \eCurrency,
        \eUSDPrice}$, holding the hotel cost information of the
      employee's travel, which might have
      been paid in some other currency than USD,
     \item $\eFlight=\tup{\eDate, \eFNumber, \ePrice,
         \eCurrency,  \eUSDPrice}$,
         holding the flight cost information.
       \end{itemize}

     The process layer is defined as $\pl=\tup{\FUNC,\aset,\rset}$ where \FUNC is a set of the following nondeterministic service
     calls, each modeling an input of an external value by the
     employee. Specifically,
    \begin{itemize}
    \item \escClientName: models the input of the employee name
      (filled in by the employee),
     \item \escHName: hotel name,
     \item \escHDate: arrival date,
     \item \escHPricec: sum paid to the hotel (possibly in foreign
       currency),
       \item \escHCurrency: currency exchange rate at that date,
       \item \escHPUSD: amount paid to the hotel in USD,
       \item \escFDate: flight date,
       \item \escFNumber: flight number,
       \item \escFprice: ticket price, possibly in foreign currency,
       \item \escFlightCurrency: currency exchange rate at date of
         ticket payment,
       \item \escFPUSD: ticket price in USD.
    \end{itemize}
    There is one additional service.
   \begin{itemize}
  \item \escMakeDecision{}: a nondeterministic service
     modeling the decision of the human monitor. It returns \eRC\ if the request is accepted, and
     returns \eRTU\ if the request needs to be updated by the
     employee.
 \end{itemize}

 The set $\aset$ of actions includes (among others):\\

$\eAReimburseRequest:$
\begin{align*}
\true &\rsa \eSysStatus(\eRTV)\\
\true &\rsa \eTravel(\escClientName)\\
\true  &\rsa \eHotel(\begin{array}[t]{@{}l}
           \escHName,\\
           \escHDate, \\ 
           \escHPricec, \\
           \escHCurrency, \\
           \escHPUSD)
\end{array}\\
\true   & \rsa  \eFlight(\begin{array}[t]{@{}l}
  \escFDate,\\
           \escFNumber,\\
           \escFprice, \\
           \escFlightCurrency, \\
          \escFPUSD)
\end{array}
\end{align*}

$\ePriliminaryCheck$:
\begin{align*}
\true & \rsa  \eSysStatus(\escMakeDecision{})\\
\eTravel(n) &\rsa  \eTravel(n)\\
\eHotel(x_1,\dots, x_5)& \rsa  \eHotel(x_1,\dots, x_5)\\
\eFlight(x_1,\dots, x_5)& \rsa  \eFlight(x_1,\dots, x_5)
\end{align*}

$\eUpdateRequest$:
\begin{align*}
\true&\rsa  \eSysStatus(\eRTV)\\
\eTravel(n) &\rsa  \eTravel(n)\\
\true  &\rsa \eHotel(\begin{array}[t]{@{}l}
           \escHName,\\
           \escHDate, \\ 
           \escHPricec, \\
           \escHCurrency, \\
           \escHPUSD)
\end{array}\\
\true   & \rsa  \eFlight(\begin{array}[t]{@{}l}
  \escFDate,\\
           \escFNumber,\\
           \escFprice, \\
           \escFlightCurrency, \\
          \escFPUSD)
\end{array}
\end{align*}

$\eAcceptRequest$:
\begin{align*}
\eSysStatus(\eRC) &\rsa \eSysStatus(\eRFR) \\
\end{align*}

When a request is initiated (modeled by the action
$\eAReimburseRequest$), (i) the system changes state ``to waiting for
verification'', (ii) a travel event is
       generated and the employee fills in his name, (iii) the employee fills in all hotel information,
and (iv) the employee fills in all flight information. 

Action $\ePriliminaryCheck$ models the preliminary check by the
monitor. Travel event, hotel and flight information are unchanged, but
the system status is set by the non-deterministic service call
$\escMakeDecision{}$, which models the monitor's decision for current
active travel information.

If the monitor rejects, then she sets the next state to $\eRTU$, which
will trigger the action $\eUpdateRequest$, which in turn collects once
again the hotel and flight information from the employee, and moves
the status to $\eRTV$.

Finally, action $\eAcceptRequest$ returns the system in the state $\eRFR$, in which it
is ready to accept a new request.

Notice the use of the always true predicate $\true$, with the evident
meaning.  A convenient way to model its meaning in the \dcds framework
is to think of it as a nullary relation, initialized to contain the
empty tuple, which is copied in perpetuity by each action ($\true$
never changes its interpretation). We omit the corresponding copy
effects, treating them as built-in.

Notice how the condition-action rules in the set \rset below guard the actions by the current
state of the system:
\[
\begin{array}{rcl}
 $\eSysStatus(\eRFR)$&\mapsto&$\eAReimburseRequest$\\
 $\eSysStatus(\eRTV)$&\mapsto&$\ePriliminaryCheck$\\
 $\eSysStatus(\eRTU)$&\mapsto&$\eUpdateRequest$\\
 $\eSysStatus(\eRC)$& \mapsto&$\eAcceptRequest$\\
\end{array}
\]

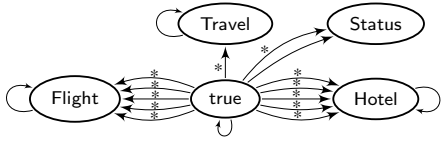
\begin{figure}[t]
\centering
\begin{tikzpicture}[node distance=1cm,font=\scriptsize, 
  every node/.style={line width=0.25mm, outer sep=0.3mm}, 
  ultra thin,>=latex']

\def \dec{decoration={markings,mark=at position 1 with {\arrow[ultra thick]{>}}},
    postaction={decorate}
}
\node[ellipse,draw](live){$\true$};
  
\node [ellipse,draw,right of=live,yshift=0cm,xshift=1cm] (hotel) {\eHotel}; 

\node [ellipse,draw,left  of=live,yshift=0cm,xshift=-1cm] (flight) {\eFlight}; 

\node [ellipse,draw,above  of=live, yshift=0cm] (travel) {\eTravel }; 

\node [ellipse,draw,right  of=travel,xshift=1cm] (sysStatus){\eSysStatus};

\path
(live) edge [->]node[auto,xshift=3,yshift=-3]{*} (travel)

(live) edge [->]node[auto,yshift=-5] {*} (hotel)
(live) edge [->,bend left=10] node[auto,yshift=-5] {*} (hotel)
(live) edge [->,bend left=-10] node[auto,yshift=-5] {*} (hotel)
(live) edge [->,bend left=20] node[auto,yshift=-5] {*} (hotel)
(live) edge [->,bend left=-20] node[auto,yshift=-5] {*} (hotel)

(live) edge [->]node[auto,yshift=7] {*} (flight)
(live) edge [->,bend left=10] node[auto,yshift=7] {*} (flight)
(live) edge [->,bend left=-10] node[auto,yshift=7] {*} (flight)
(live) edge [->,bend left=20] node[auto,yshift=7] {*} (flight)
(live) edge [->,bend left=-20] node[auto,yshift=7] {*} (flight)

(live) edge [->, bend left=20]node[auto,yshift=-8] {*} (sysStatus)
(live) edge [->, bend left=10]node[auto] {} (sysStatus)


(live) edge [loop below,->,looseness=5] (live)
(travel) edge [loop left ,->,looseness=4] (travel)
(hotel) edge [loop right ,->,looseness=4] (hotel)
(flight) edge [loop left ,->,looseness=4] (flight);
\end{tikzpicture}

\vspace{-2ex}
\caption{Dataflow graph for request system}
\label{fig:travelReimburseNonDet}
\vspace{-2ex}
\end{figure}

The dataflow graph corresponding to the request system
  is depicted in
  Figure~\ref{fig:travelReimburseNonDet}, where special edges are starred.  Notice
  that there can be multiple special edges between the same two nodes
  (these are distinguished by unique edge ids, which we omit in the
  figure to avoid clutter). 

In particular, the group of special edges
  from the $\true$ node to the $\eHotel$ node corresponds to the
  action of employee filling in the
hotel information, modeled by calls to such services as $\escHName$.
Similarly for the special edges from $\true$ to 
$\eFlight$.

The special edge between  $\true$ and $\eTravel$ is due to the employee
filling in his name into the created travel request. 
The special edge from $\true$ to $\eSysStatus$ reflects the monitor's
insertion of her decision (see the call to $\escMakeDecision{}$ in the
first effect of action $\ePriliminaryCheck$ in
Example~\ref{ex:travelReimburseSystem}), while the normal edge
corresponds to change of the status without calling a service (this
happens in other actions).
The self-loops on $\eFlight$, $\eHotel$, and $\eTravel$ are
due to the remaining (copy) effects of $\ePriliminaryCheck$. The
self-loop on node $\true$ is due to the modeling of this value by a
singleton nullary relation containing the empty tuple, which keeps being copied in
   each action.




An inspection of this dataflow graph reveals that the request system
is not GR-acyclic, since it contains several instances of two simple
cycles connected by a path that includes a special edge. For instance,
the path $\pi$ comprised of the self-loops around $\true$ and
$\eTravel$, and the special edge beetween them.  However,
the request system is GR$^+$-acyclic.  To illustrate this, notice that the
path $\pi$ 
is allowed by
GR$^+$-acyclicity because the special edge leading into the $\eTravel$
loop is caused by action $\eAReimburseRequest$, while all the
subsequent edges in $\pi$ are caused by other actions (in this case
there is only one subsequent edge in $\pi$, namely the
self-loop on $\eTravel$, caused by actions $\ePriliminaryCheck$ and
$\eUpdateRequest$).

We illustrate some $\mulpers$ properties 
pertaining to the proper operation of the request system:

%

A property of interest is that once initiated, a request will eventually be decided by the
  monitor, and the decision can only be $\eRTU$ or $\eRC$ (a liveness
  property). We show the property in the easier-to-read CTL syntactic sugar:
\begin{align*}
{\bf A G} (\forall n\ & \eTravel(n) \limp\\
                      & {\bf A} (\eTravel(n) {\bf U} (
                      \begin{array}[t]{@{}l}
                          \eSysStatus(\eRTU) \lor{}\\
                          \eSysStatus(\eRC))
                     \end{array}
 \end{align*}

 The until operator {\bf U} (for this example, it is the strong
 flavor, in which $\psi {\bf U} \phi$ means that $\phi$ is guaranteed
 to eventually hold, and until it does $\psi$ must hold in every
 step). We note that for a property to belong to \mulpers, it must
 require the bindings of quantified variables to be continuously live
 between the step when the quantification was evaluated and the step
 when the variable is used. This can be done by using $\inadom$ or by
 using any relation, in our example $\eTravel$. The \mulpers version
 of the property is given below:
\[
  \begin{array}[t]{@{}l}
    \nu X.(\forall n. \eTravel(n) \limp{}\\
    \quad \mu Y. (
    \begin{array}[t]{@{}l}
      \eSysStatus(\eRTU) \lor  \eSysStatus(\eRC)\\
      {}\lor \BOX{(\eTravel(n)\land Y)})) \land \BOX{X}
    \end{array}
  \end{array}
\]

Another property of interest is that if the flight cost is not specified, then the request is not
  accepted (a safety property). We use the special constant $\bot$ to
  denote a null value (this need not be treated specially in the
  semantics, any value of the domain can be reserved for this
  purpose):

\begin{tabbing}
${\bf G} \neg($ \= $\eSysStatus(\eRC) \land$\+\\ 
                   $\exists x_1,\dots, x_4\ \eFlight(x_1,x_2, \bot, x_3,x_4) )$.
\end{tabbing}

The \mulpers version is given below:
\[
\begin{array}{r@{}l}
\nu X.\{ \neg(&\eSysStatus(\eRC) \land{}\\
  & \exists x_1,\dots, x_4.
\eFlight(x_1,x_2, \bot, x_3,x_4) )\} \land \BOX{X}\ 
\end{array}
\]

\medskip
\noindent {\bf Audit system.} \label{ex:AuditSubsystem} After a
request is verified by the monitor in the request system, it will be
migrated to the audit system. The migration is performed by a logging
subsystem which might perform such operations as:
we extend each travel event with a freshly generated travel id, which guarantees uniqueness across the entire history of requests.
We store these tuples in a database.  We can model this migration
using the \dcds formalism, but we omit the specification and focus
directly on the audit system.

More specifically, we model the audit system by a \dcds
$\S_A=\tup{\dl_A,\pl_A}$, in which $\dl_A =
\tup{\CONST,\schema,\EC,\idb}$. $\CONST$ is a countably infinite set
of \emph{constants}. $\schema$ is a database schema as follows:
  \begin{itemize}
  \item $\eSysStatus=\tup{\eStatus}$ is a unary relation keeping the
    state of request subsystem, which can take two different values:
    \ePhaseC, and \ePhaseD, whose role is to sequence the actions of
    the audit system appropriately.

  \item $\eTravel=\tup{\underline{\eId}, \eClientName,\ePassed}$
    extends the homonymous relation of the request system with two
    fields: $\eId$ (the travel identifier), and $\ePassed$, which will be set
    by the audit system to reflect whether both the hotel and the flight
    price checks succeed.
  \item $\eHotel=\tup{\eTrId, \eHotelName, \eDate, \ePrice,
      \eCurrency, \eUSDPrice, \\ \ePassed}$, where $\eTrId$ is a
    foreign key to the travel id and $\ePassed$ is set by the
    audit system to reflect whether the claimed price and the
    calculated price match.
  \item $\eFlight=\tup{\eTrId, \eFNumber, \eDate, \ePrice,
      \eCurrency, \eUSDPrice, \\ \ePassed}$,  where $\eTrId$ and
    $\ePassed$ are analogous to
    the ones in the $\eHotel$ relation.
    \end{itemize}


     Finally, $\idb$ is the output of the logging subsystem to which
     we add the fact  $\eSysStatus(\ePhaseC)$, to initialize the audit
     system status.

The process layer is defined as $\pl=\tup{\FUNC,\aset,\rset}$ in which
\FUNC contains a deterministic service, where the call
$\escConvertAndCheck{(\mathit{price},\mathit{currency},\mathit{date},\mathit{priceInUSD})}$
perfoms the official exchange rate acquisition and computation
described above, returning true if and only if the claimed price and
the computed one match.

$\aset=\set{\eCheckPrice, \eCheckTravel}$ includes the following actions.\\

\noindent
$\eCheckPrice:$
\begin{align*}
&\true\rsa \eSysStatus(\ePhaseD)\\\\
&\eTravel(i,n,v)\rsa \eTravel(i,n,v)\\\\
&\eHotel(x_1, x_2, \eVarDate, \eVarPrice, \eVarCurrency, \eVarPriceInUSD, x_7)\rsa \\ 
&\eHotel(x_1, x_2, \eVarDate, \eVarPrice, \eVarCurrency, \eVarPriceInUSD, \\ 
              &\quad\escConvertAndCheck(\eVarDate, \eVarPrice, \eVarCurrency,\eVarPriceInUSD))\\\\
&\eFlight(x_1, x_2, \eVarDate, \eVarPrice, \eVarCurrency, \eVarPriceInUSD, x_7)\rsa\\
&\eFlight(x_1, x_2, \eVarDate, \eVarPrice, \eVarCurrency, \eVarPriceInUSD,\\
&\quad\escConvertAndCheck(\eVarDate, \eVarPrice, \eVarCurrency, \eVarPriceInUSD))
\end{align*}

\noindent
Notice that the first effect changes the audit system's state to enter
the stage in which the the two checks (for hotel and flight) are
combined.  The second effect simply copies the request information.
The third and fourth each check the claimed price (for hotel,
respectively flight),
performing the conversion described above.\\

The second action works on the result of the first (this is ensured by the appropriate status changes).\\

\noindent
$\eCheckTravel$:
\begin{align*}
\true&\rsa  \eSysStatus(\ePhaseC)\\\\
\eTravel(x_1,x_2,x_3) \land{} \\
\eHotel(x_1, y_1 \dots, y_5, p_h) \land{} &\\
\eFlight(x_1, z_1 \dots, z_7, p_f) \land  \neg (p_h \land p_f) &\rsa  \eTravel(x_1, x_2, \false) \\\\
\eTravel(x_1,x_2,x_3) \land{} & \\
\eHotel(x_1, y_1 \dots, y_5, \true) \land{} &\\
\eFlight(x_1, z_1 \dots, z_7, \true) &\rsa \eTravel(x_1, x_2, \true)\\\\
\eHotel(x_1, \dots,  x_7)&\rsa  \eHotel(x_1, \dots, x_7)\\\\
\eFlight(x_1, \dots, x_7)&\rsa  \eFlight(x_1,\dots, x_7)
\end{align*}

                 Notice that the second and third effects set the
                 \ePassed\ field for the request, computed as the
                 conjunction of the corresponding fields set by the
                 price check on flight and hotel.

The process \rset is defined as follows:
\[\begin{array}{c}
  \carule{\eSysStatus(\ePhaseC)}{\eCheckPrice}\\
 \carule{\eSysStatus(\ePhaseD)}{\eCheckTravel}
\end{array}
\]

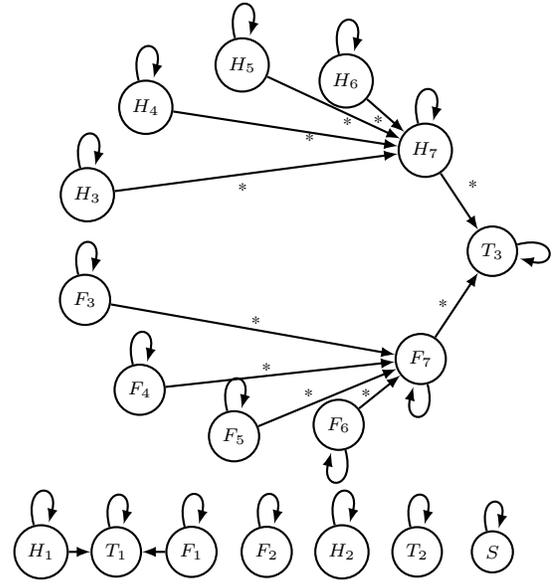
\begin{figure}[thb]
\centering
\begin{tikzpicture}[node distance=2.3 cm, >=latex,font=\scriptsize,
  rounded corners=3mm,thick]

\path (5*32.5:2.5cm) node [draw,shape=circle] (H3) {$H_3$};
\path (4*32.5:2.5cm) node [draw,shape=circle] (H4) {$H_4$};
\path (3*32.5:2.5cm) node [draw,shape=circle] (H5) {$H_5$};
\path (2*32.5:2.5cm) node [draw,shape=circle] (H6) {$H_6$};
\path (1*32.5:2.5cm) node [draw,shape=circle] (H7) {$H_7$};

\path (0:3cm) node [draw,shape=circle] (T3) {$T_3$};

\path (6*32.5:2.5cm) node [draw,shape=circle] (F3) {$F_3$};
\path (7*32.5:2.5cm) node [draw,shape=circle] (F4) {$F_4$};
\path (8*32.5:2.5cm) node [draw,shape=circle] (F5) {$F_5$};
\path (9*32.5:2.5cm) node [draw,shape=circle] (F6) {$F_6$};
\path (10*32.5:2.5cm) node [draw,shape=circle] (F7) {$F_7$};

\path (-3,-4) node [draw,shape=circle] (H1) {$H_1$};
\path (-2,-4)  node [draw,shape=circle] (T1) {$T_1$};
\path (-1,-4) node  [draw,shape=circle](F1) {$F_1$};
\path (0,-4) node [draw,shape=circle] (F2) {$F_2$};
\path (1,-4)  node  [draw,shape=circle](H2) {$H_2$};
\path (2,-4) node  [draw,shape=circle](T2) {$T_2$};
\path (3,-4) node  [draw,shape=circle](S) {$S$};

\path
(S) edge [->,loop above] (S)

(H1) edge [->,loop above] (H1)
(H2) edge [->,loop above] (H2)
(H3) edge [->,loop above] (H3)
(H4) edge [->,loop above] (H4)
(H5) edge [->,loop above] (H5)
(H6) edge [->,loop above] (H6)
(H7) edge [->,loop above] (H7)

(F1) edge [->,loop above] (F1)
(F2) edge [->,loop above] (F2)
(F3) edge [->,loop above] (F3)
(F4) edge [->,loop above] (F4)
(F5) edge [->,loop above] (F5)
(F6) edge [->,loop below] (F6)
(F7) edge [->,loop below] (F7)

(T1) edge [->,loop above] (T1)
(T2) edge [->,loop above] (T2)
(T3) edge [->,loop right] (T3);

\path
 (H3) edge [->]node[auto,yshift=-12]{*} (H7)
 (H4) edge [->]node[auto,yshift=-10,xshift=4]{*} (H7)
 (H5) edge [->]node[auto,yshift=-12]{*} (H7)
 (H6) edge [->]node[auto,yshift=-8,xshift=-8]{*} (H7)
 (H7) edge [->]node[auto]{*} (T3)
(F3) edge [->]node[auto,yshift=-3,xshift=-4]{*} (F7)
 (F4) edge [->]node[auto,yshift=-4]{*} (F7)
 (F5) edge [->]node[auto,yshift=-5,xshift=-2]{*} (F7)
 (F6) edge [->]node[auto,yshift=-7]{*} (F7)
 (F7) edge [->]node[auto,yshift=-6]{*} (T3)
(H1) edge [->] (T1)
(F1) edge [->] (T1);
\end{tikzpicture}
\caption{Weakly-acyclic dependency graph of the audit system}
\label{fig:travelRequestDet}
\end{figure}

 The corresponding dependency graph is as shown in
 Figure~\ref{fig:travelRequestDet}. In this picture nodes correspond
 to the positions of the schema. To avoid clutter, we represent each
 relation by its first letter, and denote the position number with a
 subscript. For instance, $T_1$ stands for the first ($\eId$) position
 of the relation $\eTravel$, and $S$ stands for the only position of
 the relation $\eSysStatus$. Moreover, the edges without label
 represent regular edges in the dependency graph, while the starred
 edges depict special edges. For instance, the edge $F_5
 \stackrel{*}{\longrightarrow} F_7$ is introduced due to the fourth
 effect of action \eCheckPrice. It is starred because it reflects the
 service call of $\escConvertAndCheck$, taking as argument the
 currency attribute of \eFlight (at position $5$), and storing its
 result in an \eFlight tuple at position $7$ (the \ePassed
 attribute).

 An inspection of the dependency graph reveals that the audit system
 is weakly acyclic, since there is no cycle including a special edge.

We illustrate a desirable property of the audit system: it guarantees
that the request cannot pass the audit if one of the flight or hotel checks fail:
\begin{align*}
{\bf 
A\ G} ( &\exists i,n,v,x_2,\ldots,x_6.  \eTravel(i,n,v) \land{}\\
                                & \quad (\eHotel(i,x_2,\ldots,x_6,\false) \lor \eFlight(i,x_2,\ldots,x_6,\false))\\
                                &\qquad\limp {\bf F}\  \eTravel(i,n,\false) )
\end{align*}
The \muladom version of the property is given below:
\begin{align*}
\nu X.(&\exists i,n,v,x_2,\ldots,x_6.  \eTravel(i,n,v) \land{}\\
                                &\quad (\eHotel(i,x_2,\ldots,x_6,\false) \lor \eFlight(i,x_2,\ldots,x_6,\false))\\
                              &\qquad \limp \mu Y. (\eTravel(i,n,\false) \lor
                              \DIAM{Y} ) ) \land \BOX{X}
\end{align*}

Notice that, since the audit system uses deterministic services, if we
wish to verify it in isolation from the other subsystems, we can
verify an \muladom property, which is what the above is (we are not
enforcing the liveness of the variables $i,v,n$ between the step at
which the quantification was evaluated, and the eventual step when the
\ePassed attribute of \eTravel is set to $\false$).

Recall however from Section~\ref{sec:discussion} that we can verify
mixed semantics \dcds by reduction to non-deterministic services. If
we wished to verify the above property over the collection of
subsystems, we would have to express it as an \mulpers property.  This
is easily done using an until operator {\bf U} (as illustrated above for the
request system).  Moreover, it is
actually compatible with our expectation about the system's operation:
while a request is being audited, we expect it to persist in the
system.